%% file: HIinOWLS_ref.tex
\documentclass[useAMS,usenatbib]{mn2e}

\usepackage{pict2e}
\usepackage{graphics}
\usepackage{rotating}
\usepackage{epsfig}
\usepackage{color}
\usepackage{multirow}
\usepackage{float}
\usepackage{array}

\newcommand{\hMsol}{\, h^{-1}{\rm M_{\odot}}}
\newcommand{\hMpc}{\, h^{-1}{\rm Mpc}}
\newcommand{\hkpc}{\, h^{-1}{\rm kpc}}

\newcommand{\SSNO}{\rm SS_{None}}
\newcommand{\SSDLA}{\rm SS_{DLA}}
\newcommand{\SSALF}{\rm SS_{ALFALFA}}

\DeclareGraphicsExtensions{.eps}

\input{jdefs}

\title[States of hydrogen in galaxies]{Modelling neutral hydrogen in galaxies using cosmological hydrodynamical simulations}
\author[A. R. Duffy et al.]
{Alan R. Duffy$^{1}$, Scott T. Kay$^{2}$, Richard A. Battye$^{2}$, C. M. Booth$^{3}$,\newauthor  Claudio Dalla Vecchia$^{4}$, Joop Schaye$^{3}$ \\
$^1$ICRAR, University of Western Australia, WA 6009, Australia, \\
$^2$Jodrell Bank Centre for Astrophysics, Alan Turing Building, The University of Manchester, M13 9PL, U.K.\\
$^3$Leiden Observatory, Leiden University, PO Box 9513, 2300 RA Leiden, The Netherlands \\
$^4$Max Planck Institute for Extraterrestial Physics, Giessenbachstra\ss e, D-85748 Garching, Germany}
\begin{document}

\date{19/09/2011}

\pagerange{\pageref{firstpage}--\pageref{lastpage}} \pubyear{2011}

\maketitle

\label{firstpage}

\begin{abstract}
The characterisation of the atomic and molecular hydrogen content of high-redshift galaxies is a major 
observational challenge that 
will be addressed over the coming years with a new generation of radio telescopes. 
We investigate this important issue by considering the states of hydrogen across a range of 
structures within high-resolution cosmological hydrodynamical simulations. Additionally, our
simulations allow us to investigate the sensitivity of our results to numerical resolution and to
sub-grid baryonic physics (especially feedback from supernovae and active galactic nuclei).
We find that the most significant uncertainty in modelling the neutral hydrogen distribution arises from our need to 
model a self-shielding correction in moderate density regions. Future simulations incorporating radiative transfer 
schemes will be vital to improve on our empirical self-shielding threshold.
Irrespective of the exact nature of the threshold we find that while the atomic hydrogen mass function evolves only mildly from redshift two to zero,
the molecular hydrogen mass function increases with increasing redshift, especially at the high-mass end. 
Interestingly, the weak evolution of the neutral hydrogen mass function is insensitive to the feedback scheme utilised,
but the opposite is true for the molecular gas, which is more closely associated with the star formation in the simulations.
\end{abstract}


\section{Introduction}
\label{sec:Introduction}
{Approximately 75 per cent of the baryonic material in the
Universe is hydrogen, so understanding and being able to 
reproduce the distribution of its various phases is a crucial step in 
constructing a complete theory of galaxy formation. After re-ionisation
completed at $z \simeq 6$, the majority of the hydrogen is expected to 
reside in a diffuse, ionised state outside of galaxies. However, it
is also believed that significant reservoirs of neutral (atomic and molecular)
hydrogen must exist, particularly at high-redshift, as this is the basic fuel
for star formation. 

At higher densities than the diffuse, ionised intergalactic medium surrounding the galaxies, large-scale gas clouds are 
visible in absorption as the Lyman-$\alpha$ forest. The latter can 
be probed by observations of intervening low column density HI structures in 
the spectra of quasars found, for example, in the Sloan Digital Sky Survey~\citep[e.g.][]{Prochaska:10}. 
As we probe ever closer to the galaxy disks the gas density increases, along with the HI surface density,
allowing us to directly probe the emission from neutral hydrogen. 
While a challenging observation, it greatly expands our understanding of 
galaxy formation on scales at which the galaxies themselves reside. For these reasons, great efforts have been made 
in surveying the emission of HI in the local Universe, such as the HI Parkes All 
Sky Survey, HIPASS~\citep{Barnes:01, Meyer:04} and the HI Jodrell All Sky 
Survey, HIJASS~\citep{HIJASS}.
The currently ongoing Arecibo Legacy Fast ALFA survey, ALFALFA~\citep{ALFALFA}, is 
producing the deepest all sky HI maps of the local Universe ever created~\citep[e.g.][]{Martin:10}.

The gas that resides inside the galaxy disk forms a multiphase medium in which HI gas forms $\rm H_2$ 
in the densest regions. The Atacama Large Millimeter Array (ALMA) promises 
to reveal an unprecedented wealth of information $\rm H_2$ an important tracer of star formation.
Radiation from young stars which form in these regions give rise to ionised bubbles of HII, emitting intense Lyman-$\alpha$ radiation, 
notably surveyed in SINGS~\citep{Meurer:06}. 

In addition to these impressive surveys are equally impressive plans for next generation radio telescopes. For example
Apertif\footnotemark \footnotetext{http://www.astron.nl/general/apertif/apertif}, MeerKAT\footnotemark \footnotetext{www.ska.ac.za} and 
ASKAP\footnotemark \footnotetext{www.atnf.csiro.au/projects/askap} are the
pathfinder missions for the ambitious Square Kilometer Array (SKA)
which may detect $10^9$ galaxies out to $z \approx 1.5$ yielding strong cosmological
constraints~\citep{AR}. The ASKAP all sky survey, WALLABY, will detect $5 \times 10^{5}$
galaxies in HI~\citep{Duffy:11a} which will be the largest catalogue until the completion of the 
Five hundred metre Array Spherical Telescope (FAST) which complements SKA in the Northern 
hemisphere, potentially detecting $10^6$ galaxies with mean redshift 
$\bar{z} \approx 0.15$~\citep{Duffy:08a}.

Together these planned telescopes will revolutionise our 
ability to measure both the local HI mass function and its evolution to high 
redshifts, as well as the distribution of HI within galaxies across a large fraction of the Universe.
The expected HI data is, however, far in advance of current theoretical models
due to the intrinsic complications of HI formation and destruction in galaxies.
This is in addition to the already uncertain gas treatment in the theoretical 
models for sub-grid baryonic physics.

Although the detailed distribution of HI is not well understood, the basic picture of the lifecycle of 
HI can be reasonably well described through simple physical arguments.  
The gravitational potential of the dark matter forms the framework in which gas can accumulate, 
cooling through a variety of thermal and non-thermal emission mechanisms. 
However, due to the cosmic UV background only
haloes with a virial temperature greater than $\sim 10^{4} \, \rm K$ can retain this
gas, thereby setting an effective lower mass floor for the 
haloes that can host galaxies~\citep[e.g.][]{Quinn:96}.
Eventually, as the gas density increases, it can become self-shielded against this background
UV flux and will form large collections of cold, neutral hydrogen.
In the centres of these clouds molecular hydrogen may form which will
lead to a rapid cooling of the gas. The dense core of clouds may then 
experience thermo-gravitational instabilities which can lead to star 
formation~\citep{Schaye:04}. The high-mass stars will in turn ionise the very 
regions that helped create them. This is accomplished directly via photo-ionising radiation 
from the stars, as well as in driving bulk motions of the gas from the galaxy, in turn reducing 
the self-shielding against the background radiation field~\citep[e.g.][]{PawlikSchaye:09}.

In an ideal simulation one would trace the impact of individual stars, or stellar regions,
in ionising the gas and thereby accurately track the non-local effects of star formation
on the Interstellar Medium (ISM). This is an extremely computationally expensive
process for the large number of galaxies formed in hydrodynamical simulations of cosmological volumes,
such as is considered in this work. However, by limiting the radiative transfer scheme to postprocessing existing datasets,~\citet{Altay:10} were 
able to accurately model the neutral hydrogen column density distribution function at $z=3$ (a separate study by~\citet{McQuinn:11} 
was similarly successful using a postprocessing radiative transfer scheme at $z=3$).~\citet{Altay:10} found that 
it was critical to model self-shielding above column densities of $10^{18} \rm \, cm^{-2}$
and molecular hydrogen formation above $10^{21.5} \rm \, cm^{-2}$. In particular we adopt their empirical treatment
of molecular hydrogen formation at high gas densities as discussed in Section~\ref{sec:eosgas}.

Several other works have studied the distribution of HI at $z=3$ such as~\citet{Pontzen:08},~\citet{Razoumov:08}
and~\citet{Tescari:09} but most relevant for our investigation of the low redshift Universe is~\citet{Popping:09}, henceforth P09.
We argue that our higher resolution simulations allow us to make more realistic and better resolved 
predictions for the HI, H$_2$ and cold gas mass functions\footnotemark, extending this work to the evolution of the mass functions 
and investigate the sensitivity to the physics scheme incorporated; a significant improvement on all existing hydrodynamical studies to date.

\footnotetext{Notably, we find that their mass function studies consider mass ranges we find to be unresolved, we demonstrate this
in Fig.~\ref{fig:restest_z0_hi} (\ref{fig:restest_z2_htwo}) for HI ($\rm H_{2}$).}

Other works investigating HI in galaxies across cosmological scales have taken a semi-analytic approach,
for example in~\citet{Obreschkow:09c} they calculated that the neutral hydrogen in galaxies was distributed
across atomic and molecular states with a ratio, $R = M_{\rm H_2} / M_{\rm HI}$ that increased with the
gas mass of the halo. Adopting this approach they found that semi-analytic models could reproduce the $z=0$ HI mass function 
above $M_{\rm HI} = 10^{9}\,{\rm M_{\odot}}$, below this mass they found a factor two over abundance of galaxies relative to the observations. 
This was opposite to another semi-analytic study based on the same underlying DM simulation; in~\citet{Power:10} it was noted that 
such low mass HI systems were too rare. 
Interestingly it has been suggested by~\citet{Lagos:11} that if one were to split the cold gas into HI and H$_2$ during the simulation, rather
than in postprocessing, and form stars based on the latter alone, then many more low mass HI systems survive in the simulations.

In~\citet{Obreschkow:09b} the ratio $R$ from~\citet{Obreschkow:09c} was applied at higher redshifts. With this scheme 
they found that there was little evolution in the 
cosmic density of HI in the Universe, $\Omega_{\rm HI}$, to $z=1$ beyond which the
value dramatically decreased, with only half of the local HI predicted to exist in galaxies by $z=3$. 
This is in contrast to conclusions made on the basis of observations
of damped Ly$\alpha$ systems~\citep[e.g.][]{Prochaska2005,Rao:06} which 
infer a factor two {\it increase} by $z \approx 3$.
Further observations from DLA systems~\citep{Prochaska:09} indicate that the comoving
HI density at $z=2$ is, within the errors, consistent with the result at $z=0$ 
as measured by HIPASS~\citep{HIPASS} (with a best-fit value only $5\%$ higher). This local value is similar to the 
partly completed ALFALFA survey of the local Universe, although they find a value $16\%$ higher~\citep{Martin:10}. 
Tantalising evidence that the HI cosmic density at $z=0.8$ is also similar to that at $z=0$ 
is given by~\citet{Chang:10}, who cross-correlated galaxies from optical catalogues with brightness temperature radio maps 
to measure a statistical `stacked' estimate of the HI emission.
Furthermore,~\citet{Prochaska:09} found that this value increases weakly beyond $z=1$, only doubling 
over the redshift range $z\approx 2 \,-\, 4$. 
Therefore, although the integrated HI mass function is observed to be roughly constant 
in time, the distribution of HI in mass may vary. This is the main subject of our work and the goal
of upcoming deep radio surveys.

This paper is organised as follows. In Section~\ref{sec:HIsimulations} we describe the physics models and
range of resolutions that we have simulated as well as the numerical tests
we have performed to remove artefacts. In Section~\ref{sec:hydrogen_estimation} we 
detail our methodology of converting the hydrogen abundances of the simulation into HII, HI, $\rm H_{2}$.
We analyse the HI properties of haloes in Section~\ref{sec:HIinOWLS}.
First we consider the distribution of the stellar mass and 
gaseous components as a function of halo mass in Section~\ref{sec:hi_in_haloes}.
We then probe the HI and $\rm H_{2}$ as a function of stellar mass in Section~\ref{sec:stars_vs_hi}.
As a last probe of the distribution of the baryons within haloes we examine their half-mass radii
for various halo masses and redshifts in Section~\ref{sec:halfmass}.
We then consider the HI and $\rm H_{2}$ mass functions and their evolution with 
redshift in Section~\ref{sec:HImassfn}. In Section~\ref{sec:HImassfn_multi} 
we vary the sub-grid physics schemes to probe the sensitivity of the mass function 
predictions to the particular physics prescription.
Finally, we conclude in Section~\ref{sec:HI_conclusion}.

\section{Simulation Details}
\label{sec:HIsimulations}
We have made use of a number of high resolution, large-volume
hydrodynamical runs created as part of the OverWhelmingly Large Simulations
project ({\sc OWLS}; \citealt{Schaye:10}). This series of simulations all share the same initial conditions, i.e. are 
identical simulation volumes, but are resimulated using $\sim 50$ different physics models which allows robust 
testing of the relative importance of each scheme. 
OWLS has seen a number of successes in reproducing observed galaxy properties.  
For example, the cosmic star formation history~\citep[e.g.][]{Madau:96} was 
shown to be reproduced in~\citet{Schaye:10}. 
In~\citet{Booth:09} and~\citet{Booth:10} the AGN implementation was found to recover
 a number of empirical properties of super-massive black holes, 
such as the black hole mass - halo mass relation~\citep{Bandara:09};
the black hole mass - stellar mass relation~\citep[e.g.][]{HaringRix:04}; and
the black hole mass - bulge velocity dispersion correlation~\citep[e.g.][]{Tremaine:02}. 
Furthermore, this AGN implementation is capable of 
reproducing a number of galaxy group properties such as their X-ray temperature and density 
profiles as well as galaxy stellar masses and age distributions~\citep{Mccarthy:10}. Of particular note
for this work is the study by~\citet{Altay:10} who found that OWLS can reproduce the observed 
$z=3$ HI column density distribution function, when post-processed with a radiative transfer model.
More generally, a particular strength of the OWLS suite is that the variety of models 
allows us to test the sensitivity of a particular observable phenomenon to the 
details of the physical processes included (e.g. metal enrichment, feedback mechanisms etc).

The OWLS simulations were run with an updated version of the publicly-available {\sc gadget-2} $N$-body/hydrodynamics 
code~\citep{Springel2005b} with new modules for radiative cooling~\citep{Wiersma:09a}, star formation~\citep{Schaye:08},
stellar evolution and mass loss~\citep{Wiersma:09b}, galactic winds driven by SNe~\citep{DallaVecchia:08} and, for one
model, AGN~\citep{Booth:09}. Each simulation employed $N^3$ dark matter (DM) particles, and 
$N^3$ gas particles,  where $N = 128$, $256$ and $512$, within cubic volumes of comoving length, 
$L = 25$, $50$ and $100\hMpc$. For all cases, glass-like cosmological initial conditions were generated at 
$z=127$ using the Zeldovich approximation and a transfer function generated using {\sc cmbfast} (v.~4.1, \citealt{CMBFAST}). 

Simulations with the smallest box were run to $z=2$ and the two larger volumes were simulated down to $z=0$. 
For the present study the complete range of box sizes exist for only one of the physics models, that 
termed `\emph{ZC\_WFB}' in~\citet{Duffy:10}. Table~\ref{tab:sims} lists the available box sizes and other 
pertinent numerical parameters, including the Plummer-equivalent softening lengths and particle masses. (Note that
the dark matter particles have a mass that is a factor $(1-f^{\rm univ}_{\rm b})/f^{\rm univ}_{\rm b}$ greater than the 
mass of the gas particles, where $f^{\rm univ}_{\rm b} =  \Omega_{\rm b} / \Omega_{\rm m}$ is the Universal baryon fraction.)

We use the Wilkinson Microwave Anisotropy Probe 3 year results~\citep{WMAP3}, 
henceforth known as WMAP3, to set the cosmological parameters 
$[\Omega_{\rm m},$ $\Omega_{\rm b},$ $\Omega_{\Lambda},$ $h,$ $\sigma_{8},$ $n_{\rm s}]$
to [0.238, 0.0418, 0.762, 0.73, 0.74, 0.95] and 
$f^{\rm univ}_{\rm b} = 0.176$. Although we use a cosmology
that has a $\sigma_8 $ about $10\%$ lower than the current year 7 {\it WMAP} result~\citep{Komatsu:11}
we believe that the effects on our predictions for the various phases of hydrogen will be small. This is because
only the high-mass end ($M \ge 10^{14} \hMsol$) of the mass function is exponentially sensitive to a change 
in $\sigma_8$ whereas the majority of the HI signal in our simulations is due to lower mass DM haloes near the `knee'
at $M \approx 10^{12} \hMsol$ which are much less sensitive to $\sigma_8$. When comparing DM only simulations with $\sigma_8$ 
values for WMAP3 and 7, the overall number counts agree within 5 per cent for haloes of mass $M=10^{11}-10^{12}\hMsol$, 
while clusters ($M=10^{14}-10^{15}\hMsol$) are nearly 70 per cent more common in the higher $\sigma_8$ universe. 
We note that the study of the dependence of the hydrogen phases on the sub-grid physics will be insensitive to the cosmology used,
as all simulations have the same initial conditions.

\begin{table*}
\begin{center}
\caption{Details of the various simulation parameters used in this study. From left to right the columns show: 
simulation identifier; the physics scheme(s) available; comoving box-size; number of dark matter particles 
(there are equally many baryonic particles); initial baryonic particle mass; dark matter particle mass; comoving 
(Plummer-equivalent) gravitational softening; maximum physical softening; final redshift. Note that we 
have adapted the {\sc OWLS} simulation list from~\citet{Schaye:10} such that  \emph{REF} in that work is referred
to as \emph{ZC\_WFB} here, while \emph{All} corresponds to the full range of simulations listed in Table~\ref{tab:physics}.}
\label{tab:sims}
\begin{tabular}{rrrrrrrrr}
\hline
Simulation & Physics & $L$ & $N$ & $m_{\rm b}$ & $m_{\rm dm}$ &
$\epsilon_{\rm com}$ & $\epsilon_{\rm prop}$ & $z_{\rm end}$\\
& & $(\hMpc)$ & & $(\hMsol)$ & $(\hMsol)$ & $(\hkpc)$ &
$(\hkpc)$ & \\
\hline
\emph{L025N128} & \emph{ZC\_WFB} &  25 & $128^3$ & $8.7 \times 10^7$
& $ 4.1 \times 10^8$ & 7.81 & 2.0 & 0 \\
\emph{L025N256} & \emph{ZC\_WFB} &  25 & $256^3$ & $1.1 \times 10^7$
& $ 5.1 \times 10^7$ & 3.91 & 1.0 & 2 \\
\emph{L025N512} & \emph{All} & 25 & $512^3$ & $1.4 \times 10^6$
& $ 6.3 \times 10^6$ & 1.95 & 0.5 & 1.45 \\
\emph{L050N128} & \emph{ZC\_WFB} & 50 & $128^3$ & $6.9 \times 10^8$
& $ 3.2 \times 10^9$ & 15.62 & 4.0 & 0 \\
\emph{L050N256} & \emph{ZC\_WFB} & 50 & $256^3$ & $8.7 \times 10^7$
& $ 4.1 \times 10^8$ & 7.81 & 2.0 & 0 \\
\emph{L050N512} & \emph{All} & 50 & $512^3$ & $1.1 \times 10^7$
& $ 5.1 \times 10^7$ & 3.91 & 1.0 & 0 \\
\emph{L100N256} & \emph{ZC\_WFB} & 100 & $256^3$ & $6.9 \times 10^8$
& $ 3.2 \times 10^9$ & 15.62 & 4.0 & 0 \\
\emph{L100N512} & \emph{All} & 100 & $512^3$ & $8.7 \times 10^7$
& $ 4.1 \times 10^8$ & 7.81 & 2.0 & 0 \\
\hline
\end{tabular}
\end{center}
\end{table*}

\subsection{Simulation models}
We briefly recap the physics prescriptions used in this work but direct the
reader to~\citet{Schaye:10} and the references therein for a more extensive explanation.
We are primarily interested in the sensitivity of the HI abundance to differing 
feedback schemes and identify such simulations with the labels `\emph{WFB}', `\emph{SFB}' or 
`\emph{NFB}' to denote stellar feedback that is weak, strong or absent, respectively.  
Furthermore we denote a simulation that includes AGN feedback by `\emph{AGN}'. 
Lastly, we want to test the sensitivity of the results to the
additional cooling of the gas due to the presence of metals and denote these
simulations with `\emph{ZC}', while those without are labeled `\emph{PrimC}'. Refer to Table~\ref{tab:physics}
for an overview.

For the simulations labeled `\emph{PrimC}' we model the cooling of gas via line 
emission from primordial elements, H and He. Those labeled `\emph{ZC}' have 
additional contributions from nine metals: C, N, O, Ne, Mg, Si, S, Ca and Fe. 
In both cases, additional cooling via free-free Bremsstrahlung emission is also present.
The cooling rates for the different elements are taken from 
3-D {\sc cloudy}~\citep{Cloudy} tables which assume collisional
equilibrium before reionisation ($z>9$) and photoionisaton equilibrium afterwards, 
due to the presence of an evolving meta-galactic UV/X-ray background~\citep{HaardtMadau:01}.
For more details of the implementation of gas cooling (and heating) rates in the simulations,
see~\citet{Wiersma:09a}.

Gas particles are converted stochastically into stellar particles 
(such that particle numbers are conserved) at a pressure-dependent 
rate in accordance with the Kennicutt-Schmidt law (for a review see~\citealt{Kennicutt:98a}) using
the prescriptions of~\citet{Schaye:08}.
We assume a Chabrier initial mass function (IMF; see~\citealt{Chabrier:03}) 
within each star particle and track the metal mass released by the stars using the method
described in~\citet{Wiersma:09b}.

The stars formed in the simulations inject energy in neighbouring gas particles via kinetic feedback. 
For our assumed IMF, the energy per solar mass of material available from
supernovae (SNe) is $1.8 \times 10^{49}\, {\rm erg} \,{\rm M}^{-1}_{\odot}$. 
This is distributed amongst the local gas according to the dimensionless
mass-loading parameter $\eta$, in which a total gas mass of 
$\eta {M}_{\rm SN}$ is given an initial wind velocity of ${v_{\rm w}}$. 
This results in a fraction
\begin{equation}
\label{eqn:snfeedback}
f_{w} \approx 0.4 \left(\frac{\eta}{2}\right)\left(\frac{v_{\rm w}}{600 \, {\rm km\,s^{-1}}} \right)^{2} \left( \frac{\epsilon_{\rm SN}}{1.8 \times 10^{49}\,{\rm erg}\,{\rm M}^{-1}_{\odot}} \right)\,,
\end{equation}
of the energy of the SNe being imparted in the form of kinetic 
feedback~\citep{DallaVecchia:08}; the values $\eta = 2$ and 
$v_{\rm w} = 600 \,{\rm km\,s}^{-1}$ were adopted for the `\emph{WFB}' model. For the `\emph{SFB}' 
simulations $v_{\rm w}$ ($\eta$) increases (decreases) with the gas density in the region of the star
while keeping $f_{w}$ fixed.
Note that the kinetically-heated gas particles are never decoupled from the hydrodynamics 
and therefore will tend to collide and `drag' a larger number of gas particles 
with them, resulting in an effective mass loading which is higher than implied by $\eta$. 

In the simulations labeled `\emph{AGN}' the black hole at the centre of a halo
grows at the expense of the surrounding gaseous medium by gas accretion
as well as through mergers between black holes. 
The maximum accretion rate is capped at the Eddington limit and 1.5\% of the rest mass energy 
of the accreted matter is distributed thermally to the nearby gas particles as 
explained in~\citet{Booth:09}.~\citet{Booth:09} demonstrated that for this value of the feedback
efficiency the observed scaling relations for supermassive black holes are reproduced.

\begin{table*}
\begin{center}
\caption{A list of the simulated sub-grid baryon physics schemes utilised in this study. From left to right, the 
columns list the simulation name used in this study; the name as defined by~\citet{Schaye:10};
and a brief the description of the physics modelled. The list of available simulations run with each model (box-size and resolution)
is given in Table~\ref{tab:sims}.}
\label{tab:physics}
\begin{tabular}{lll}
\hline
Simulation & {\sc OWLS} name & Brief description\\
\hline
\emph{PrimC\_NFB} & \emph{NOSN\_NOZCOOL} & No energy feedback from SNe and  cooling assumes primordial abundances\\
\emph{PrimC\_WFB} & \emph{NOZCOOL}  & Cooling as \emph{PrimC\_NFB} but with fixed mass loading of winds from SNe feedback \\
\emph{ZC\_WFB} & \emph{REF} & SNe feedback as \emph{PrimC\_WFB} but now cooling rates include metal-line emission \\
\emph{ZC\_SFB} & \emph{WDENS} & As \emph{ZC\_WFB} but wind mass loading and velocity depend on gas density\\
\emph{ZC\_WFB\_AGN} & \emph{AGN} & As \emph{ZC\_WFB} but with AGN feedback \\
\hline
\end{tabular}
\end{center}
\end{table*}

\subsection{Identifying Galaxies and Haloes}
\label{sec:halofinder}
In this work we will consider the HI content of galaxies and the associated HI gas
lying between them, which is particularly important in groups and clusters. 
We initially determine the halo in a simulation using a Friends-of-Friends (FoF) 
algorithm~\citep{Davis85} with a linking length $b$. This will link all 
DM particles together that are within a distance 
$b \bar{l}$, where $\bar{l}$ is the mean inter-particle distance in the simulation. 
The gas is then attached to the nearest DM particle.
The halo can thus have arbitrary shape, defined by an iso-density curve of overdensity 
relative to the background
$\rho/\bar{\rho} \approx 3/(2\pi b^{3})$~\citep{ColeLacey96}. 
We choose a value $b=0.2$ which, for an isothermal density profile, will have a halo mean over-density of 
$\langle \rho \rangle /\bar{\rho} \approx 180$~\citep{LaceyCole94}. This is 
close to the spherical top-hat collapse model prediction
for a virialised object $\approx 18\pi^2 \approx 178$.

In the simulations with $L=100\hMpc$ we typically resolve 230 (15) haloes of virial mass $10^{13} - 10^{14} \hMsol$ 
($ > 10^{14} \hMsol$) at $z=0$ and for $L=25\hMpc$ there are 840 (13) haloes of $10^{12} - 10^{13} \hMsol$ ($ > 10^{13} \hMsol$)
at $z=2$. As all simulations of the same volume share the same initial conditions, the haloes formed are identical,
except for the specific choice of sub-grid baryonic physics. This means that the number of haloes
between simulations typically differs by less than 5 per cent, essentially due to slight variations in mass of specific haloes
which can affect whether they meet mass cuts or not.

To identify gravitationally bound structures within haloes we use {\sc subfind}~\citep{Dolag:09}. 
This allows us, for example, to identify gas that has been 
stripped from an infalling galaxy but which is still bound to the background cluster, as well as satellite galaxies.
Unless stated otherwise, the masses and sizes which we will quote 
are for objects found with {\sc subfind} (and we will refer to such objects as {\it galaxies}), 
but when we discuss the halo in full (including satellites) we will use that
which comes from the FoF algorithm and will add the `Halo' superscript to the variable.

\section{Modelling the Neutral Hydrogen}\label{sec:hydrogen_estimation}

\begin{figure}
  \begin{center}
   \epsfig{figure=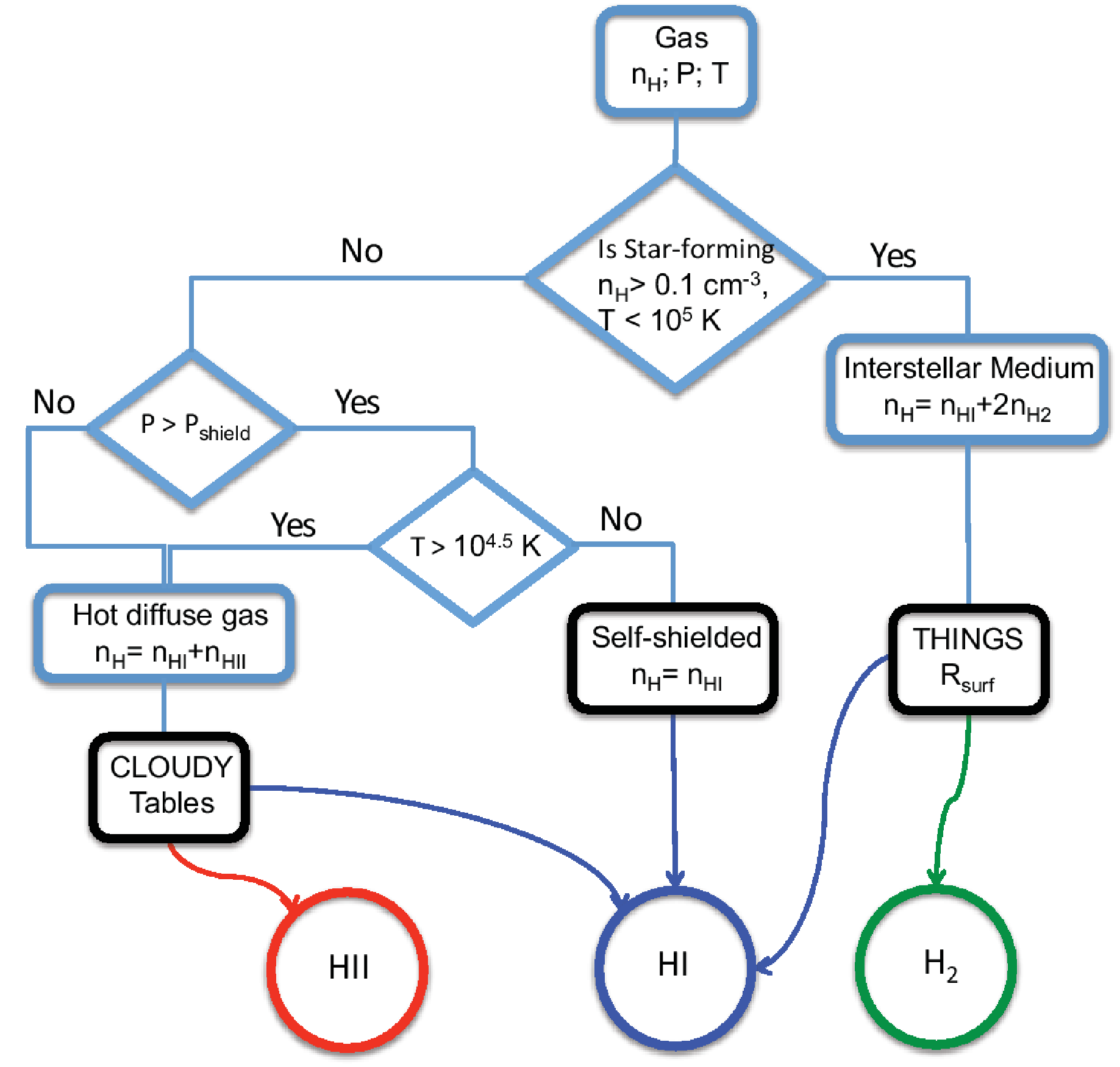, scale=0.4} 
    \caption[HI/H ratio phase diagram]
            {A simple flow diagram indicating the methodology to attribute the states of hydrogen to the
            gas particles of the simulation. In our model, the HII and $\rm H_2$ components arise from 
            the diffuse (intergalactic) and star-forming (interstellar) components respectively, whereas the HI signal is found in all regimes. Note that we only
            ever consider particles that belong to gravitationally bound structure.}
    \label{fig:flowchart}
  \end{center}
\end{figure}

In this section we discuss our method for calculating the mass fraction of hydrogen
in neutral molecular (H$_2$), atomic (HI) and ionised (HII) phases, for each gas
particle. The total hydrogen mass fraction is first estimated
for each particle by taking a smoothed average over its neighbours, using the SPH kernel 
(a value for the hydrogen abundance is carried by individual particles). The methodology 
for splitting the gas into the 3 phases of hydrogen depends on whether it is low-density
intergalactic gas or high-density interstellar gas. We consider each one in turn below, with the
overall procedure summarised as a flow diagram in Fig.~\ref{fig:flowchart}. 

\subsection{Intergalactic Gas}\label{sec:noneosgas}
Most of the gas resides outside, or in the outskirts, of galaxies and will have a significant HII component,
due to both the photo-ionising UV/X-ray background and collisional ionisation resulting from gravitational shock heating. 
{\sc OWLS} tracks an evolving UV/X-ray background~\citep{HaardtMadau:01} after reionisation at $z=9$,
that primarily affects the cold, low-density intergalactic medium. Gas at higher density can cool more efficiently,
leading to larger neutral (HI) fractions (for full details see~\citealt{Wiersma:09a}).
A significant fraction of gas is shock-heated to much higher temperatures
at low-redshift, creating the warm-hot intergalactic medium and the intracluster medium.
We use pre-calculated ionisation tables for hydrogen from {\sc cloudy}~\citep{Cloudy}, 
to compute the optically thin HI and HII fractions of gas particles as a function of their current temperature and (SPH smoothed) density.

\subsubsection{Optically Thick Gas}\label{sec:opticallythick}
For gas that resides at moderate densities (in the range $n_{\rm H} \approx 10^{-4}\,-\,10^{-1}\,{\rm cm^{-3}}$) 
assumptions of optically thin gas exposed to a uniform UV background start to break down. 
Firstly, the local galactic UV intensity can be significantly 
stronger than the global average, from the presence of young massive stars or a quasar (the so-called 
proximity effect). In this case, the {\sc CLOUDY} tables will overestimate the amount of HI present as
the assumed photo-ionisation rate will be lower than it should be. Secondly, near galaxies the gas density 
can reach levels which are sufficient to self-shield the gas from the global UV background, which in turn would
lead us to underestimate the fraction of HI.  As we have previously discussed, ideally we would perform a full 
radiative transfer calculation to distribute ionising flux from nearby stellar sources through the gas. 
Such a technique would also lead to a more realistic calculation of the self-shielding in dense 
regions. We turn to simpler prescriptions that can be more easily applied to large volume simulations. In this study, we assume that
only the self-shielding is important and attempt to correct for it using a simple method guided by the observational data.

\subsubsection{Self-shielding Correction}
We follow a methodology similar to that of P09 and choose to approximate the onset of self-shielding by adopting a pressure-based 
threshold. It is reasonable to expect that self-shielding occurs above a given surface density. 
As, shown by e.g.~\citet{Schaye:08}, the pressure in the disk ought to scale quadratically with the surface
density as the thickness of the disk is set by the local Jeans length. We can even use similar arguments to define an order of magnitude
estimate for the pressure threshold value $P_{\rm shield}$ by considering the local volume density at the threshold, $n_{\rm shield}$, related 
to the pressure by $P_{\rm shield}/k = (n_{\rm shield} / k) 10^{4}\,\rm K$ with a calculation of $n_{\rm shield}$ from Equation 11 in~\citet{Schaye:01a} 
\begin{eqnarray}
n_{\rm H, \, shield} & = & 10^{-4} {\, \rm cm^{-3}} (1+\delta)^{-1.5} \Gamma \left(\frac{T}{10^4 \,\rm K}\right)^{-1} \nonumber \\
& & \times \left(\frac{1+z}{4}\right)^{-9/2}\left(\frac{\Omega_{\rm b}h^{2}}{0.02}\right)^{-3/2}\left(\frac{f_{\rm g}}{0.16}\right)^{-1/2} \nonumber \\
& & \times \frac{N_{\rm HI}}{2.7 \times 10^{13} {\, \rm cm^{-2}}}\,,
\end{eqnarray}
giving $n_{\rm H, \, shield} \simeq 10^{-3} - 10^{-2} \, \rm cm^{-3}$ for $z=3$ and assuming that 
overdensities relative to the background density, $\delta$, of order $10^2$ represent
self-shielded galaxies, which have a fraction $f_{\rm g}$ of their disk mass in HI. This is equivalent to a pressure
$P_{\rm shield}/k \simeq 10 - 10^{2} \, \rm K cm^{-3}$. This simple estimate has been confirmed using a radiative transfer simulation in~\citet{Altay:10}.

The pressure threshold used in P09 was tuned to reproduce the mean HI density from the 1000 Brightest HIPASS galaxy 
catalogue~\citep{Zwaan:03}, finding a value $P_{\rm shield}/k =155\,\rm K cm^{-3}$. They only applied this to gas
cool (or dense) enough to have a recombination time less than or equal to the sound crossing time (see Fig. 2 in P09). 
While it is clear that hot, collisionally ionised gas should be excluded, the physical motivation for the particular criterion chosen
by P09 is unclear to us. 
We simply use an upper temperature limit for gas that is considered to be self-shielded, essentially forming a `cold' 
gas population with $T \le 10^{4.5} \,\rm K$. Our approach is similar to that
adopted by~\citet{Power:10} for their investigation of cold ($T \le 10^{4} \,\rm K$) gas mass functions from semi-analytic simulations.

\begin{table}
\begin{center}
\caption{A list of the pressure thresholds 
used to approximate the onset of self-shielding, $P_{\rm shield}/k$,
for gas particles that have temperatures below $10^{4.5} \, \rm K$ in methods $\SSALF$ (2nd column, normalised to 
low-redshift ALFALFA observations) and $\SSDLA$ (3rd column, normalised to high-redshift Damped Lyman-$\alpha$
observations). 
Values in brackets denote pressure thresholds that produce HI densities 10 per cent higher and lower than observed, respectively.
Zero values correspond to cases where even making all cold gas in the galaxy neutral gives insufficient HI to match the observations 
(typically $\sim$ 20 per cent below the desired value).
The neutral interstellar gas is still separated into atomic and molecular neutral components.}
\label{tab:pressure}
\begin{tabular}{rrr}
\hline
Simulation & $\SSALF$ & $\SSDLA$ \\
\hline
\emph{PrimC\_NFB} & 0 [0, 0] & 159 [132, 193] \\
\emph{PrimC\_WFB} &  0 [0, 0] & 47 [30, 68] \\
\emph{ZC\_WFB} & 150 [133,175] & 98 [80, 123] \\
\emph{ZC\_SFB} & 74 [60, 93] & 17 [3, 28] \\
\emph{ZC\_WFB\_AGN} & 27 [21, 35] & 76 [62, 93] \\
\hline
\end{tabular}
\end{center}
\end{table}

In what follows we have treated the self-shielding regime in three different ways. In what we shall call $\SSNO$ we shall presume
that there is no self-shielding. This is clearly unphysical but it can be helpful to compare this with the other two methods. 
For method $\SSALF$ we match the measured cosmic HI density in the HI mass range, $M_{\rm HI} = 10^{10} - 10^{11}  \hMsol$, from the $z=0$ 
ALFALFA survey~\citep{Martin:10}. This normalisation is computed for the $100 \hMpc$ simulation as this volume has been run with all sub-grid physics 
models. Cold gas above the pressure limit is then considered to be completely neutral, while gas below the threshold has HI and HII fractions taken from
the optically-thin {\sc cloudy} tables discussed previously. We find that a pressure of $P_{\rm shield}/k \approx 150 \,\rm K cm^{-3}$ matches 
the ALFALFA result for our default model (\emph{ZC\_WFB}, which is also the simulation closest to that studied by P09 in terms of the sub-grid physics 
modelled). 
The pressure thresholds used for all simulation models are listed in Table~\ref{tab:pressure}; typically in runs with stronger feedback we require a lower 
pressure threshold to match the HI density while in the runs with primordial cooling and weak or no feedback, there is insufficient ($\sim$ 20 per cent too 
low) HI gas in simulated galaxies to match the observations at low-redshift (the reverse is true at high-redshift).
  
There is no reason to believe that the pressure threshold will be independent of redshift. In fact, we expect it to depend on the intensity of the
evolving UV background~\citep{Schaye:01a}. For method $\SSDLA$ we therefore normalise 
$P_{\rm shield}$ to agree with observations at higher redshift, namely the cosmic HI density inferred from Damped Lyman-$\alpha$ 
absorption systems. In particular, we will use measurements presented in~\citet{Prochaska:09} for $z\approx 2$. These results are close
to the $z=0$ HI cosmic density inferred by ALFALFA (e.g. $\Omega_{\rm HI,DLA} = 1.05 \, \Omega_{\rm HI,ALFALFA} = 3.75 \times 10^{-4}$),
but lead to somewhat higher values for $P_{\rm shield}$. We tune the pressure value until the integrated HI mass function 
(where available, as shown in Fig.~\ref{fig:himassfn_multi_hicut}) matches $\Omega_{\rm HI,DLA}$.

\subsection{Star-forming Interstellar gas}\label{sec:eosgas}
As gas cools, its density will eventually increase until its structure
can no longer be modelled reliably in our simulations. Such gas, associated with the
star-forming interstellar medium in galaxies, is expected to be 
multiphase for hydrogen particle number densities above 
${n}^{*}_{\rm H} \approx 10^{-2}$ - $10^{-1} \,{\rm cm}^{-3}$~\citep{Schaye:04}. 
In {\sc OWLS}, such gas particles are identified when they exceed a critical particle density,
${n}^{*}_{\rm H} = 0.1 \,{\rm cm}^{-3}$ and have a temperature smaller than $10^{5} \,\rm K$, 
after which the gas density and pressure is evolved according to an effective polytropic Equation of state 
\begin{equation}\label{eqn:pressurelaw}
P = P_{\rm crit}\left( \frac{n_{\rm H}}{{n}^{*}_{\rm H}} \right)^{\gamma_{\rm eff}}\,,
\end{equation}
where $P_{\rm crit}/k = 2.3 \times 10^{3}\,{\rm K}\,{\rm cm}^{-3}$.
All runs considered assume $\gamma_{\rm eff} = 4/3$ which has the
advantage that the Jeans mass is independent of density~\citep{Schaye:08}, avoiding the situation
where particle masses are greater than the Jeans mass in dense regions, a 
numerical artefact known to result in artificial fragmentation of the gas 
cloud~\citep{Bate:97}. 

The Equation of state gas, which we will refer to as interstellar gas, lies in dense 
regions of the simulation where self-shielding effects are important.
We therefore assume that the hydrogen exists as either HI or $\rm H_2$ (we ignore
interstellar HII regions as they will contribute little to the hydrogen mass in galaxies). 
To separate the two phases, we utilise the empirical ratio relating $\rm H_{2}$ and HI surface 
densities, $R_{\rm surf} = \Sigma_{\rm H_2}/ \Sigma_{\rm HI}$, to the local ISM 
pressure measured in the THINGS survey (Table 6 and Fig. 17 of~\citealt{Leroy:08}), 
given by 
\begin{equation}\label{eqn:rsurf}
R_{\rm surf} =  \left( \frac{P/k}{10^{4.23} \,{\rm K cm^{-3}}} \right)^{0.8}\,,
\end{equation}
which was observed for a pressure range of $10^{3} \le P/k \, [\rm K\,cm^{-3}] \le 10^{6}$. 
It is then a trivial step to compute the mass of interstellar gas in atomic HI,
\begin{equation}\label{eqn:mhimh2}
{M_{\rm HI}} = {M_{\rm H}} / (1 + R_{\rm surf}) \,.
\end{equation}
Note that we have assumed that the HI and H$_2$ in the same disk are measured locally.

In using Equation~\ref{eqn:mhimh2} we must consider whether the pressure 
used in the empirical law of Equation~\ref{eqn:rsurf} corresponds to the pressure averaged over the 
physical length scales that the simulations resolve.
The THINGS survey reduces its sample of 23 dwarf and gas 
spiral galaxies to a standard resolution of $\sim 20$ arc seconds. However,
the dwarf galaxies are typically closer than the spirals, as they are less 
luminous. The resolved physical lengthscales for the dwarf and spiral galaxies
are around 400 and 800 pc respectively. 
This is close to the Plummer-equivalent softening length-scale (in proper units) 
of the $N=512^3$ simulations, as given in Table~\ref{tab:sims}.
Only the $100\,h^{-1}\,{\rm Mpc}$ boxsize may be cause for concern when we 
convert the pressure to HI mass, but as shown in Appendix~\ref{appendix:restest_hi},
this simulation appears sufficiently well resolved.

We also note that Equation~\ref{eqn:rsurf} is for the local Universe only and the assumption that it holds at higher
redshifts is potentially an issue. It is believed that this ratio will depend on 
metallicity~\citep{Schaye:01b,Schaye:04,Krumholz:08,Krumholz:09a} which 
decreases with redshift~\citep[e.g.][]{Tremonti:04,Savaglio:05,Erb:06}. 
We therefore caution the reader that the molecular fraction is likely to be a
more complicated function of galactic properties than assumed here, given the
limited physical information we have for galaxies in our simulations.

\subsection{Resultant phase diagram for a specific case}

\begin{figure*}
  \begin{center}
      \begin{tabular}{cc}
    \epsfig{figure=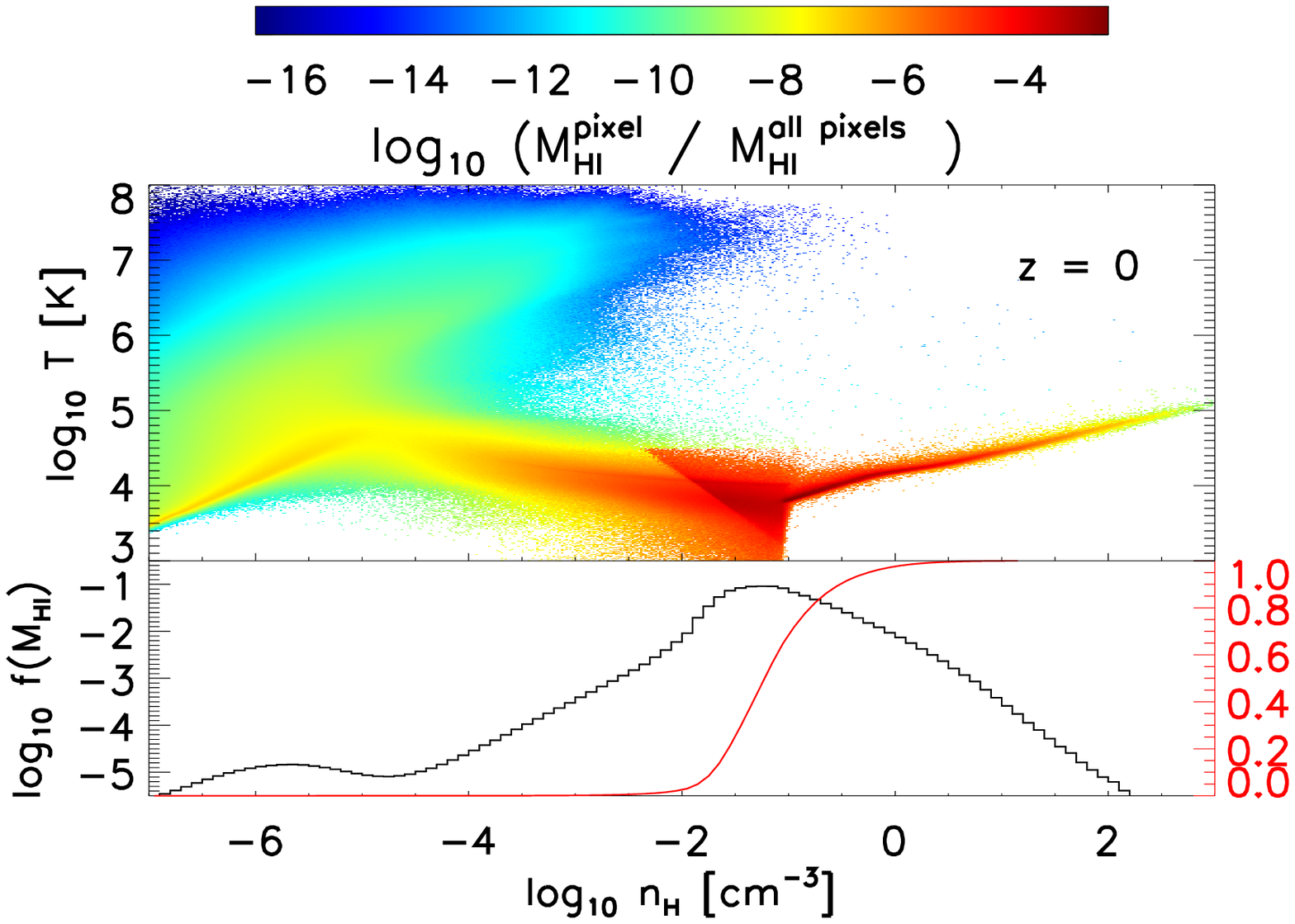, scale=0.4} &
    \epsfig{figure=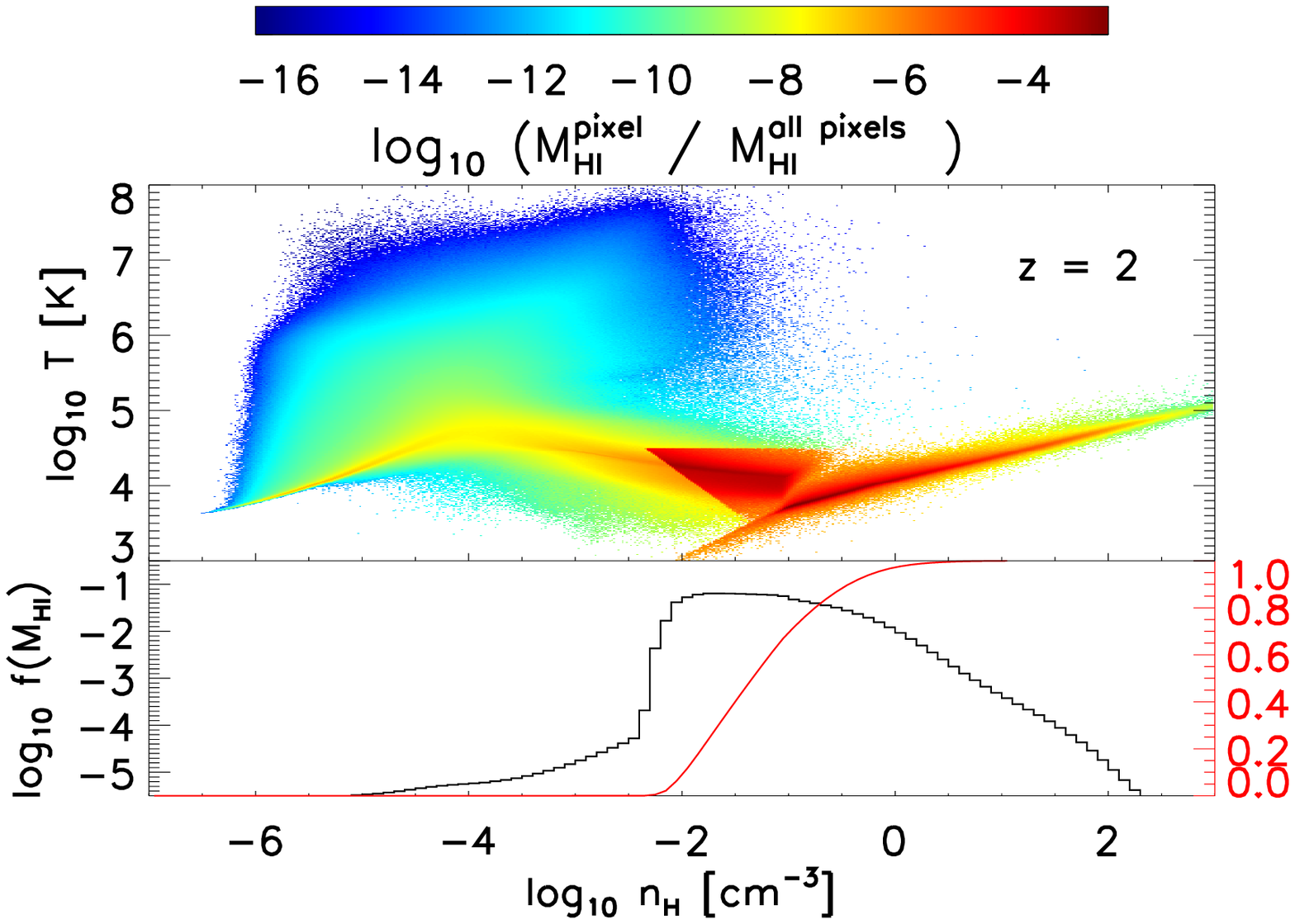, scale=0.4} \\
    \epsfig{figure=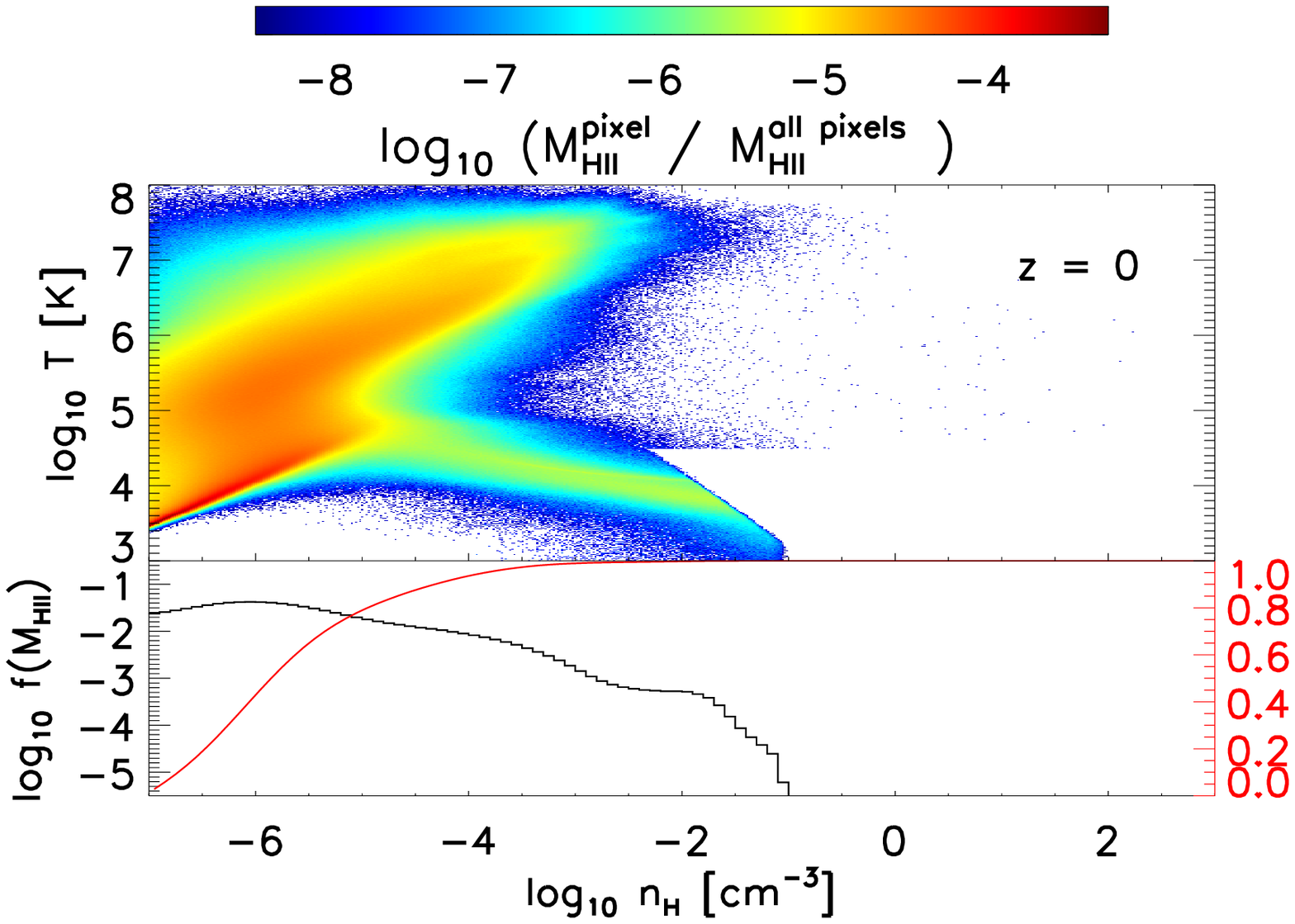, scale=0.4} &
    \epsfig{figure=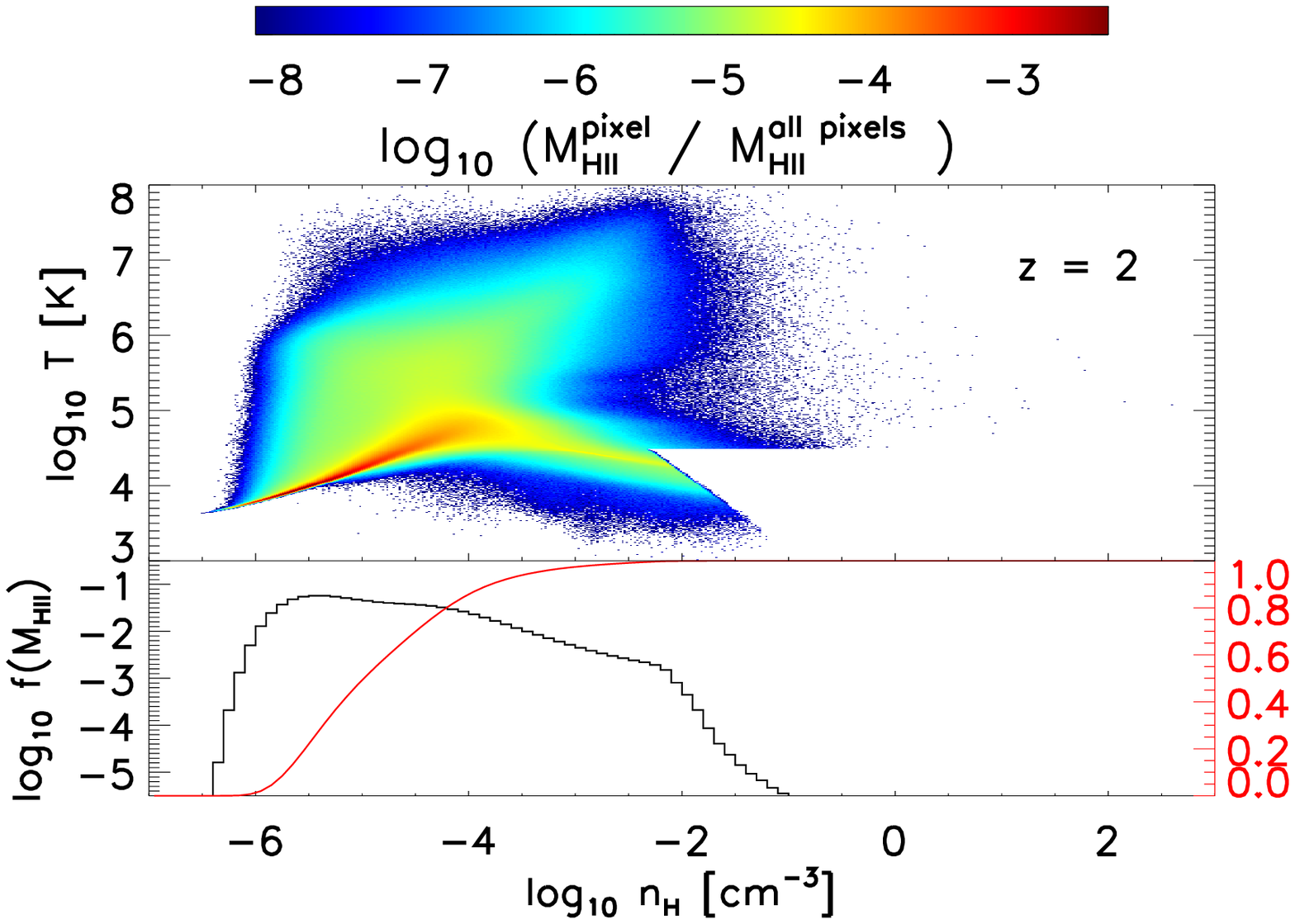, scale=0.4} \\
    \epsfig{figure=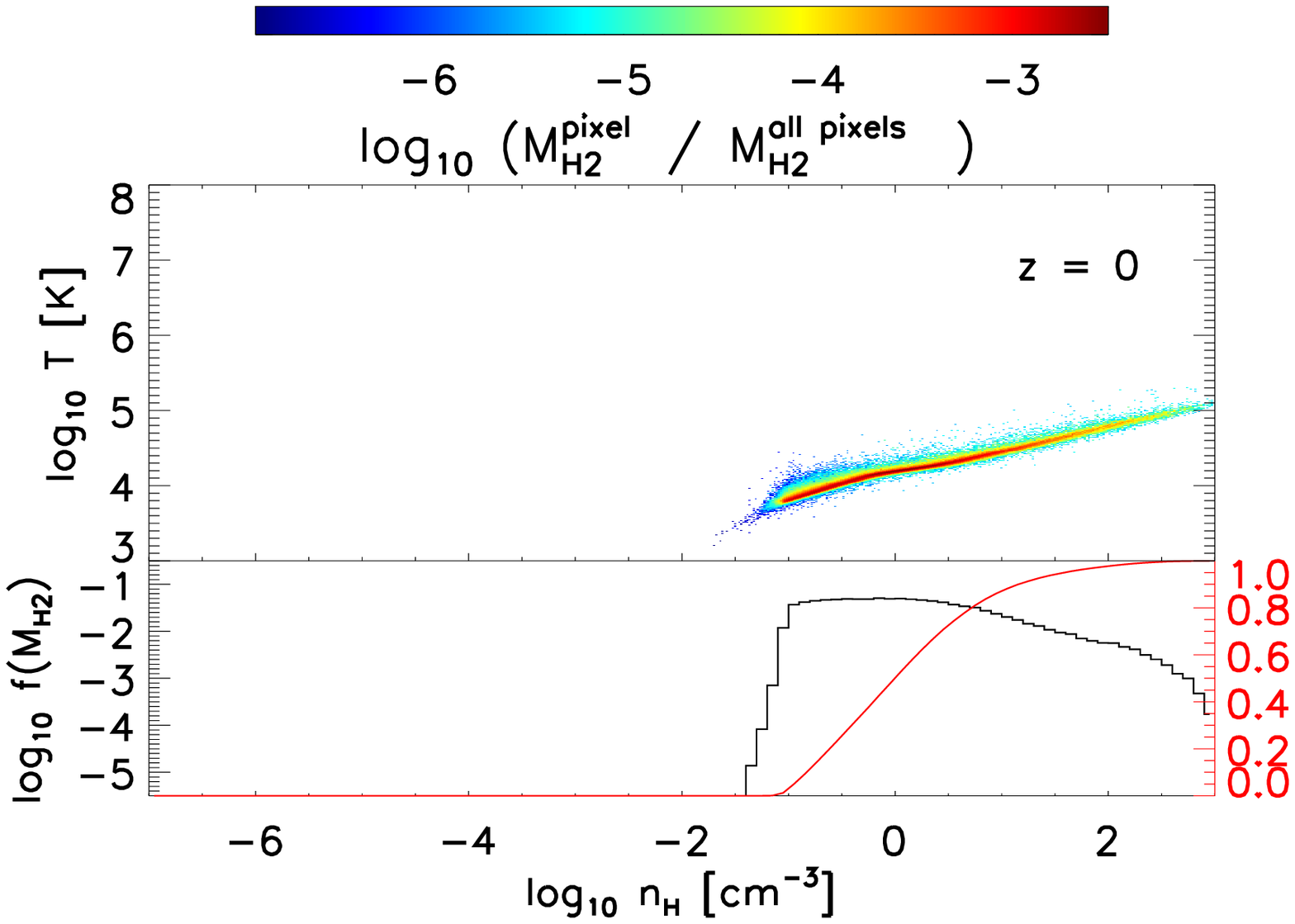, scale=0.4} &
    \epsfig{figure=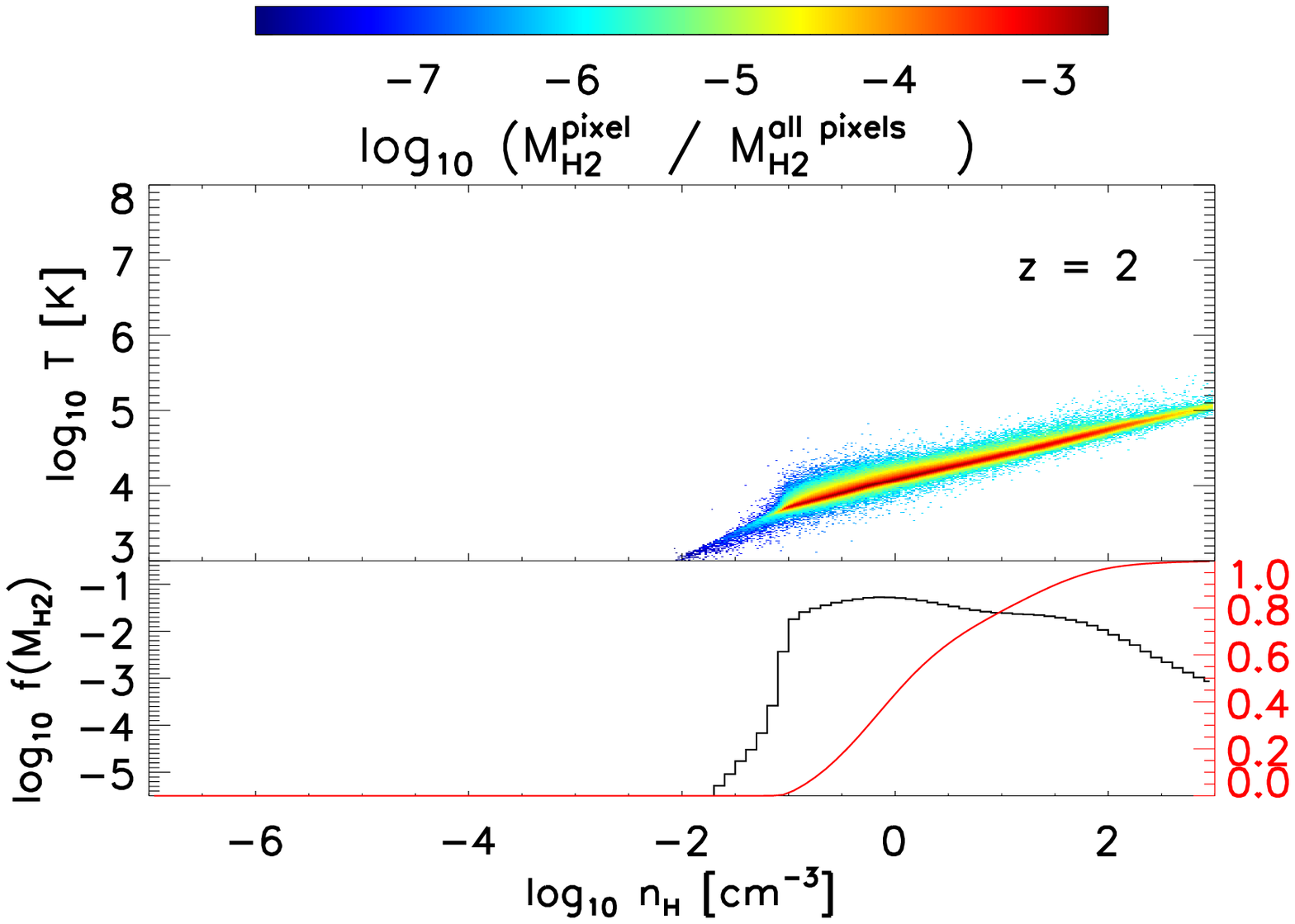, scale=0.4} \\
            \end{tabular}
    \caption[HI/H ratio phase diagram]
            {From top to bottom, we present the mass fraction of HI, HII and $\rm H_{2}$ for all gas in a simulation for a given 
            gas temperature and hydrogen particle number density, in the \emph{ZC\_WFB} simulation.
            Left panels show results at $z=0$ and right panels for $z=2$. Method $\SSALF$ is assumed for the 
            self-shielding correction (tuned to the ALFALFA HI mass function at low-redshift). 
            Additionally, we have projected each fraction as a function of density 
            in the histogram below each image (black line), with the red line a simple
            cumulative distribution function of this histogram. The pressure threshold for the onset of self-shielding is 
	    especially clear in the top panels as a discontinuity in the HI fraction. 
             The $\rm H_2$ is constrained, by design, to reside in the high-pressure multiphase ISM, modelled as a 
             polytropic Equation of state.}
    \label{fig:hi}
  \end{center}
\end{figure*}

In Fig.~\ref{fig:hi} we show the overall neutral fraction in atomic 
(HI; top panel) and molecular ($\rm H_{2}$; bottom panel) states as well as the 
ionised HII component (middle panel) in the temperature - density phase diagram
at $z=0$ and $2$ (left and right columns respectively) for all gas particles in a simulation. We show results from the 
\emph{ZC\_WFB} $100\hMpc$ simulation, using method $\SSALF$ for the self-shielding correction.
(The equivalent phase diagrams for the other simulation schemes are broadly similar.)

The main result from the top panels is that the majority of HI resides near the self-shielding limit,
demonstrating the importance of such a correction.
Furthermore, even though at $z=0$ the majority of hydrogen resides in the hot plume of 
intergalactic material (middle panels), this gas is typically too highly ionised 
to be a large contributor to the overall HI mass. By design, the molecular 
component exists only in the densest material, which obeys the imposed 
Equation of state power law. 
Differences in the phases between low and high redshift are most apparent
in the HI distribution below the self-shielding limit. At $z=2$ (right
column) the HI content of the gas drops more sharply as the more
intense global UV background ionises more of the gas.

The relative importance of each phase in gravitationally bound structure (and overall in the Universe) is
summarised in Table~\ref{tab:hvalues} as a function of redshift for the $50\hMpc$ simulation with the \emph{ZC\_WFB} physics scheme
and $\SSALF$ self-shielding model.

\begin{table}
\begin{center}
\caption{A list of relative fractions of the hydrogen phases (HII, HI and H$_2$) for the simulation model \emph{ZC\_WFB}
in the $50\hMpc$ simulation volume using self-shielding threshold $\SSALF$. As we are unable to resolve the low mass
end of the galaxy distribution the HI and H$_2$ values are lower estimates. We consider the two cases
for the hydrogen phases in both gravitationally bound structures and overall in the Universe (latter values are in parentheses).}
\label{tab:hvalues}
\begin{tabular}{rrrr}
\hline
Redshift & HII & HI & H$_2$ \\
\hline
0 & 0.964 (0.993) & 0.026 (0.005) & 0.010 (0.002) \\
1 & 0.862 (0.983) & 0.085 (0.010) & 0.053 (0.007) \\
2 & 0.736 (0.974) & 0.167 (0.016) & 0.097 (0.010) \\
\hline
\end{tabular}
\end{center}
\end{table}

\section{Neutral Gas in Galaxies and Haloes}\label{sec:HIinOWLS}

{\sc OWLS} provides a wide range in dynamical mass over which the distribution of baryons
can be measured. Additionally, different feedback and cooling schemes can be tested~\citep[e.g.][]{Schaye:10,Duffy:10}. 
In considering how the HII, HI and $\rm H_2$, as well as the stars formed from the dense hydrogen gas, are distributed across
the simulation, we must separate the intertwined effects of baryon physics with mass scales and evolution.
The results in this section concentrate on one simulation, \emph{ZC\_WFB}, but we will also make comparisons with 
 \emph{ZC\_WFB\_AGN} (arguably our most realistic model) to understand the possible effects of AGN feedback on the phases of hydrogen.
 
\begin{figure*}
 \begin{center}
 \begin{tabular}{m{0.12in}>{\centering}m{50mm}>{\centering}m{50mm}>{\centering\arraybackslash}m{50mm}}
 & $M_{\rm vir} = 10^{12} \hMsol,\,z=2$& $M_{\rm vir} = 10^{13} \hMsol,\,z=0$& $M_{\rm vir} = 10^{14} \hMsol,\,z=0$ \\
 HII & \epsfig{figure=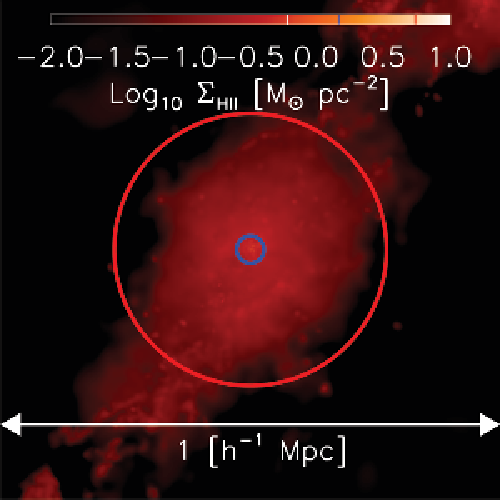} &
  	 \epsfig{figure=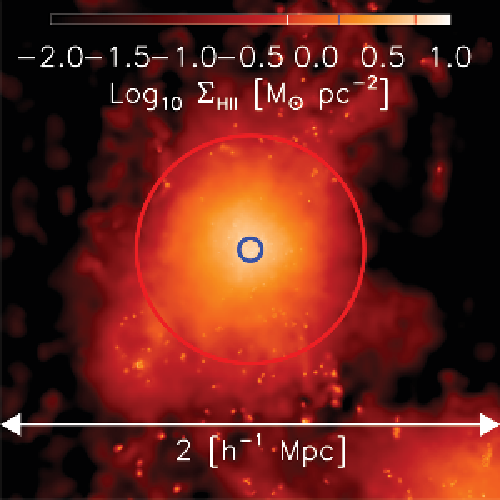} &
 	 \epsfig{figure=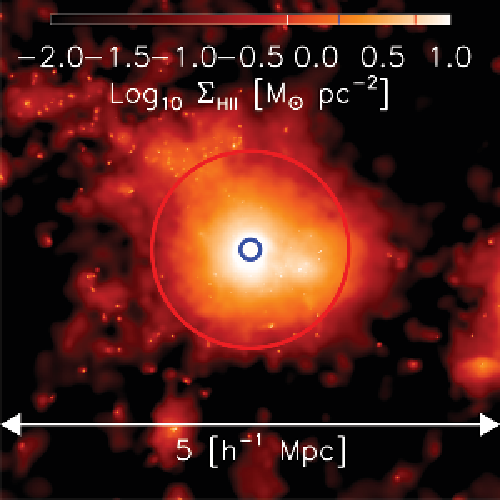} \\
	
  HI & \epsfig{figure=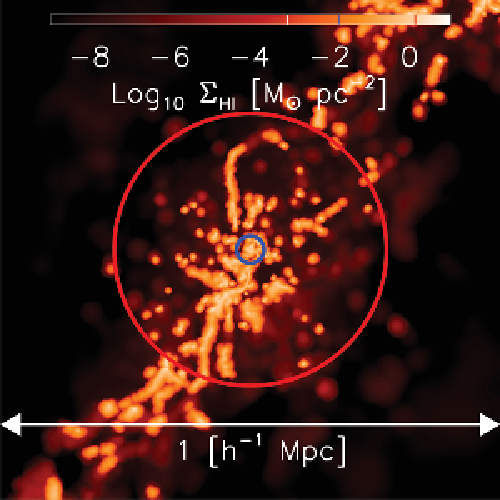} &
   	 \epsfig{figure=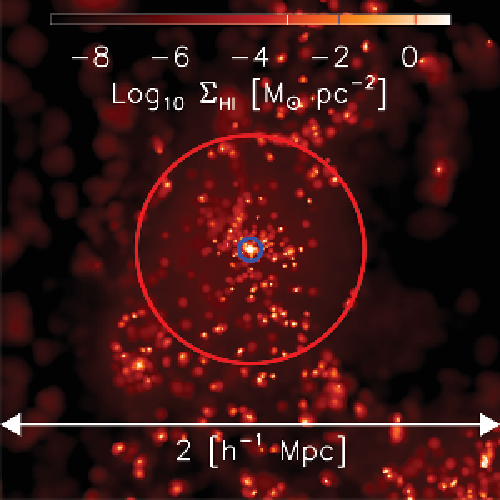} &
     	 \epsfig{figure=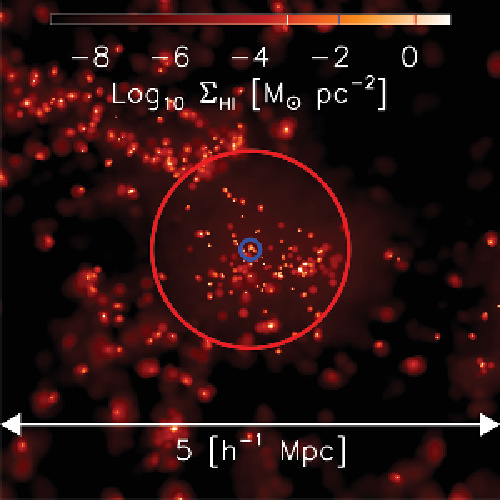} \\

  H$_2$ & \epsfig{figure=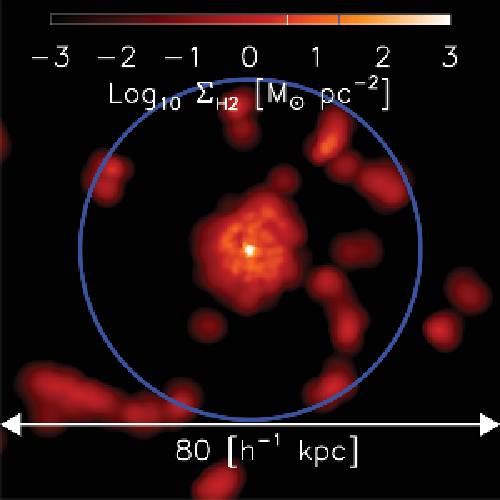} &
     		\epsfig{figure=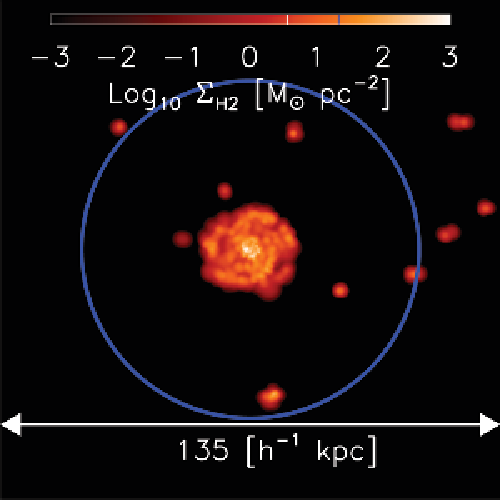} &
     		\epsfig{figure=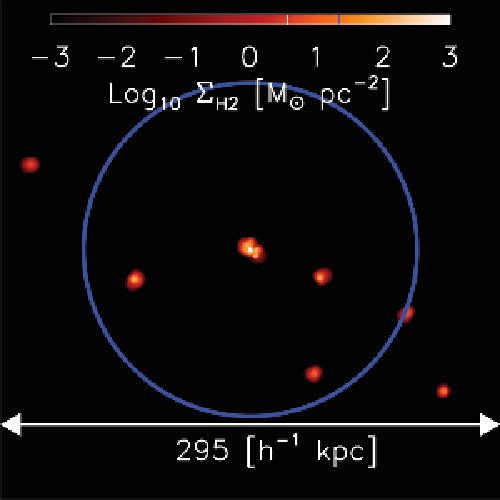} \\

  Stars & \epsfig{figure=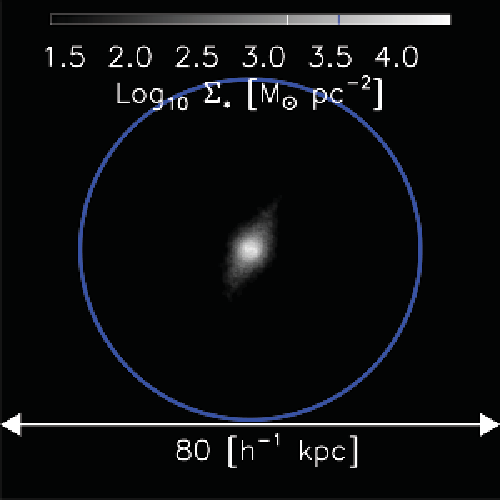} &
	       \epsfig{figure=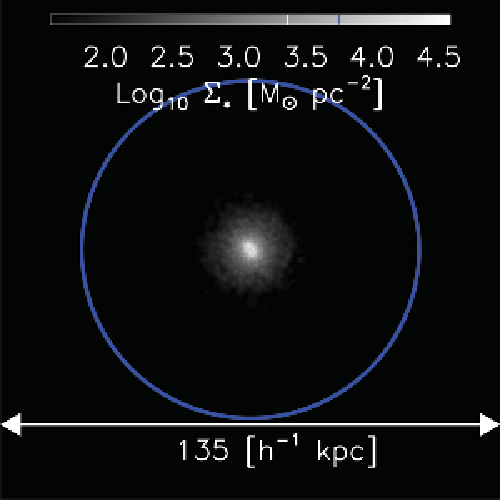} &
     	       \epsfig{figure=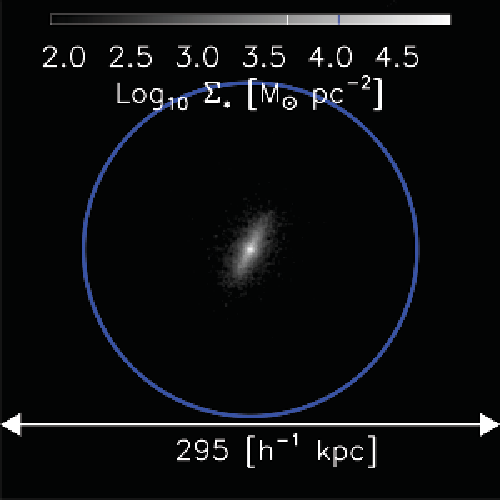} \\
    \end{tabular}
    \caption[H phases map]
            {The distribution of various baryonic components in a typical $z=2$ galaxy (left), $z=0$ group (middle) and $z=0$ cluster halo (right), 
	     with $M_{\rm vir} \approx 10^{12}$, $10^{13}$ and 
              $10^{14}\hMsol$ respectively.   We show the column
              densities of different baryonic components in the rows; from top 
	      to bottom these are HII, HI, $\rm H_2$ and stars. Results are taken  
              from the $N=512$ simulation with $L=25$, $50$ and 
              $100\hMpc$, left to right columns respectively. Note that different panels
              show different size regions, as indicated in each panel. The coordinates (column densities) are in
              comoving (proper) units. Red and blue circles denotes $R_{\rm vir}$ and $0.1\, R_{\rm vir}$, respectively.
              The boxes have the same length along the line of sight as across, as indicated by the arrow. All images have been rotated to present
              the object within $0.1\, R_{\rm vir}$ to be face-on.}
    \label{fig:himap}
  \end{center}
\end{figure*}

In Fig.~\ref{fig:himap} we provide a visual overview of the effects of halo mass and 
redshift on the various distributions of the hydrogen phases.
We compare haloes of virial mass $M_{\rm vir} \approx 10^{12}$,  $10^{13}$ and 
$ 10^{14} \hMsol$ (approximately corresponding to a galaxy, group and cluster, respectively), 
in columns left to right respectively. From top to bottom, the rows show density maps for the
HII, HI, $\rm H_2$ and stellar components. The galaxy halo is at $z=2$ while the
group and cluster haloes are at $z=0$.

Immediately apparent is the diffuse and filamentary nature of the HII in comparison
with the clumpiness of the $\rm H_2$ / HI components,
due to the latter two being associated with the cold and dense gas.

\subsection{Resolution Effects}
\label{sec:reseffects}

Before we study the neutral gas content of galaxies and haloes in detail, we first investigate the
effects of resolution on our HI and H$_2$ mass determinations. This is presented in 
Appendix~\ref{appendix:restest} where we test the effect of finite mass resolution
and simulation box size on the predicted mass of HI (Appendix \ref{appendix:restest_hi}), the HI mass function 
(Appendix \ref{appendix:restest_himassfn}) and the molecular hydrogen mass function (Appendix \ref{appendix:restest_htwomassfn}). 
In summary, we find that the properties of the baryonic species in haloes
are converged above a certain limiting mass, with each species, X, as
\begin{equation}
\label{eqn:hires}
M_{\rm res,\, X} =M_{\rm lim,\, X} \left(\frac{L}{50 {\rm h^{-1} Mpc}}\frac{512}{N}\right)^{3}\,,
\end{equation}
where $N$ is the cube-root of the number of Dark Matter particles and $L$ is the comoving length of the simulation cube.
The co-efficient $M_{\rm lim}$ depends on the species X; in ascending mass order these are 
$M_{\rm lim} = [1, 5, 5, 10, 500]  \times 10^9 \hMsol$ for X$=[{\rm HI},\, {\rm Cold \,Gas},\, {\rm Stellar},\, {\rm H_{2}}, \, {\rm Virial}]$ respectively.
This stellar mass limit implies that approximately 500 star particles are required for convergence, while the Virial (or Total) mass limit is dominated by the Dark Matter 
component and corresponds to nearly 7000 particles; from~\citet{Duffy:10} we are confident that the masses and profiles of such systems are well converged.
For HI, this value is more conservative than assumed by P09 who did not present convergence tests
(see Fig.~\ref{fig:restest_z0_hi} for the comparison of the P09 results and our own resolution test).

\subsection{Halo mass fractions}\label{sec:hi_in_haloes}
We now proceed to investigate the mass fractions of the various baryonic species in haloes. All available simulation volumes 
listed in Table~\ref{tab:sims} are used for the two sub-grid physics models considered, \emph{ZC\_WFB} and \emph{ZC\_WFB\_AGN}. 
Taken together, we are able to probe a virial mass range of $5\times10^{10}\, - \,5\times10^{14} \hMsol$,  representing the largest 
dynamic range in mass yet considered in such a study. Such a large dynamic range allows us to probe whether 
there exists a characteristic mass scale at which HI is particularly abundant. This is an important issue for
understanding the selection effects that a blind, flux-limited HI sample would suffer from,
as well as the prospects of using HI surveys for cosmology. We only consider self-shielding method $\SSALF$ in this section as we are
primarily interested in the sensitivity of the various components to halo mass, redshift and galaxy formation modelling
(specifically whether we include AGN or not).

\begin{figure*}
  \begin{center}
  \begin{tabular}{ccc}
      \epsfig{figure=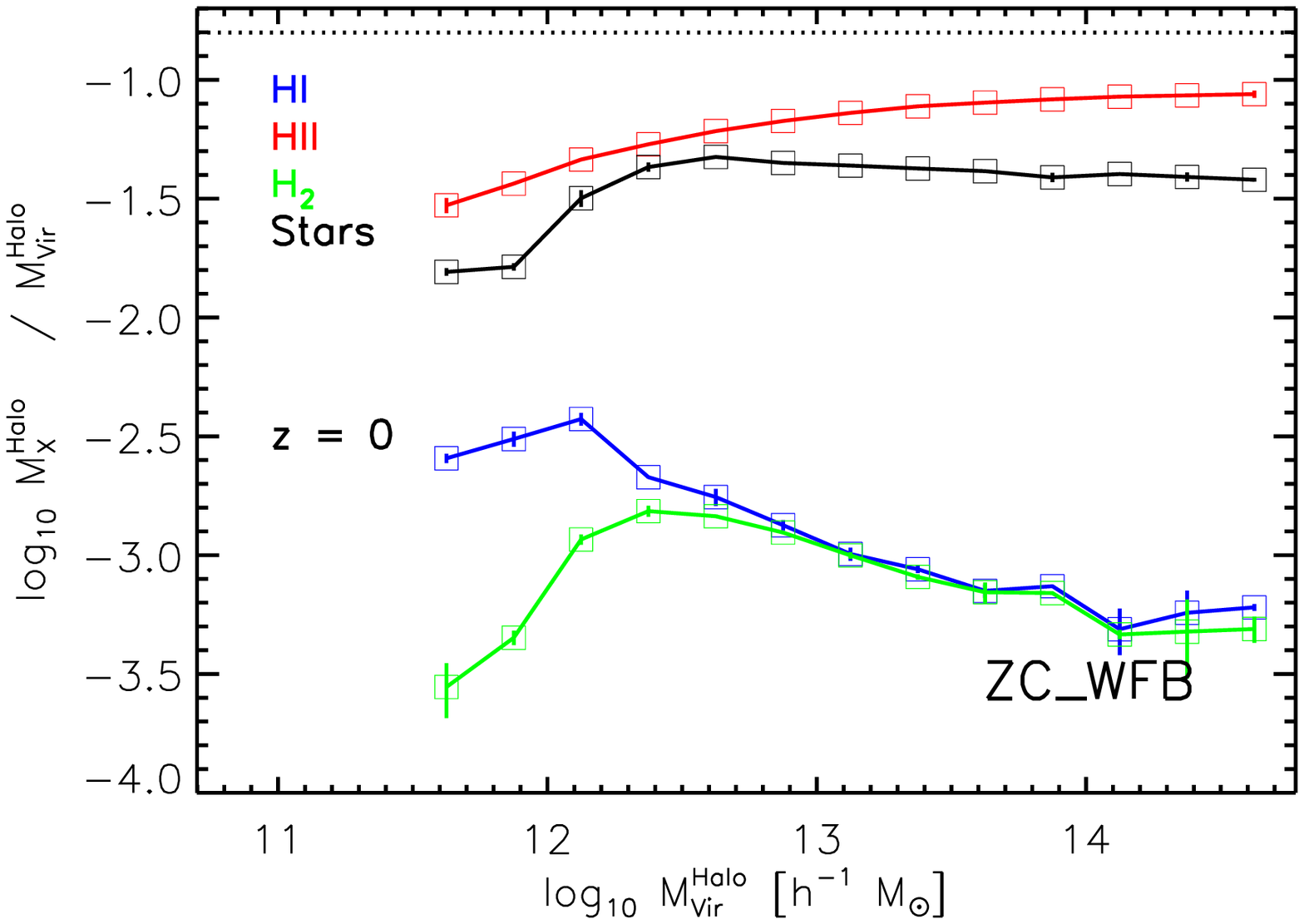, scale = 0.3} &
      \epsfig{figure=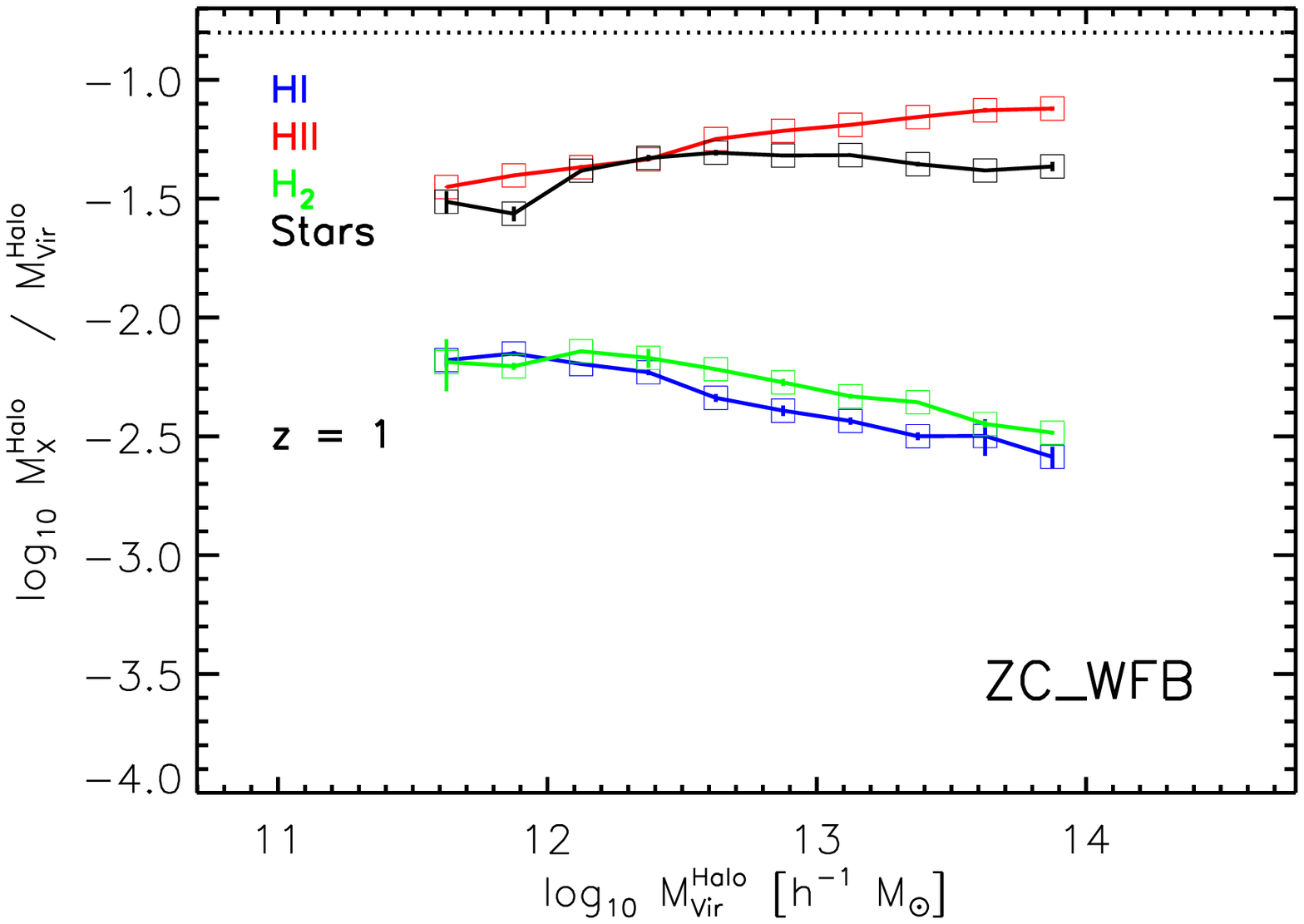, scale = 0.3} &
      \epsfig{figure=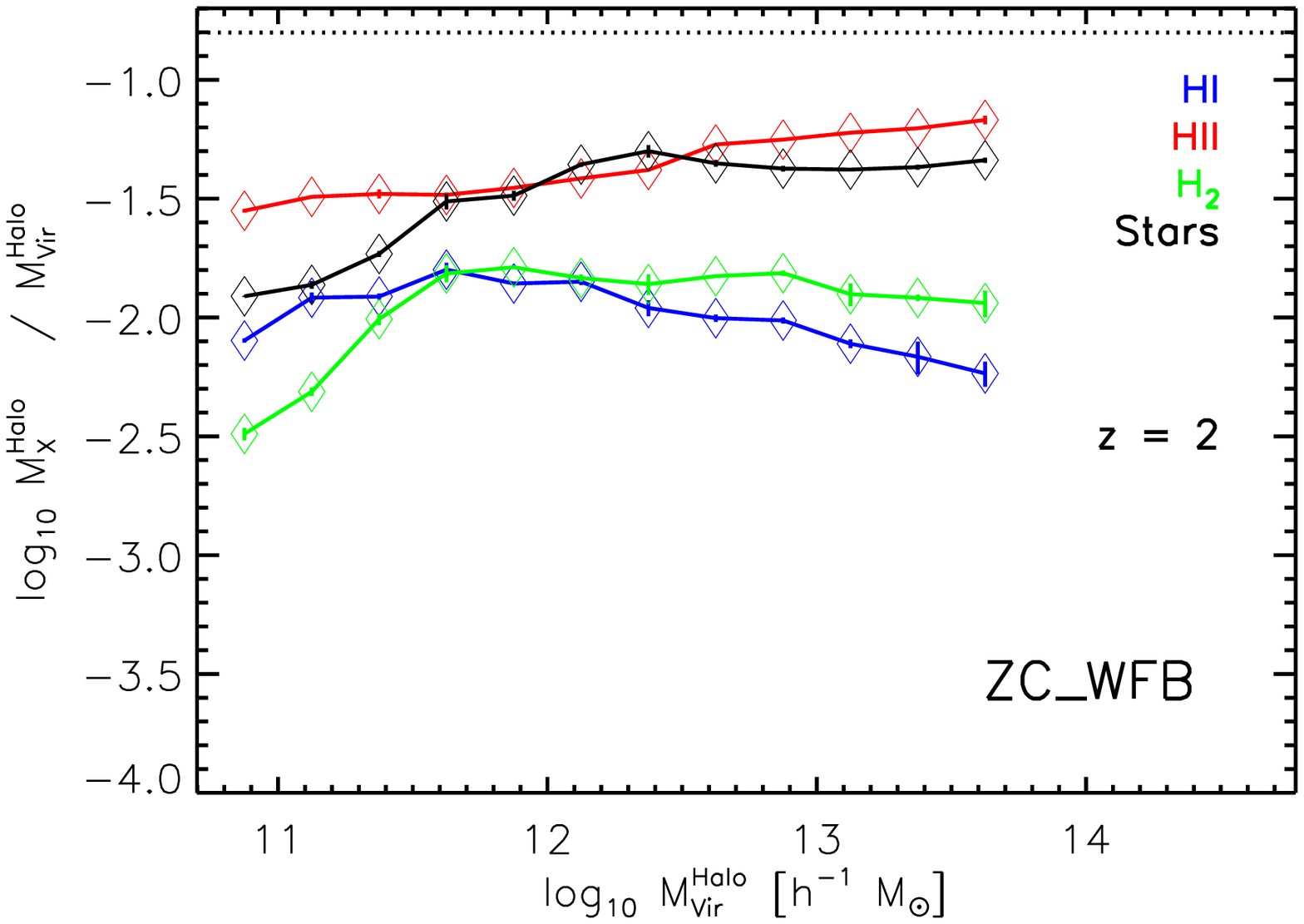, scale = 0.3} \\
      \epsfig{figure=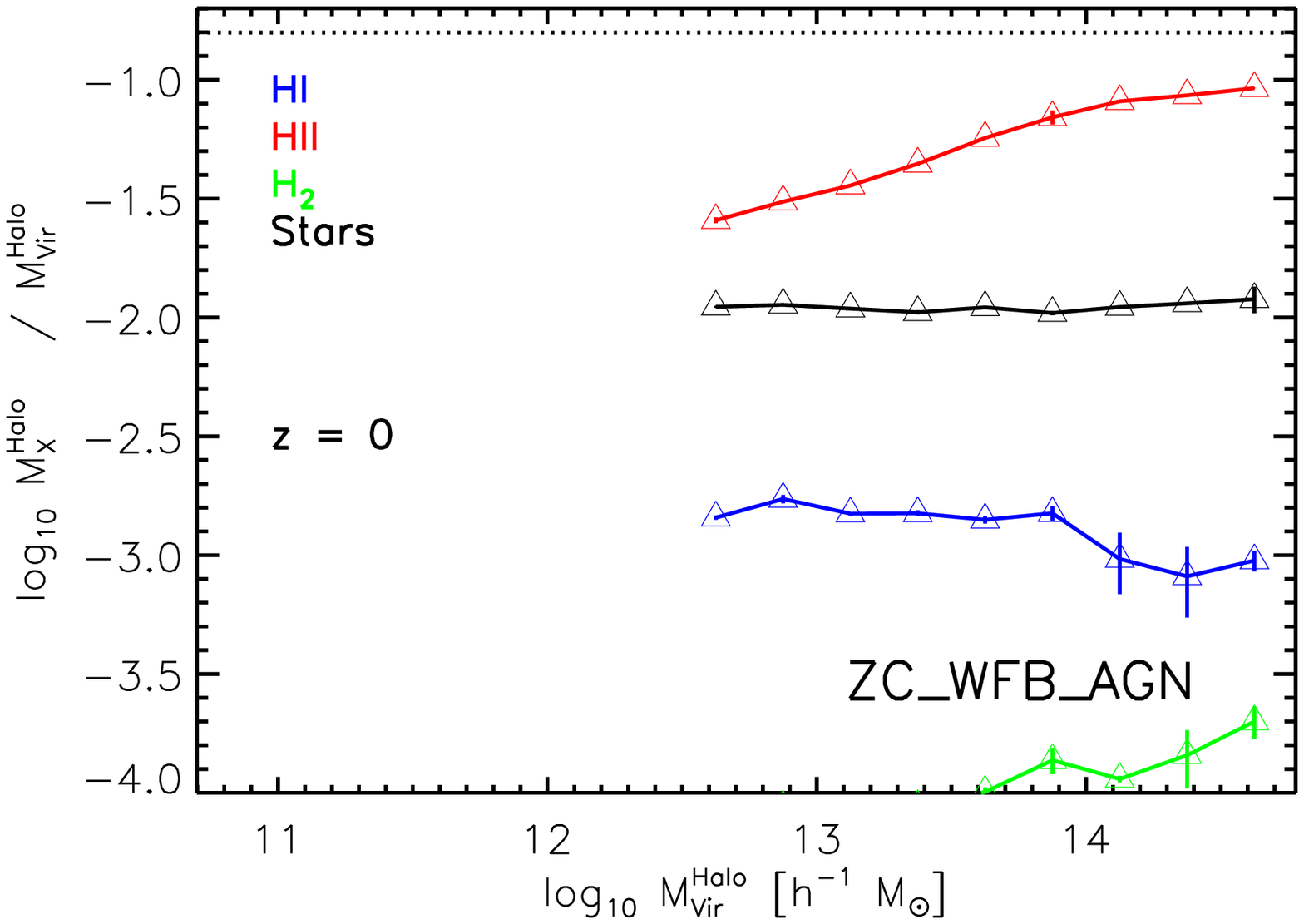, scale = 0.3} &
      \epsfig{figure=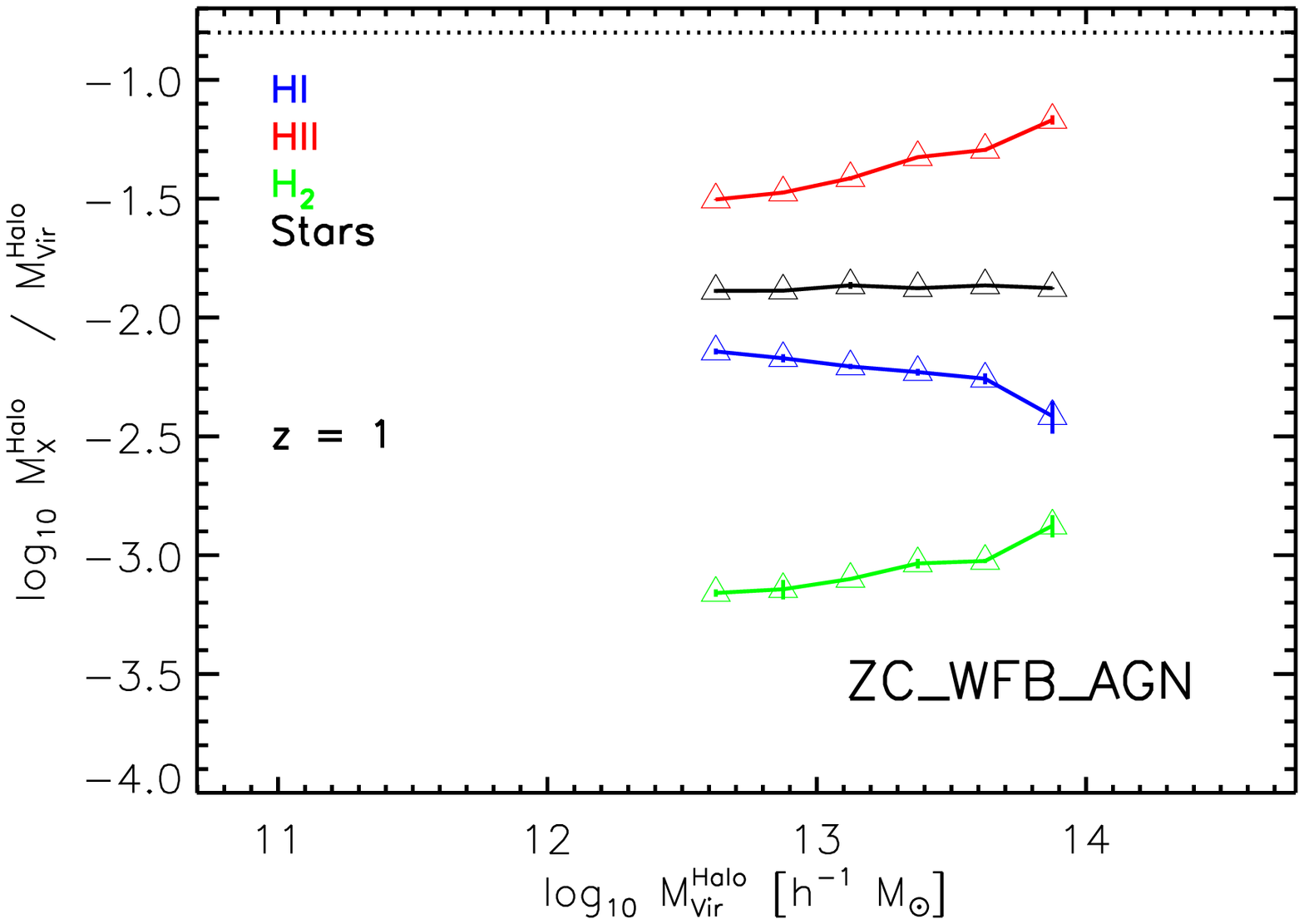, scale = 0.3} &
      \epsfig{figure=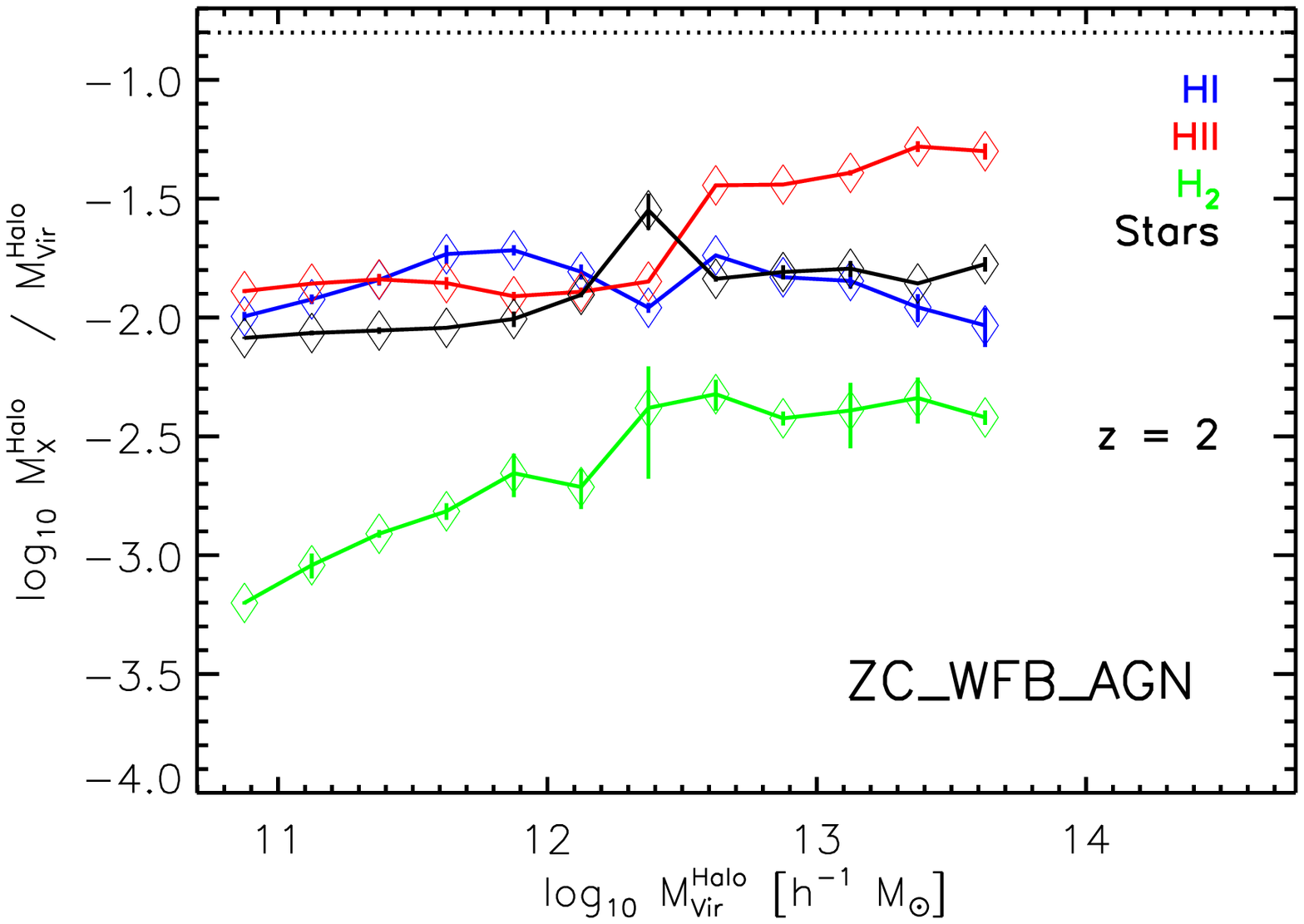, scale = 0.3} \\
      \end{tabular}
    \caption{The HI, HII, $\rm H_2 $ and stellar mass fractions of haloes as a 
      function of the total halo mass for $z=0$, $1$ and $2$, left to right 
      respectively. The horizontal dotted line is $0.9\,f^{\rm univ}_{\rm b}$,
      the value found in non-radiative gas simulations of~\citet{Crain:07}.
      We consider runs with metal line emission cooling and
      supernovae feedback, \emph{ZC\_WFB}, in the top row and then the runs with the
      AGN scheme added, \emph{ZC\_WFB\_AGN}, in the bottom row. This latter case is particularly important
      when considering the stellar fractions in massive haloes. Method $\SSALF$ (using ALFALFA data)
      is used for the self-shielding approximation. We only consider haloes with a virial mass above the limit in Equation~\ref{eqn:hires},
      these haloes typically have $10^{4}$ dark matter particles. Note the smaller dynamic range 
      in the AGN case at $z=0$ and 1 where we do not have an \emph{L050N12} simulation to probe galaxy scales.}
    \label{fig:fullmassfn_mhi_mfof}
  \end{center}
\end{figure*}

In the top row of Fig.~\ref{fig:fullmassfn_mhi_mfof} we present the mass fraction of the baryonic species for \emph{ZC\_WFB},
as a function of halo mass at redshifts $z=0, 1$ and 2 (left, middle and right panels, respectively).
The bottom row shows the corresponding results for the simulation including AGN feedback (\emph{ZC\_WFB\_AGN}).

At $z=0$ the ionised gas is the largest baryonic mass component of all haloes, followed by stars, although the two are comparable
for $M_{\rm vir} \approx 10^{12} - 10^{13} \hMsol$.
This pattern is seen at all redshifts both with and without AGN, although the ionised component is less dominant at higher redshifts. 

For the simulation without AGN feedback (top row)
the HI has a more complex distribution. At $z=0$, top left panel, there is an initial rise and then a clear decrease 
in the HI as a fraction of the total halo mass, beginning at $0.5$ percent at 
$M^{\rm Halo}_{\rm vir} \approx 10^{12} \hMsol$ and decreasing to
 nearly an order of magnitude less at 
$M^{\rm Halo}_{\rm vir} \approx 5 \times 10^{14} \hMsol$.
The pattern is similar at $z=2$ where the increased 
dynamic range allows us to identify a broader peak around a smaller mass,
$M^{\rm Halo}_{\rm vir} \simeq 5 \times 10^{11} \hMsol$. {\it However, at $z=2$ the HI mass fractions are typically an order of 
magnitude higher than at $z=0$.} The $\rm H_2$ fractions broadly track the HI fractions although they are
noticeably lower in the lowest mass haloes. 
 In such systems, the mass fraction of baryons
as a whole decreases as the gas is expelled by supernova feedback, leading to less
efficient star formation. At higher masses
(massive galaxies, groups and clusters) the decline in neutral fractions is due to the gas being
used up and turned into stars, as the supernovae are no longer able to blow significant amounts of
gas out of the halo. Ram pressure stripping of the cold gas in the hot halo atmospheres and the larger cooling times also
contribute. 

We also see evidence of a form of  `downsizing'~\citep[e.g.][]{Pozzetti:03,Fontana:04,Drory:03,Drory:05} 
in the HI and $\rm H_2$ fractions as a function of redshift, which decrease faster for higher halo masses. 
We attribute this to a runaway conversion of this gas into stars for the \emph{ZC\_WFB} simulation, as it was 
noted in~\citet{Duffy:10} that this particular physics scheme overproduces the stellar fraction in groups.
This can be seen in the bottom panels of Fig.~\ref{fig:fullmassfn_mhi_mfof} where we include AGN which 
prevents the cooling flows, giving a low stellar (and cold gas) mass fraction. AGN feedback strongly suppresses the molecular
mass fractions.
The HI fraction, however, is typically slightly greater than in the simulation without
AGN feedback.
This is due to the cold gas typically being at lower density in the AGN
model due to the stronger feedback.

\subsection{Stars and Neutral Hydrogen}\label{sec:stars_vs_hi}
In Fig.~\ref{fig:stars_vs_hi} we show the HI mass as a function of 
the stellar mass of the galaxies in \emph{ZC\_WFB} at $z=0$, $1$ and $2$, top to bottom 
panels respectively. There is a clear positive correlation between the two 
parameters, namely that
larger stellar mass systems also have more HI, as shown in the left column. 
We probe this correlation over approximately 2 decades of stellar mass at $z=2$.

\begin{figure*}
  \begin{center}
    \begin{tabular}{ccc}
      \epsfysize=2in \epsfxsize=4in
      \epsfig{figure=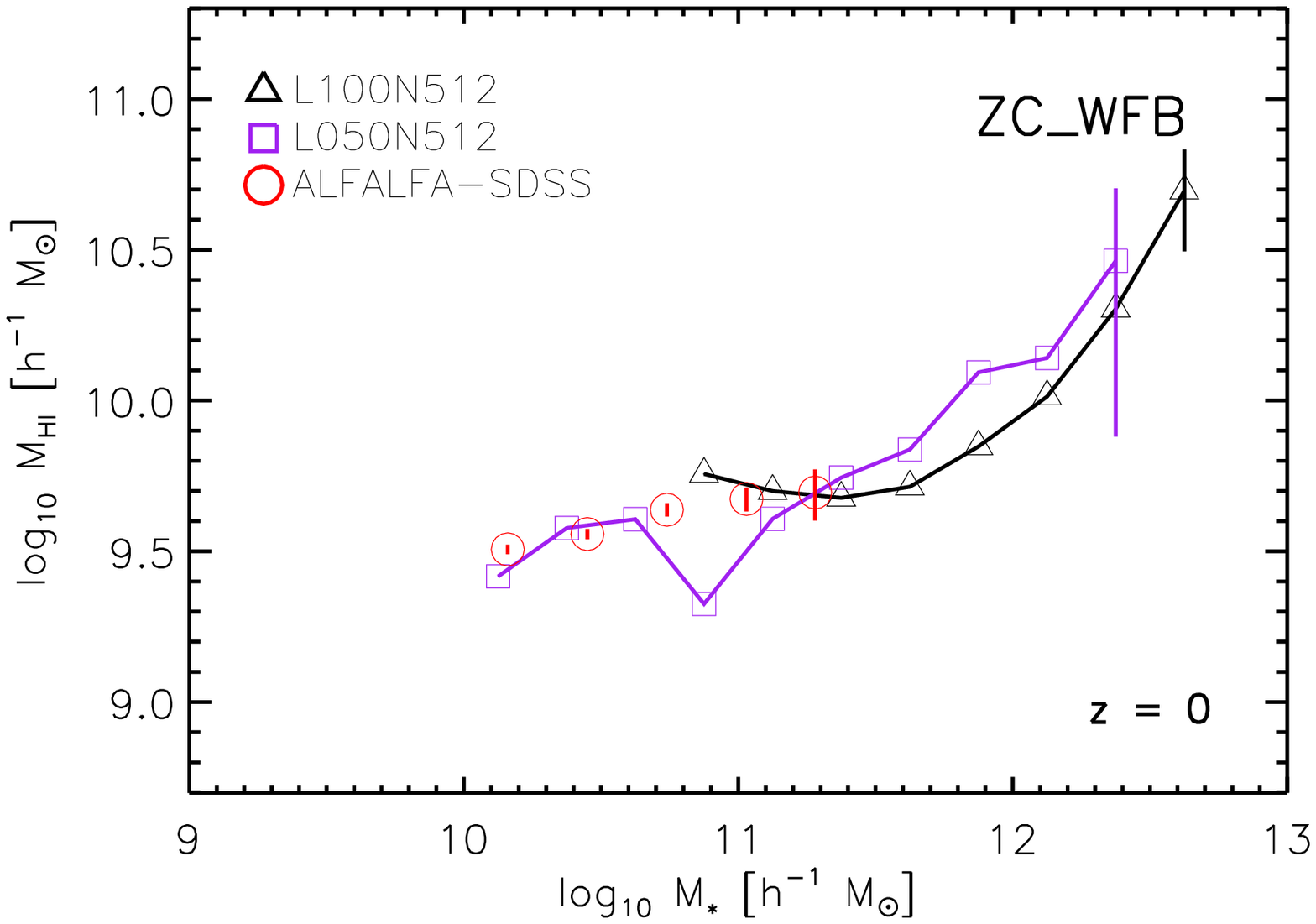, scale=0.3} &
      \epsfig{figure=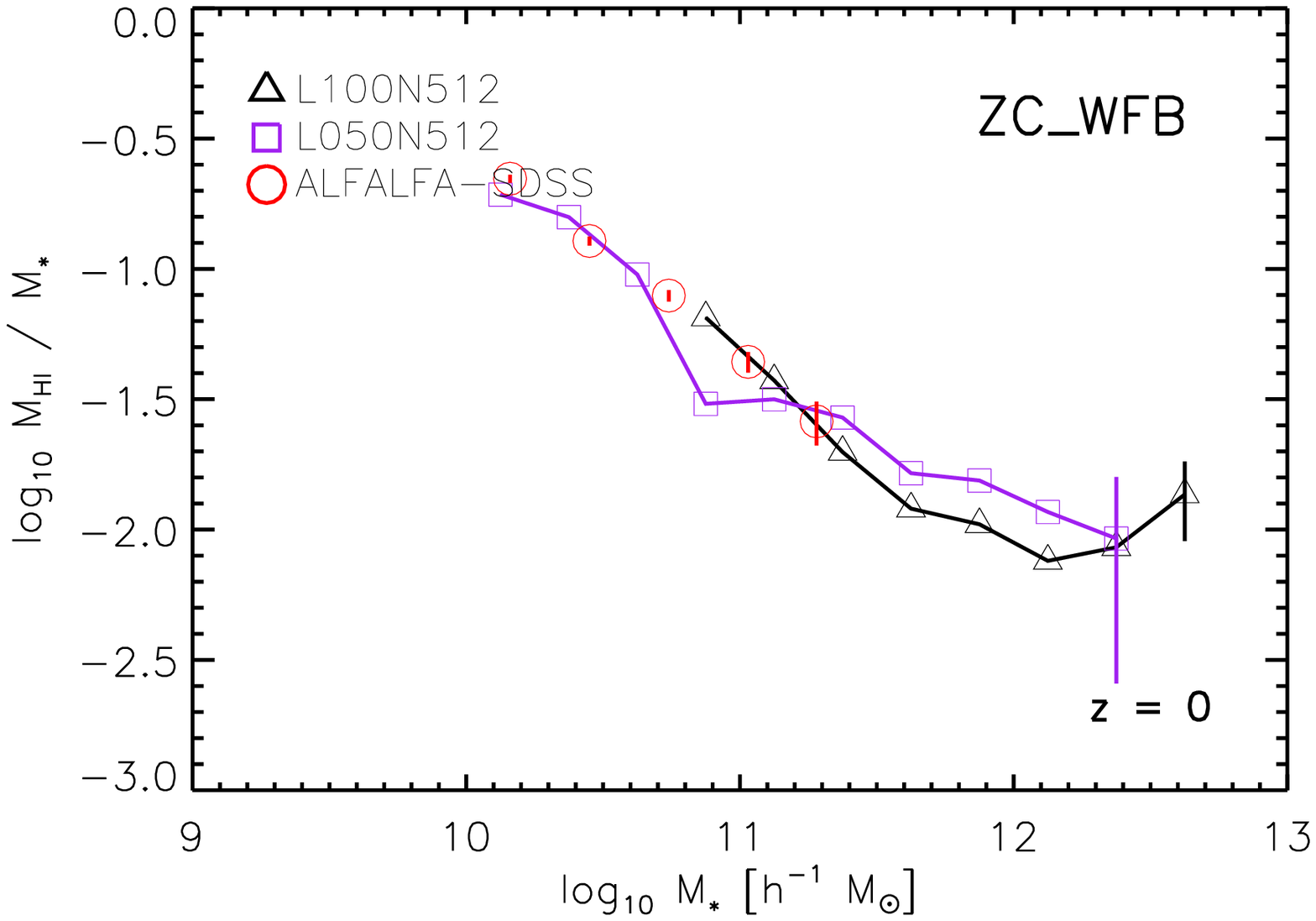, scale=0.3} &
      \epsfig{figure=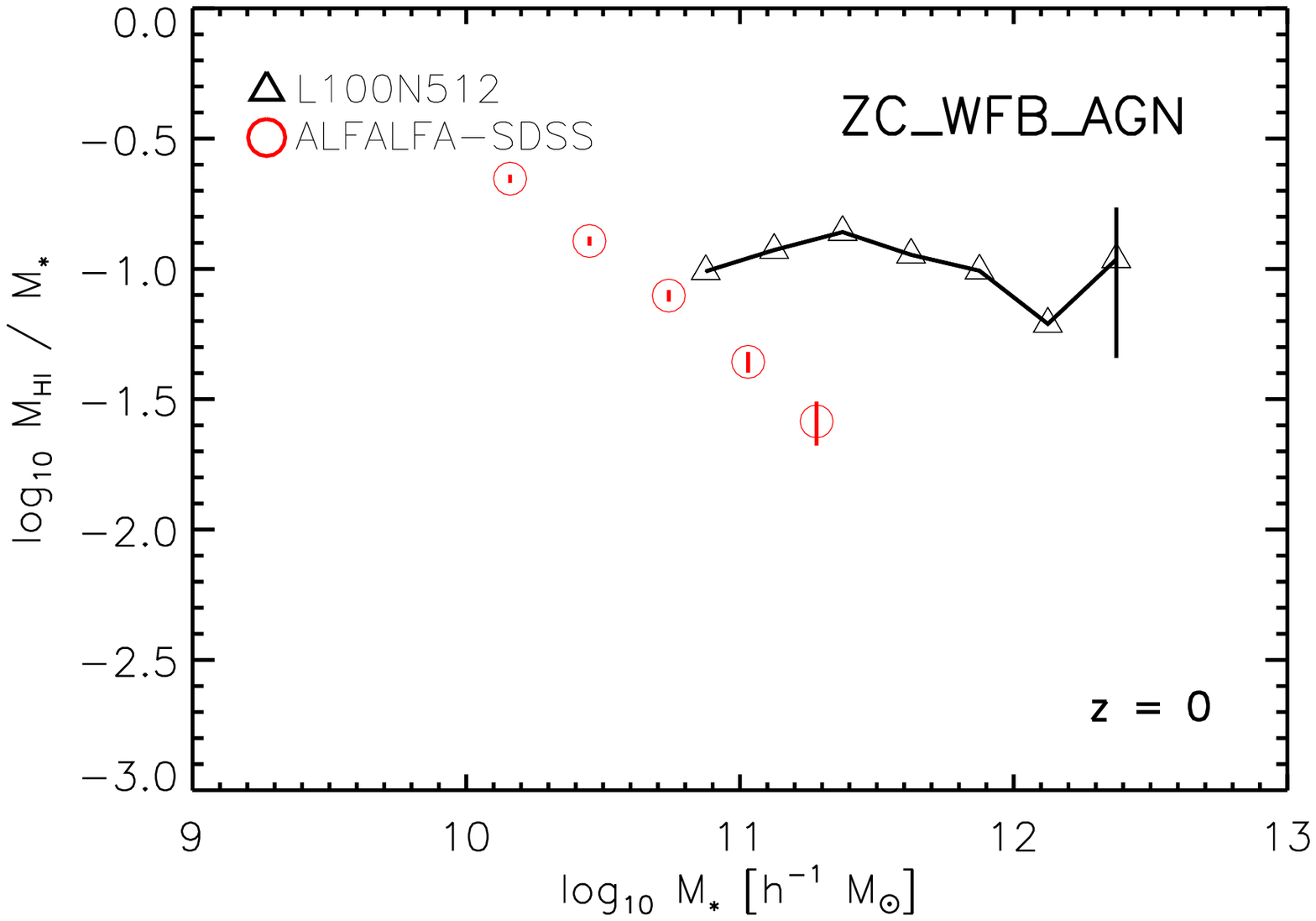, scale=0.3} \\      
      \epsfig{figure=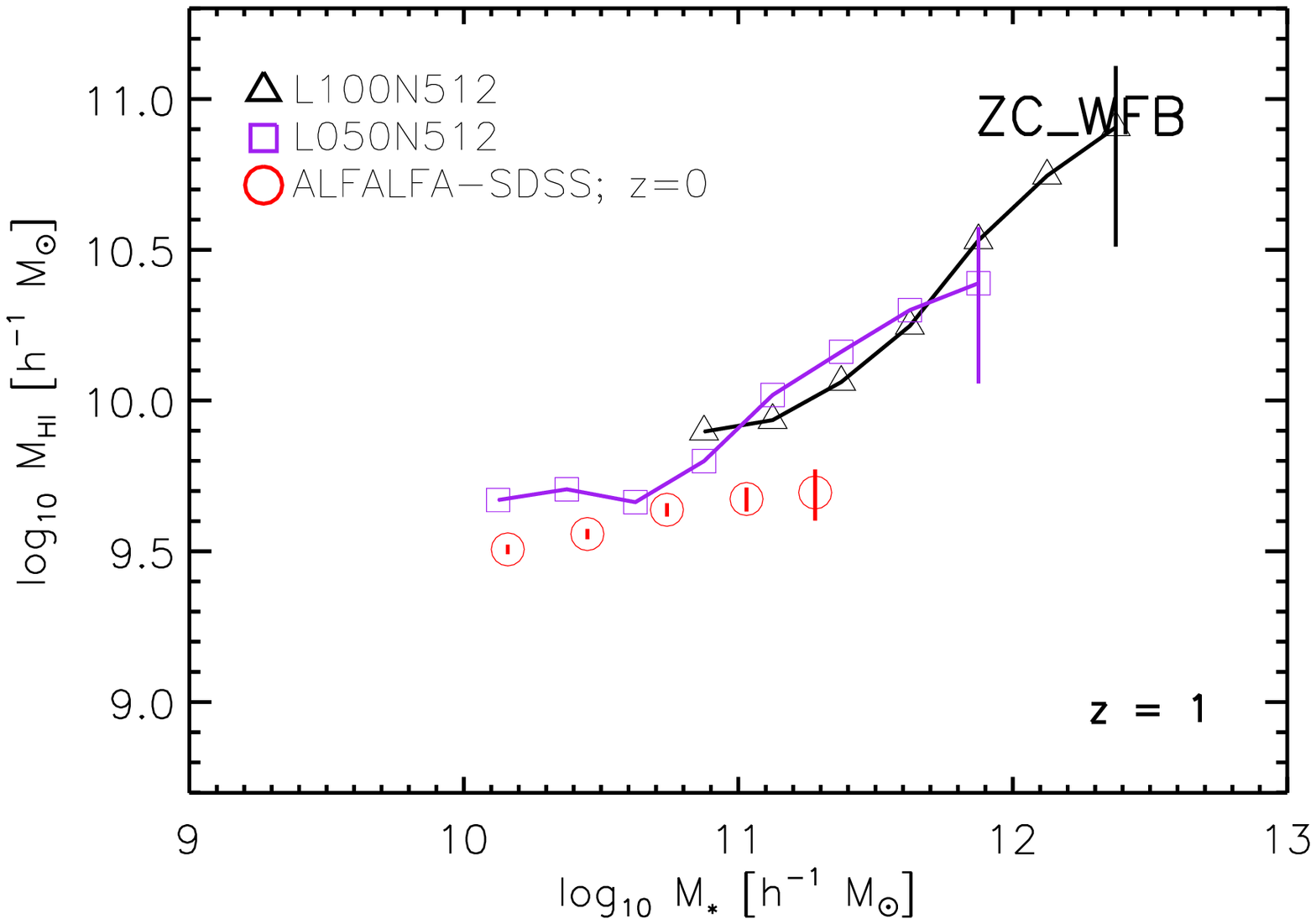, scale=0.3} &
      \epsfig{figure=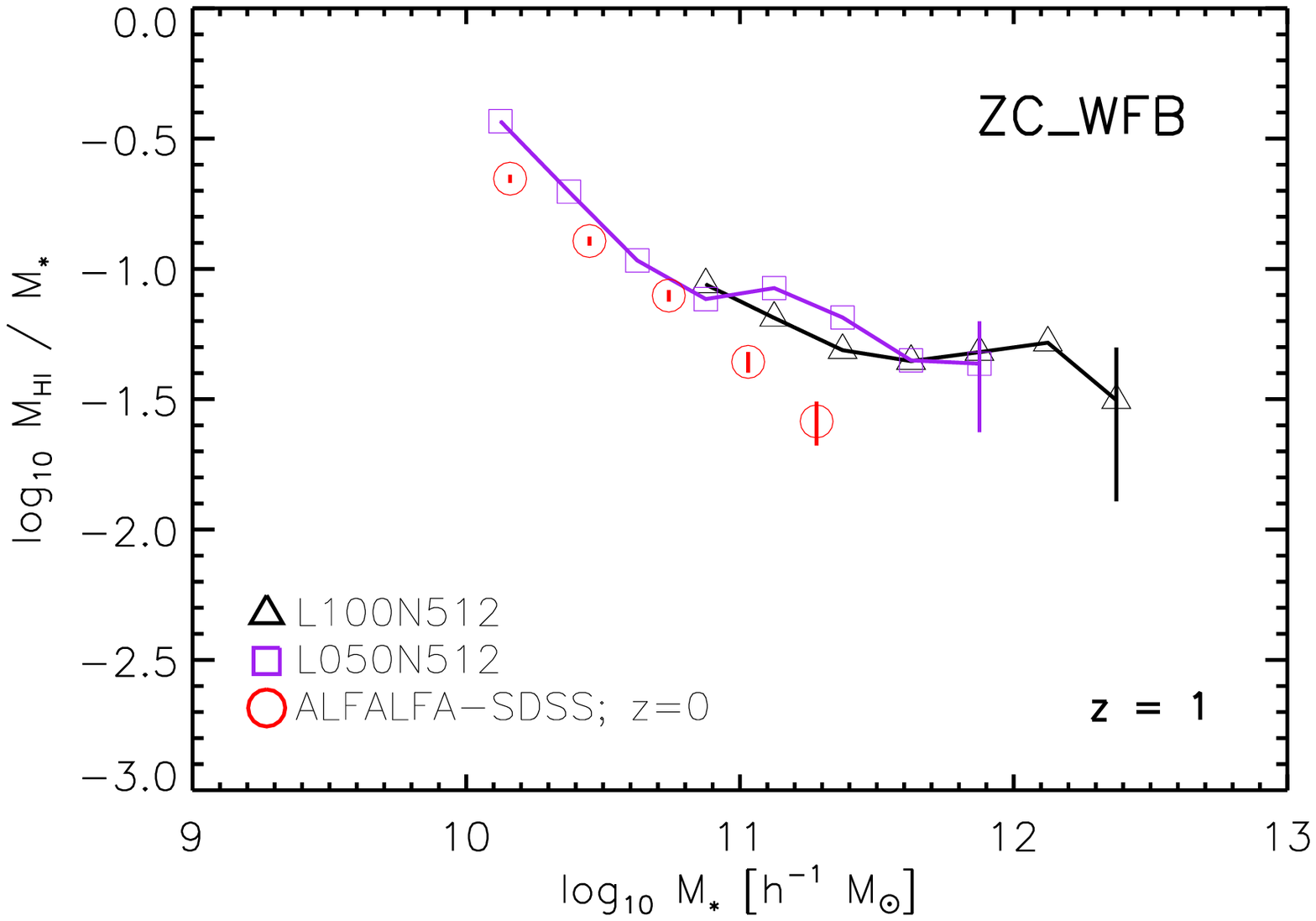, scale=0.3} &
      \epsfig{figure=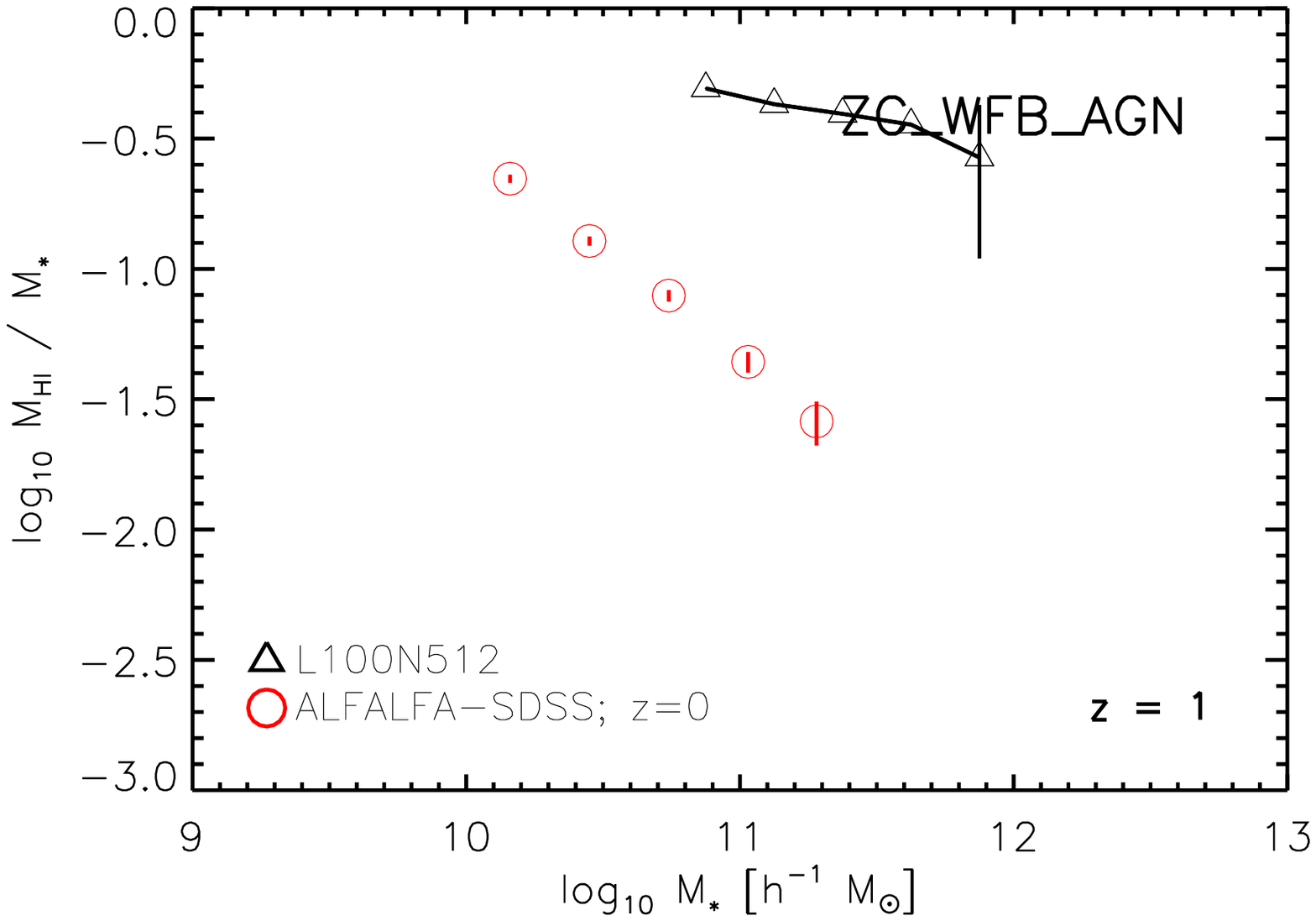, scale=0.3} \\
      \epsfig{figure=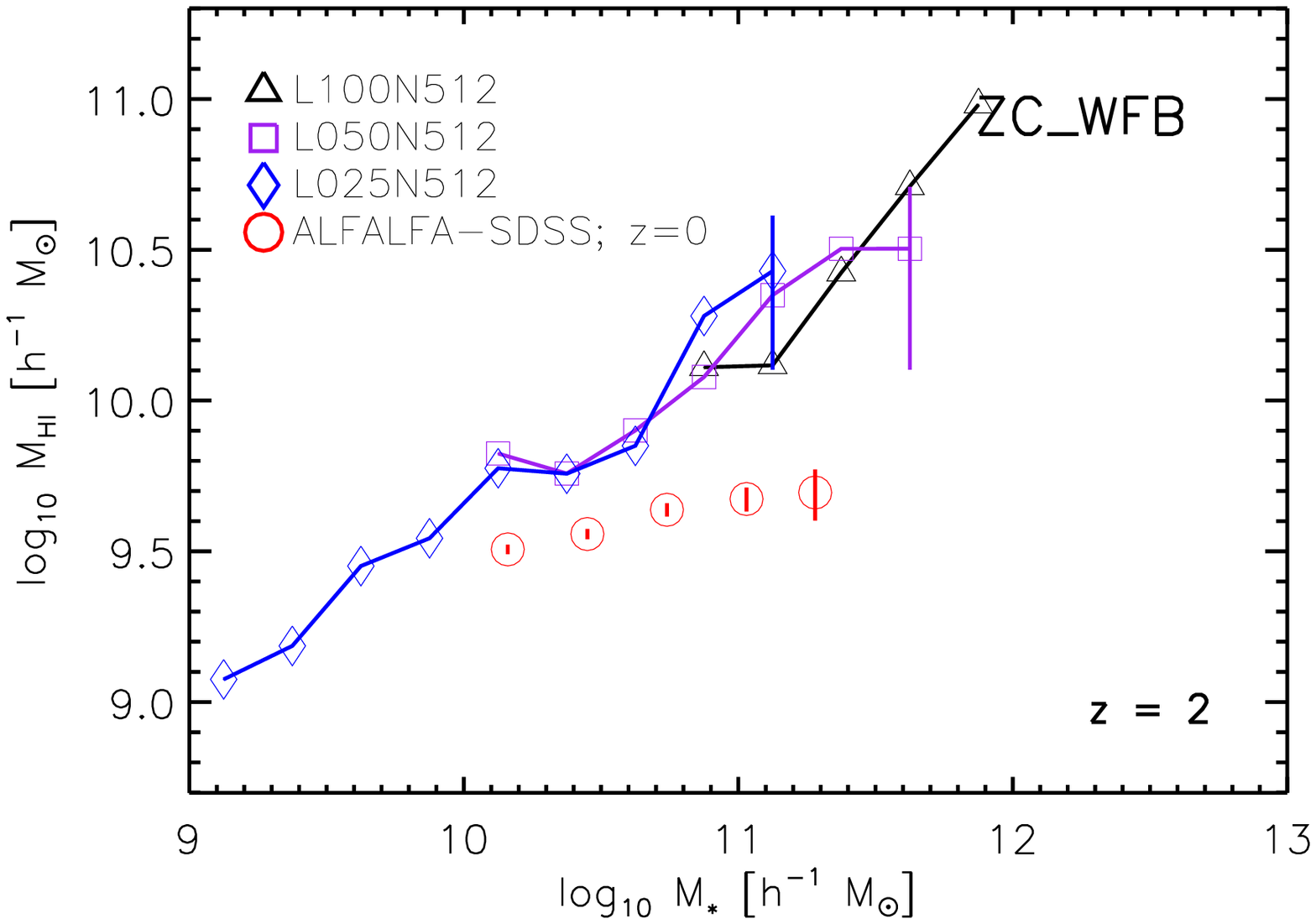, scale=0.3} &
      \epsfig{figure=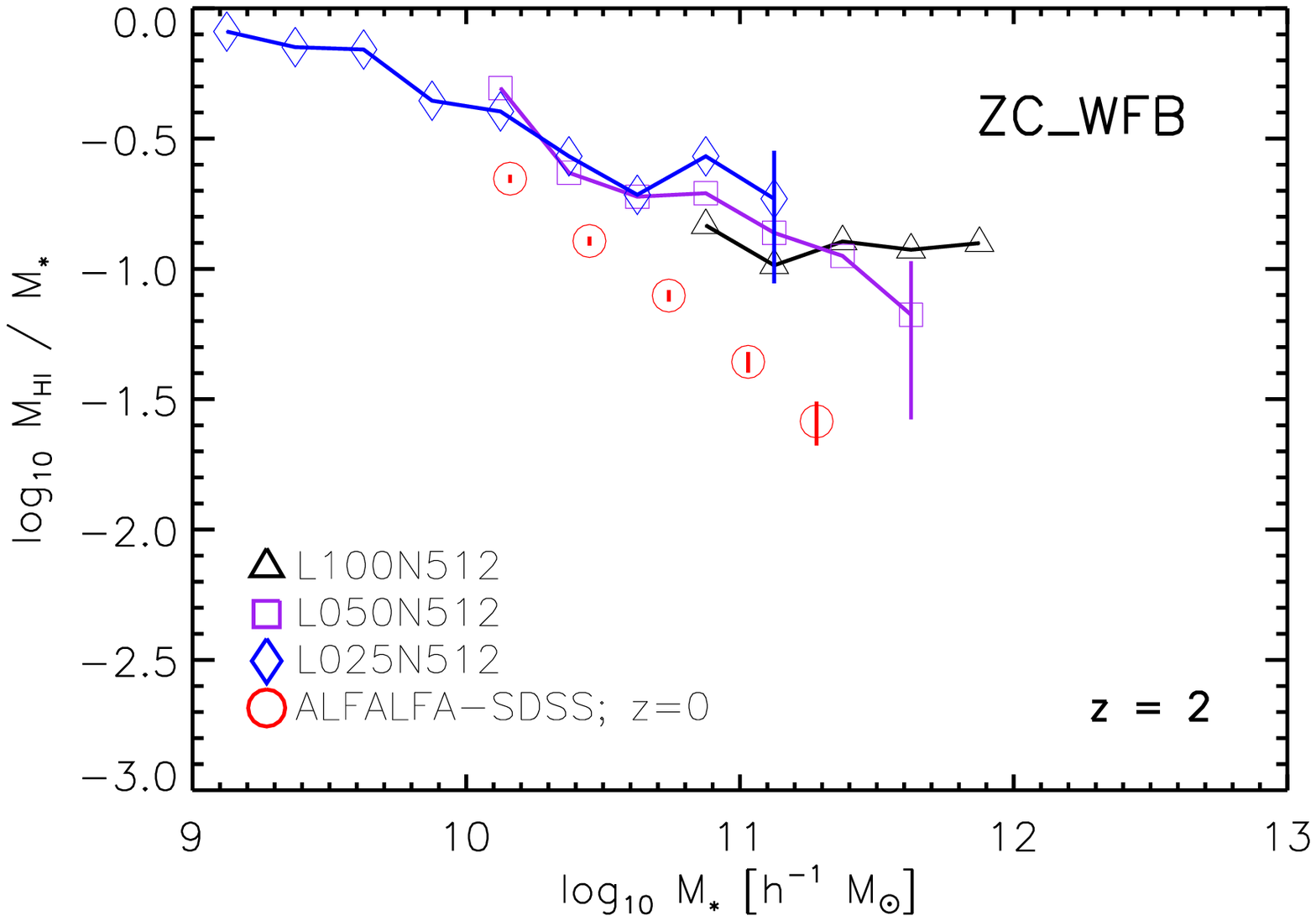, scale=0.3} &
      \epsfig{figure=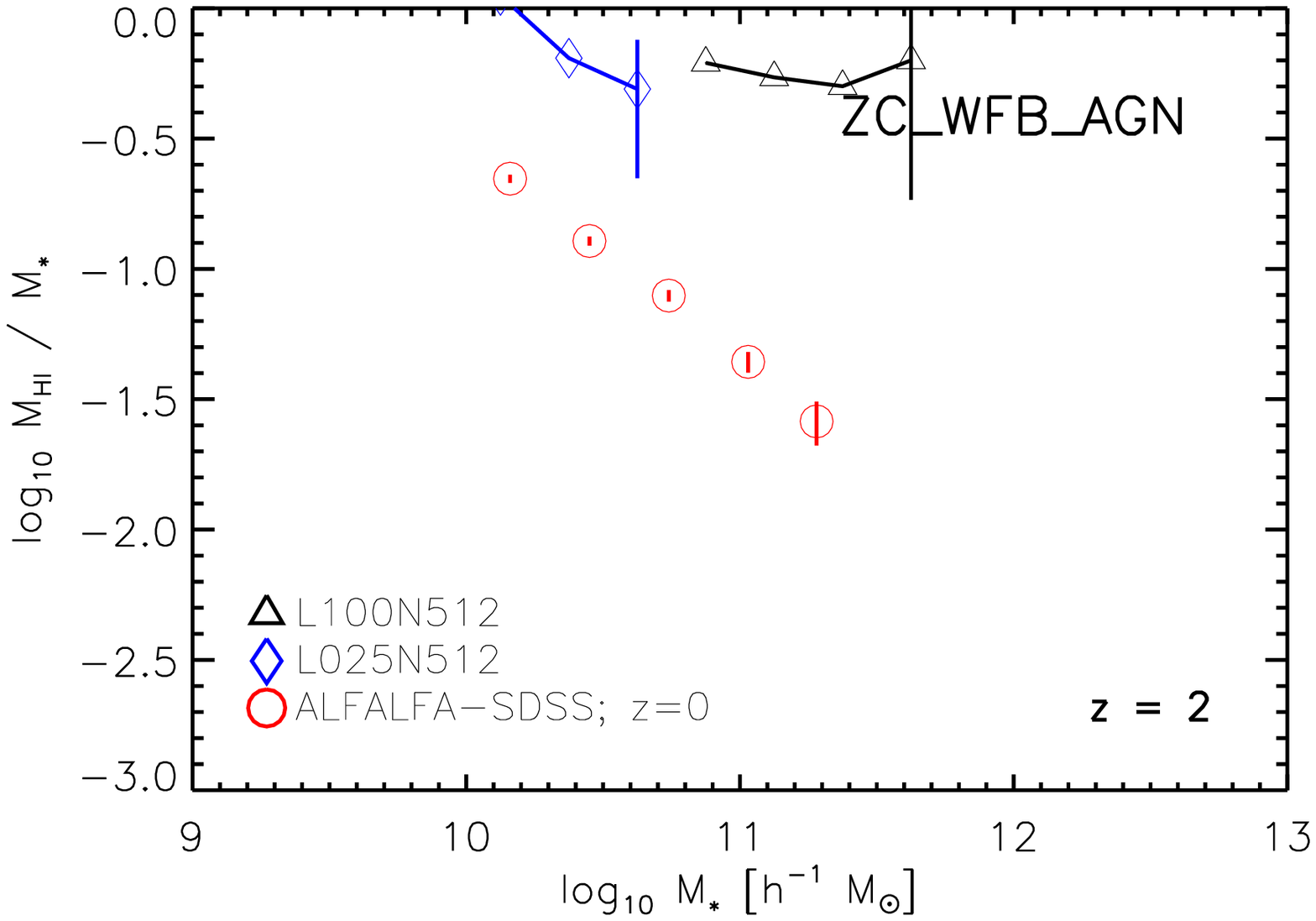, scale=0.3} \\      
      \end{tabular}
    \caption{In the left column we present HI masses of all subhaloes in 
      logarithmic bins of stellar mass $z=0$, $1$ and 
      $2$, top to bottom respectively. The middle and right columns show the ratio of 
      HI mass to stellar mass of each halo as a function of stellar mass.
      Results are for the \emph{ZC\_WFB} (left column) and the AGN model (right column, \emph{ZC\_WFB}\_AGN). 
      We have included a range of simulation volumes to attain the maximum 
      dynamic range. We find that $5 \times 10^{3}$ stellar particles is sufficient to attain a well resolved value.
      The errors are the $1\sigma$ errors estimated by bootstrap analysis and
      added in quadrature with Poisson error estimates (typically only important for the highest mass bins in a simulation). The red circles are from 
      a joint ALFALFA-Sloan Digital Sky Survey comparison at low-redshift, compiled by~\citet{Fabello:11}, which
      nicely agree with our \emph{ZC\_WFB} simulation. Intriguingly the more realistic model which includes AGN overproduces the observed
      relation; to attain similar stellar masses the haloes with AGN are more massive. This also means that the haloes with have a higher HI mass
      as shown in Fig.~\ref{fig:fullmassfn_mhi_mfof}.}
    \label{fig:stars_vs_hi}
  \end{center}
\end{figure*}

\begin{figure*}
  \begin{center}
    \begin{tabular}{ccc}
      \epsfysize=2in \epsfxsize=4in
      \epsfig{figure=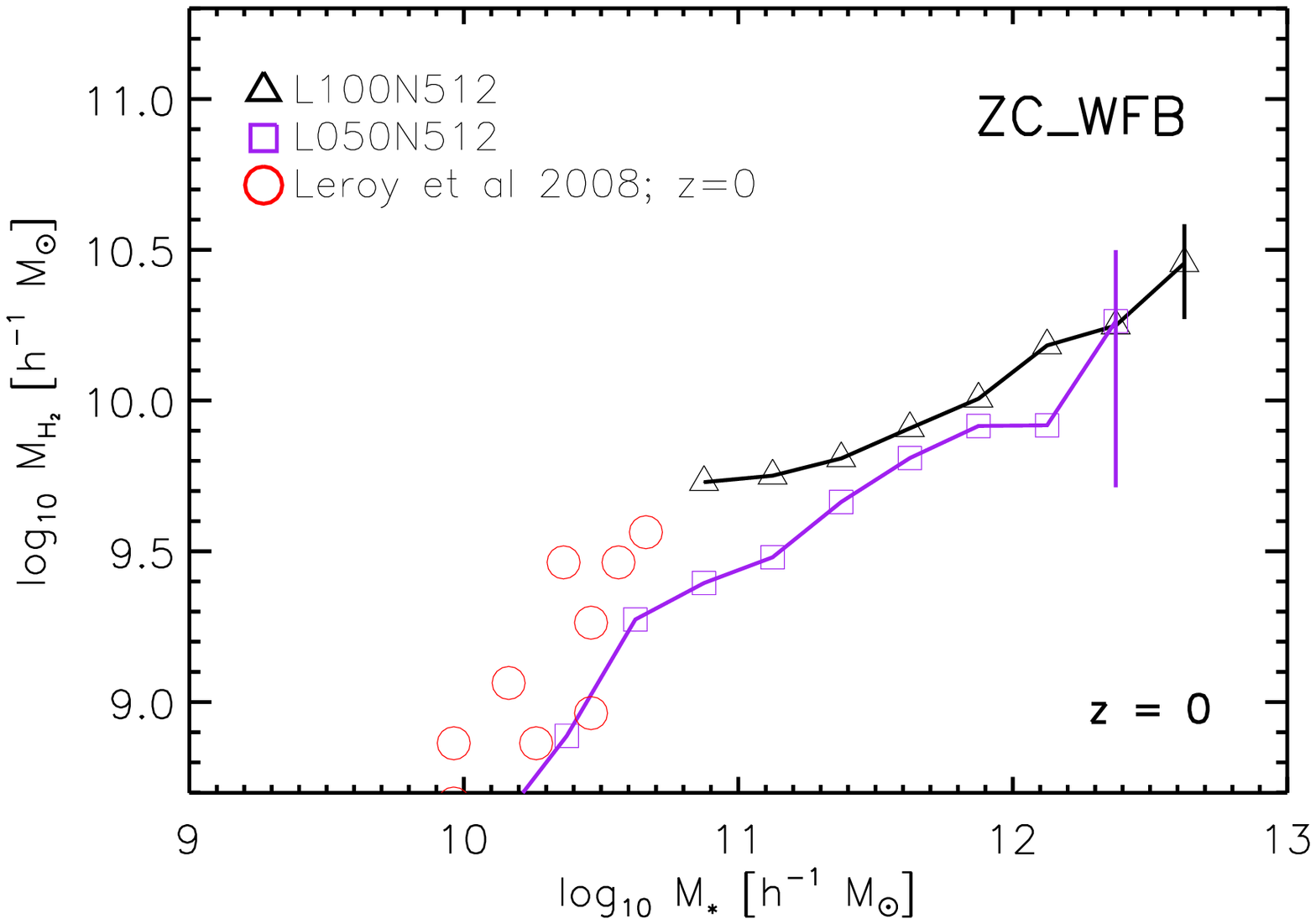, scale=0.3} &
      \epsfig{figure=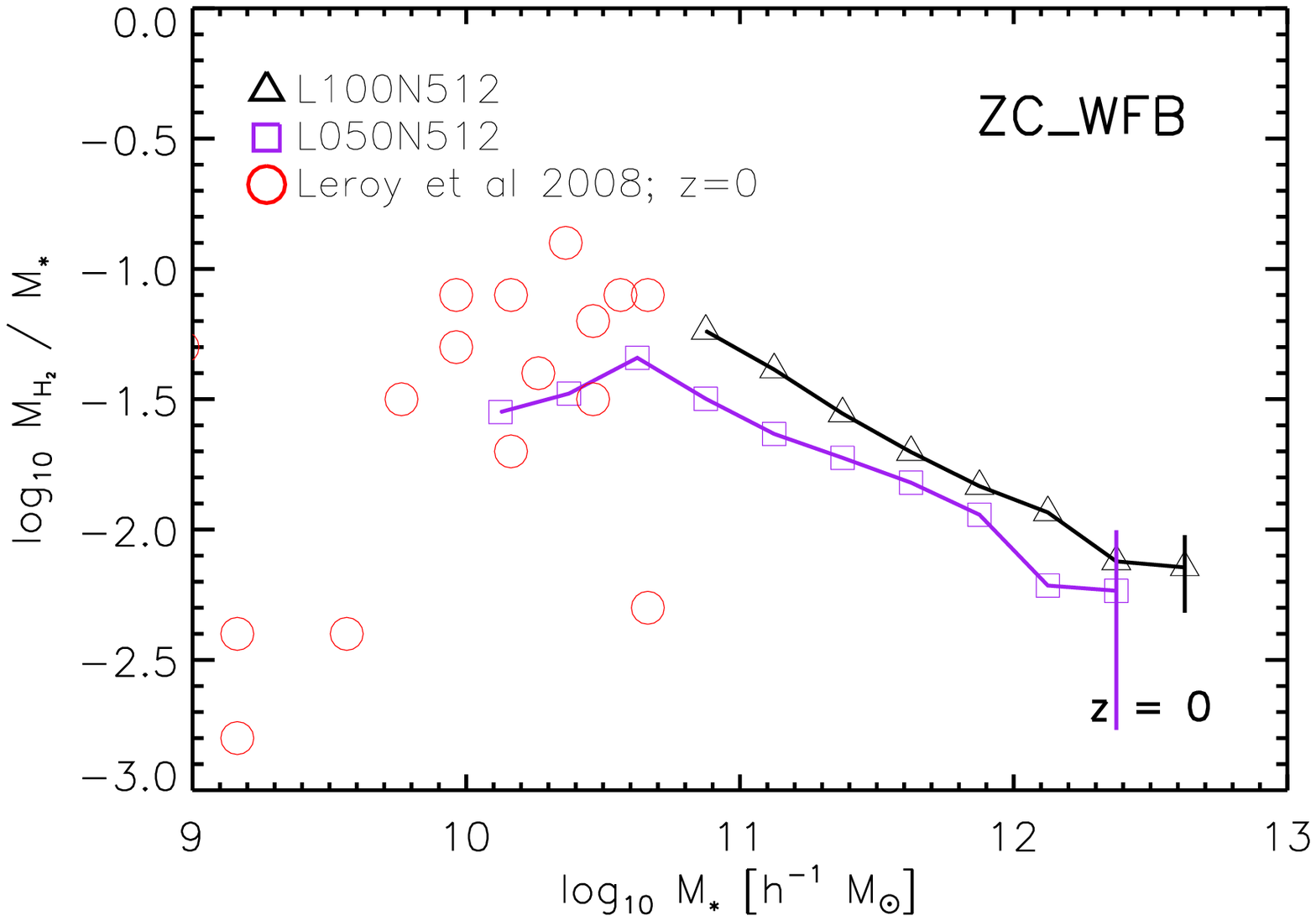, scale=0.3} &
      \epsfig{figure=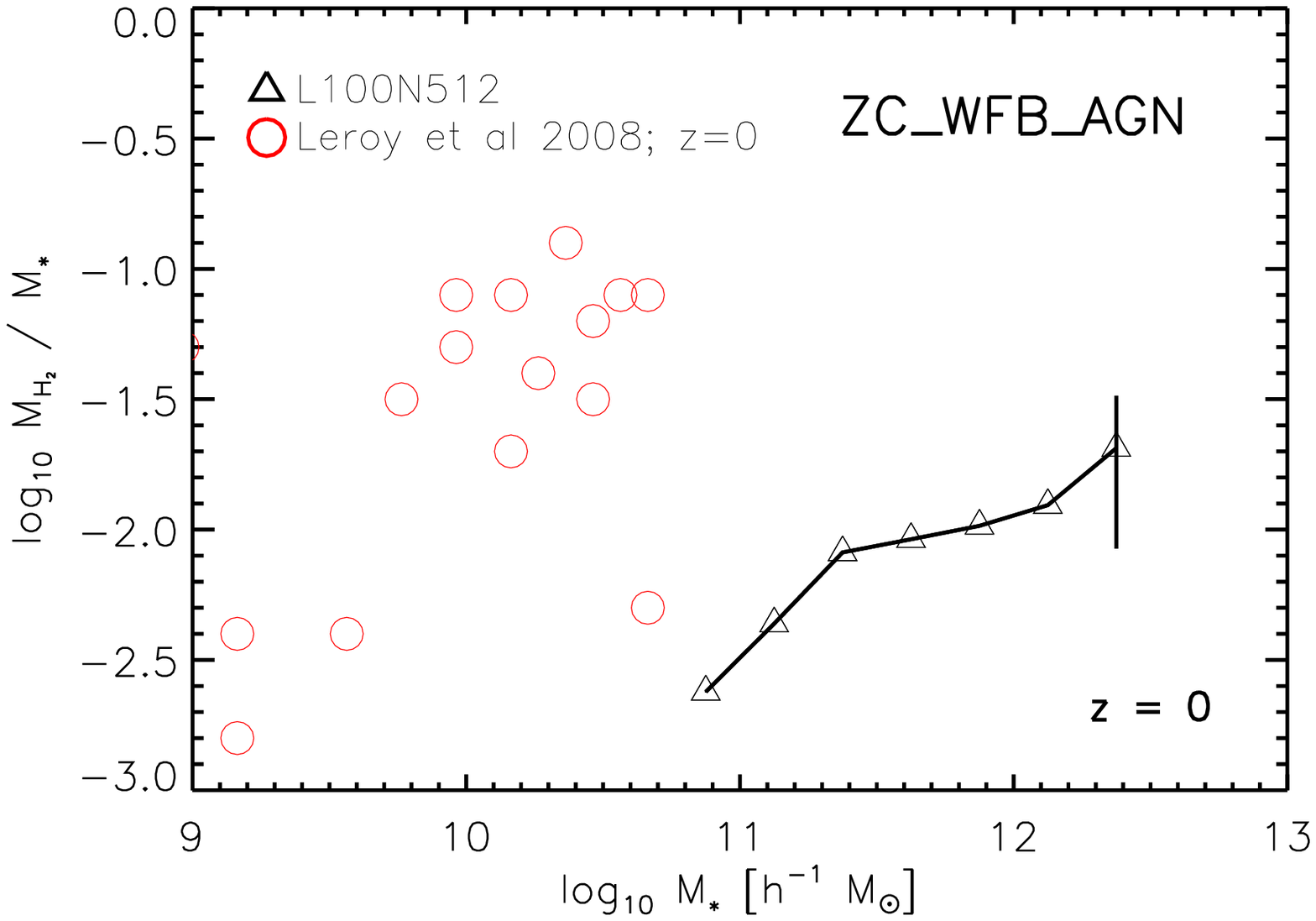, scale=0.3} \\      
      \epsfig{figure=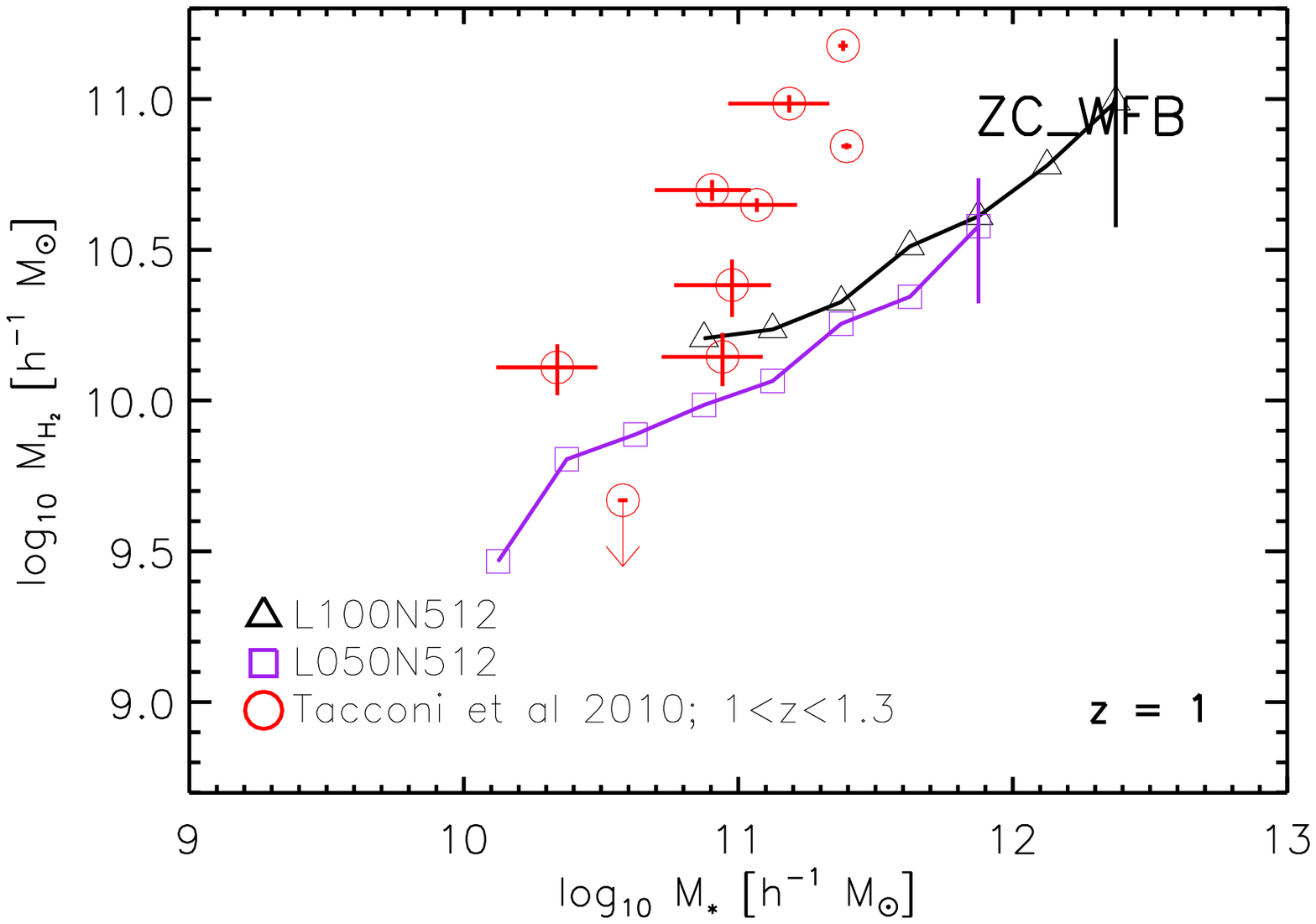, scale=0.3} &
      \epsfig{figure=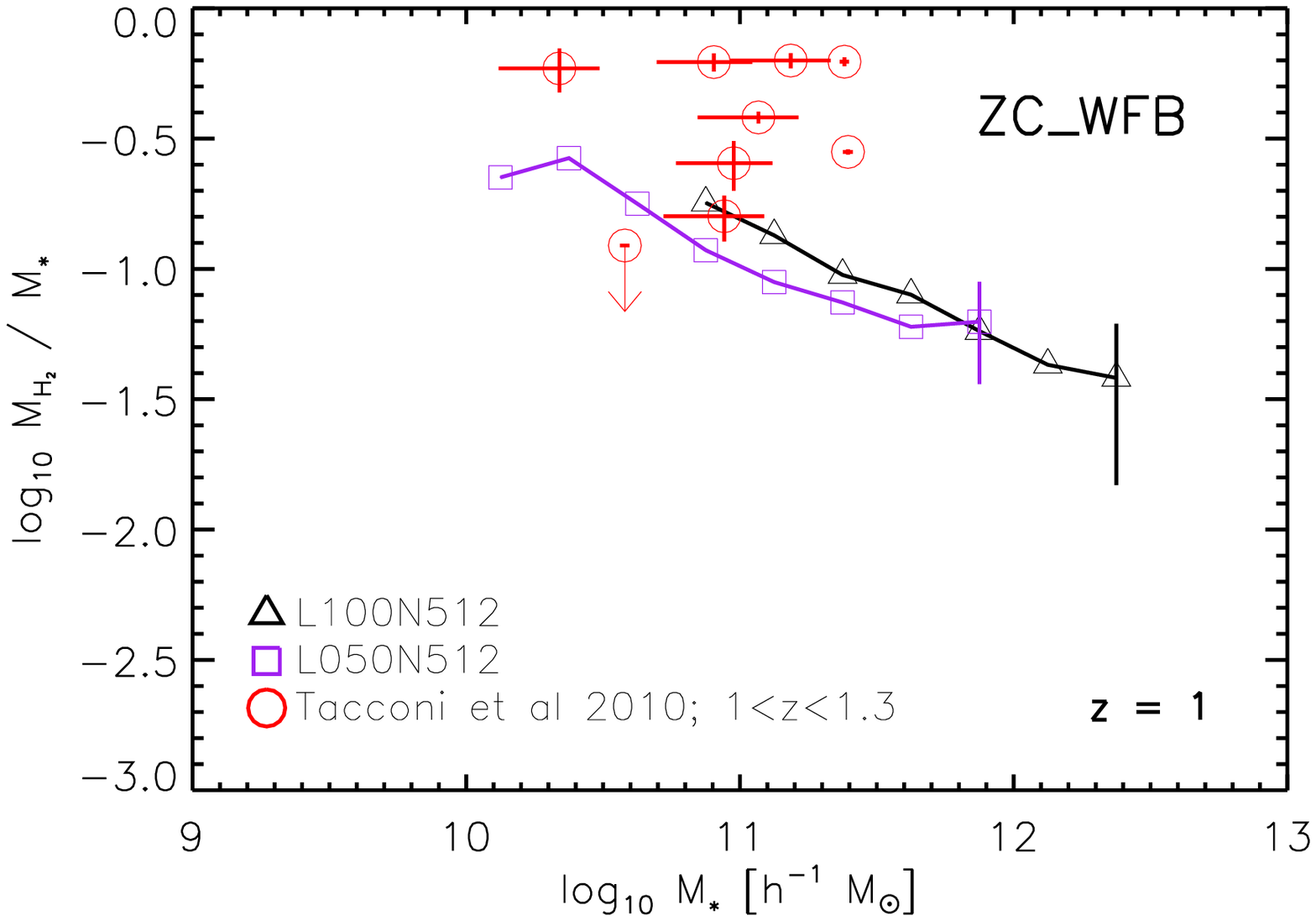, scale=0.3} &
      \epsfig{figure=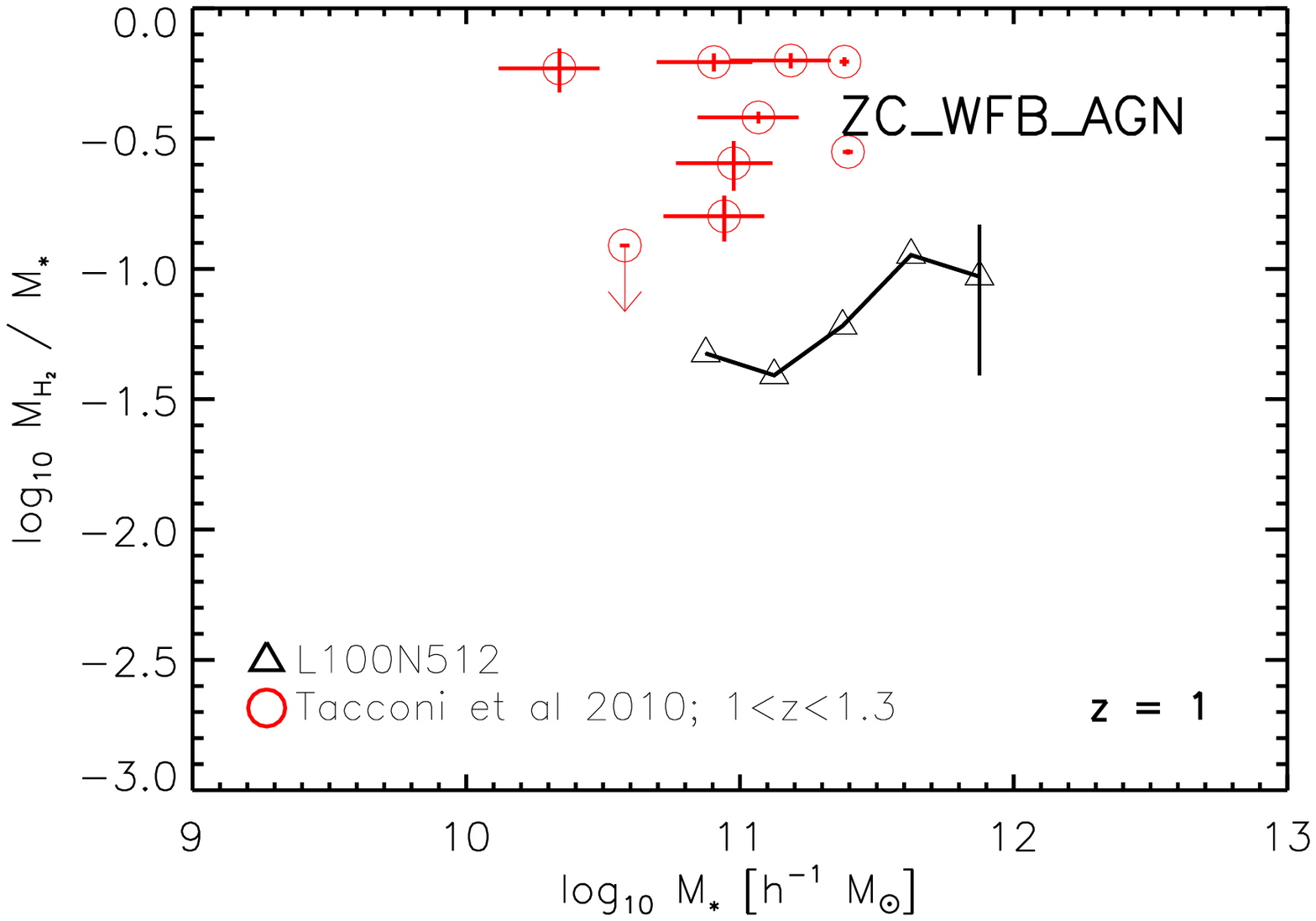, scale=0.3} \\
      \epsfig{figure=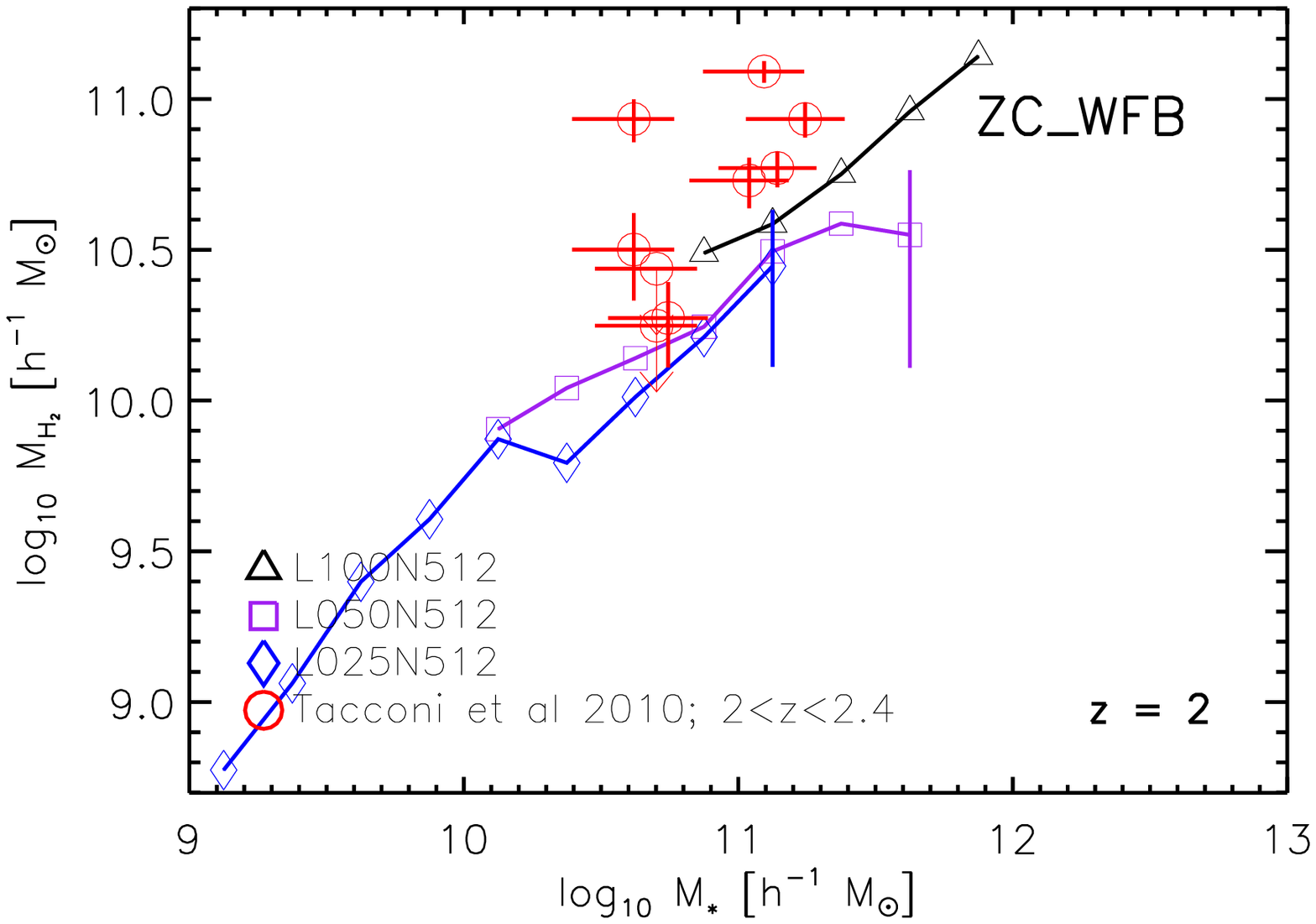, scale=0.3} &
      \epsfig{figure=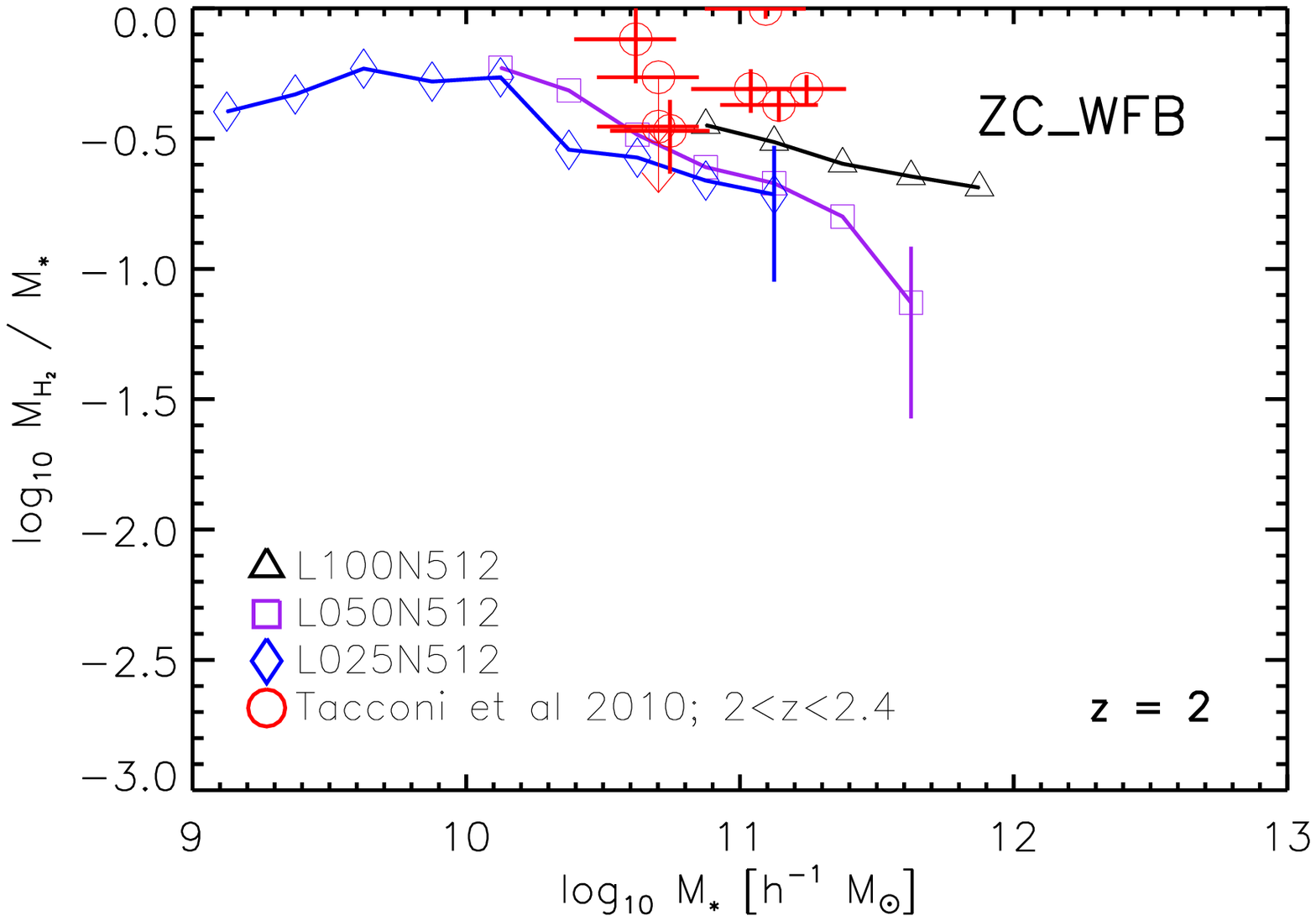, scale=0.3} &
      \epsfig{figure=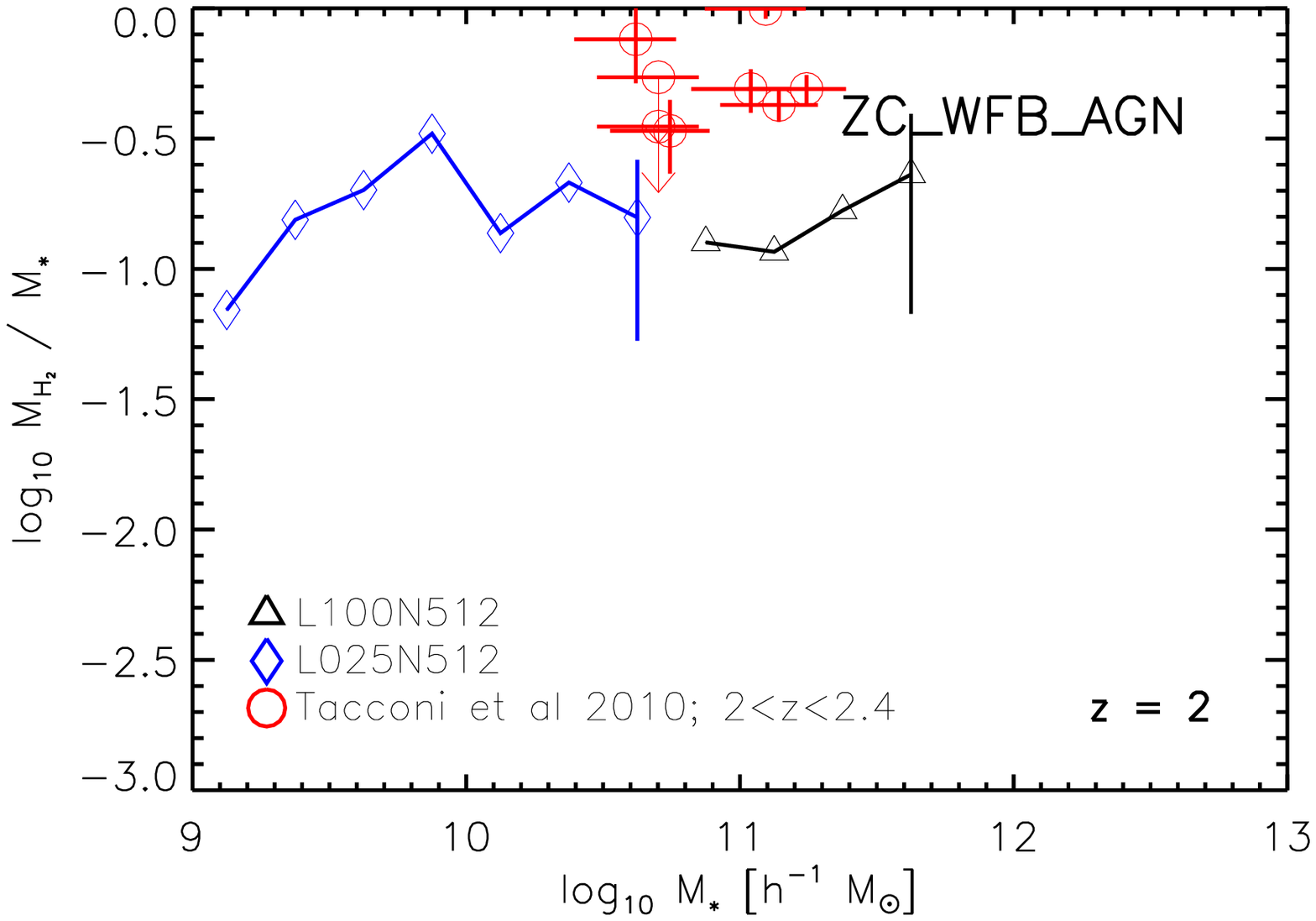, scale=0.3} \\      
      \end{tabular}
    \caption{As in Fig.~\ref{fig:stars_vs_hi} but now we consider the molecular mass as a function of stellar mass in the objects 
	for the \emph{ZC\_WFB} (left column) and the AGN model (right column, \emph{ZC\_WFB\_AGN}). 
            The errors are the $1\sigma$ errors estimated by bootstrap analysis and
      added in quadrature with Poisson error estimates (typically only important for the highest mass bins in a simulation). The red circles 
      in the top row are local galaxies from~\citet{Leroy:08} and at higher redshifts subsamples (close to the relevant redshift) from~\citet{Tacconi:10}.}
    \label{fig:stars_vs_h2}
  \end{center}
\end{figure*}

The correlation seen in the left column of Fig.~\ref{fig:stars_vs_hi}  
is less than linear however, as is clear from the middle column where the
ratio of HI to stellar mass is presented as a function of the stellar
mass. The ratio is a decreasing function of stellar mass,
which means that, for their size, massive galaxies are relatively deficient in
HI when compared with smaller galaxies. This is in very good agreement
with observations at low-redshift from~\citet{Fabello:11}, who combined HI and stellar 
data from the ALFALA and SDSS surveys, respectively (their results are 
plotted in Fig.~\ref{fig:stars_vs_hi} as red circles). We find that the trend continues
at higher masses with the final data point indicating perhaps a turnover in the
power law, a regime which is, however, influenced by the lack of AGN in this model. 
However, as seen in the right column of Fig.~\ref{fig:stars_vs_hi}, model AGN
typically has more HI at a given stellar mass than observed. 

The ratio of HI to stellar mass increases with redshift. For example, at $z=2$ a galaxy of stellar mass $10^{11} \hMsol$ will
have nearly an order of magnitude more HI than a similar galaxy at the present day.
Such an evolutionary trend is expected as the accretion of cold gas on to 
galaxies slows down with time and is no longer sufficient to replace the gas used in
star formation~\citep[e.g.][]{vdVoort:11a,vdVoort:11b}.

In Fig.~\ref{fig:stars_vs_h2} we consider the case of the molecular hydrogen as a function of the stellar mass of objects,
for $z=0-2$ for a simulation with SNe feedback, and one with additional AGN feedback (\emph{ZC\_WFB} and \emph{ZC\_WFB\_AGN}).
For the former simulation at $z=0$ we compare to local galaxies from~\citet{Leroy:08} we find that the simulations
match the low stellar mass end well over the stellar mass range of the observations.

The situation at $z=2$ is more complex. At the low mass end the \emph{ZC\_WFB} simulation matches the observations of~\citet{Tacconi:10},
however for systems with stellar mass $\ge 10^{11} \hMsol$ the observed galaxies typically have 0.5 dex greater molecular mass
than the simulations. This situation becomes even more extreme at $z=1$ with the disparity at the high stellar mass end now an 
order of magnitude. We can attempt to match these observations by suppressing the conversion of cold gas into stars through additional
feedback with an AGN but the discrepancy increases, with the molecular mass of simulated galaxies lying below the observations at all redshifts. 	
If we compare this with the over-prediction of the HI mass for a given stellar mass (bottom right column of Fig.~\ref{fig:stars_vs_hi}) 
we see that for a given stellar mass, the simulation with AGN feedback is probing haloes with a \emph{much} greater halo mass and hence will
have more shock-heated gas. In our formalism this will move gas from a dense regime in which it is dominated by molecular hydrogen into a lower
density, hotter gas that has more atomic hydrogen, this redistribution of gas to larger radii is demonstrated in Fig.~\ref{fig:halfmassfn_mhi_mfof}.

\subsection{Spatial distributions}
\label{sec:halfmass}

\begin{figure*}
  \begin{center}
  \begin{tabular}{ccc}
      \epsfig{figure=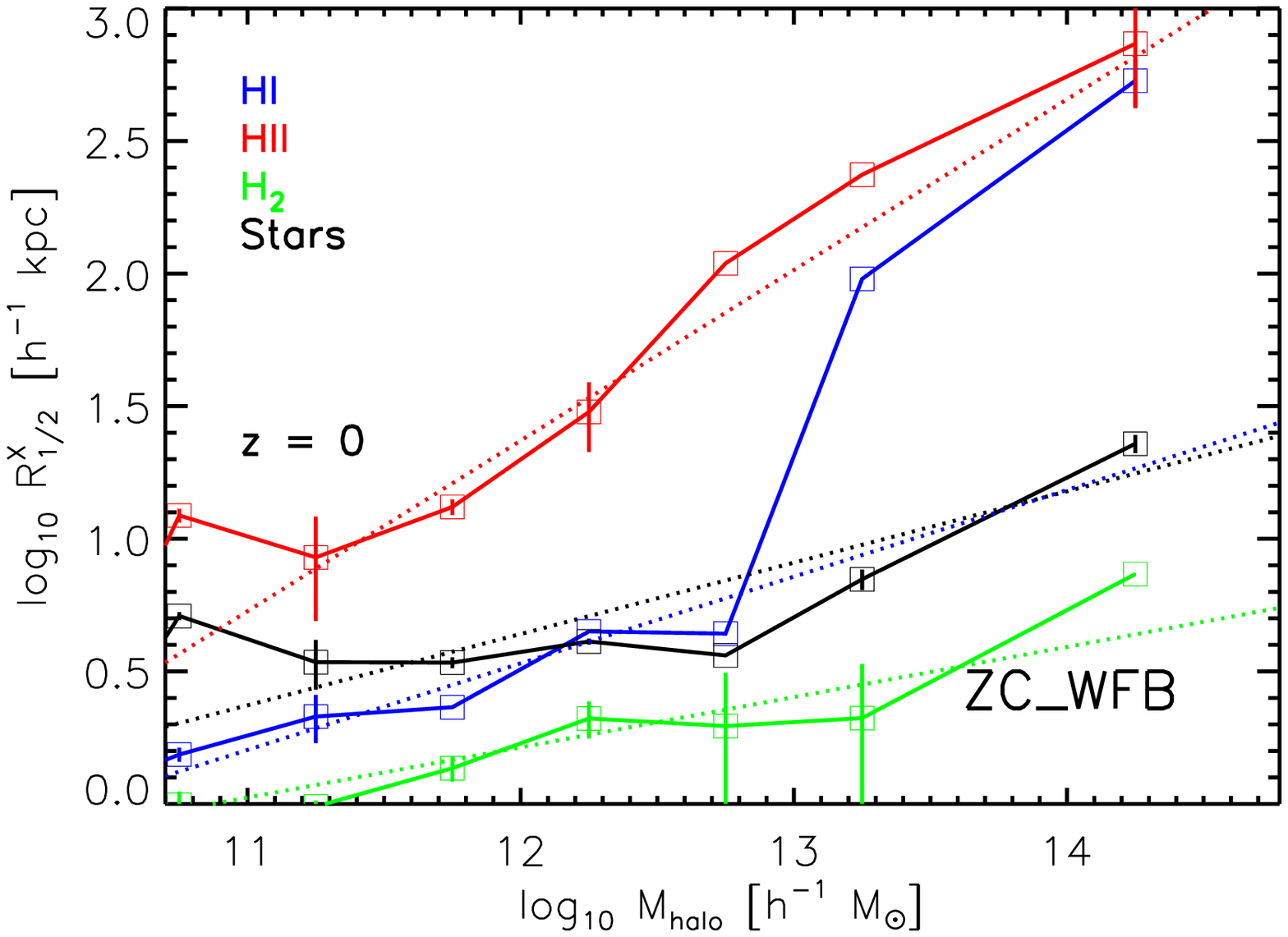, scale = 0.3} &
      \epsfig{figure=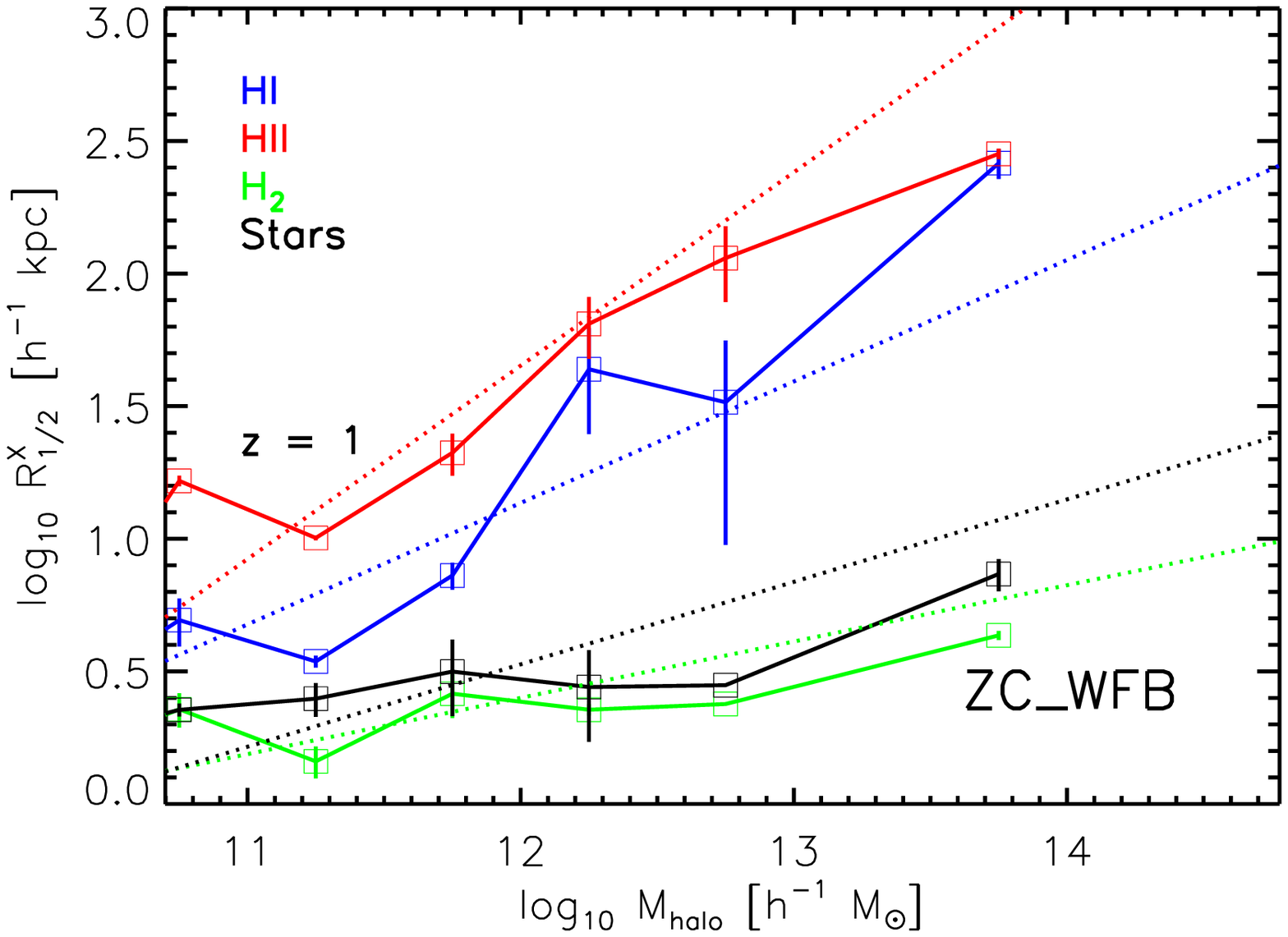, scale = 0.3} &
      \epsfig{figure=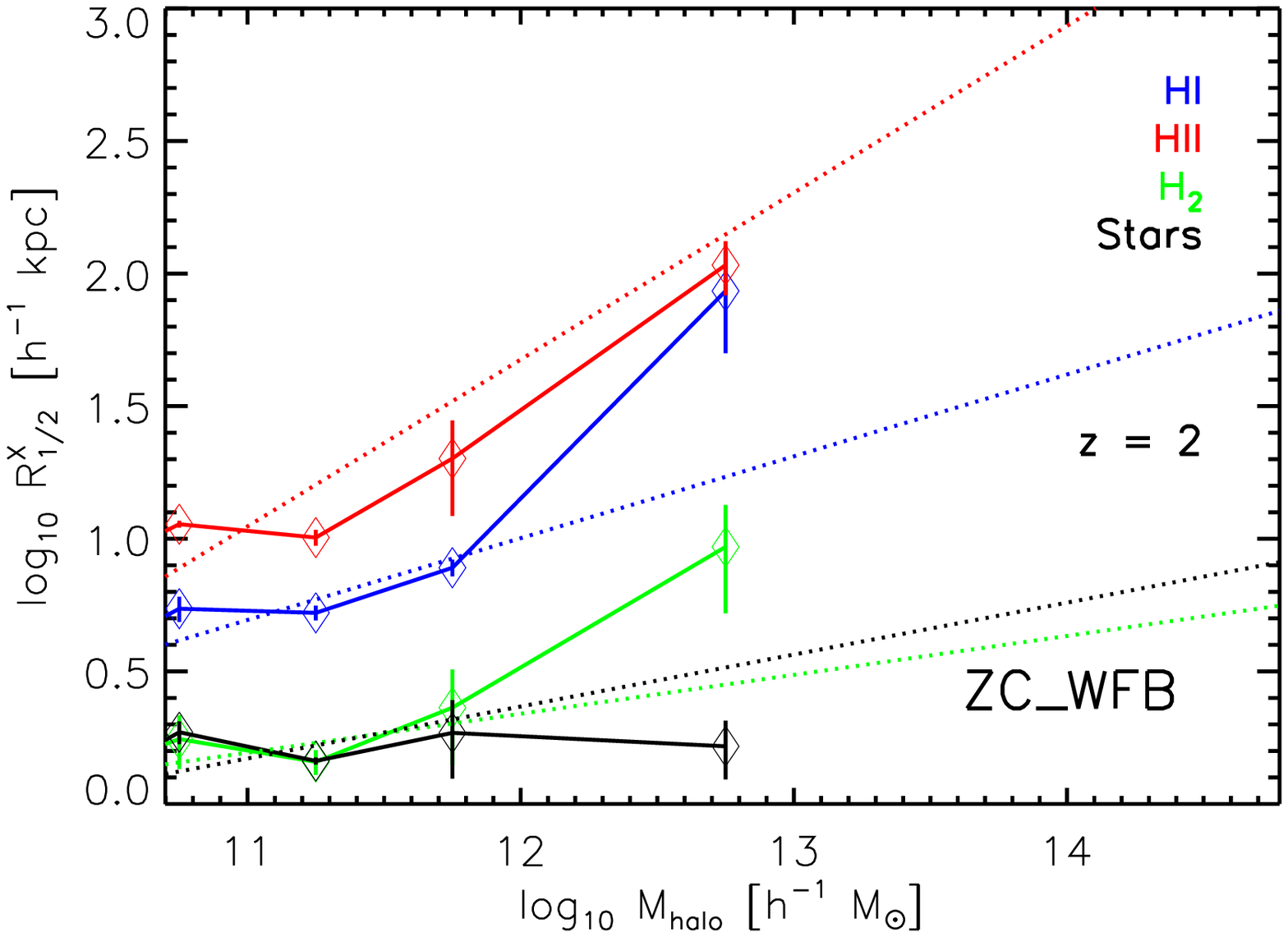, scale = 0.3} \\
      \epsfig{figure=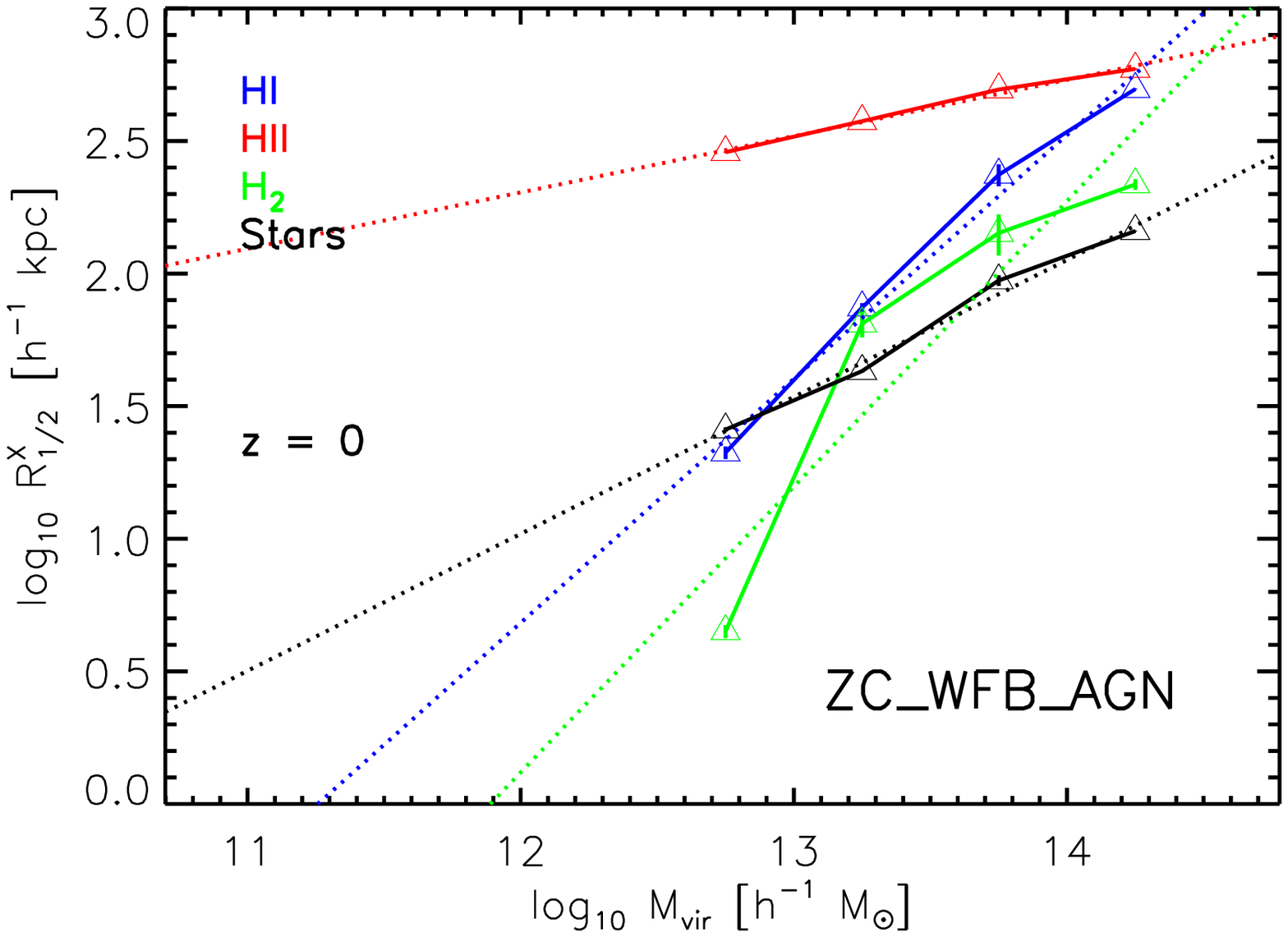, scale = 0.3} &
      \epsfig{figure=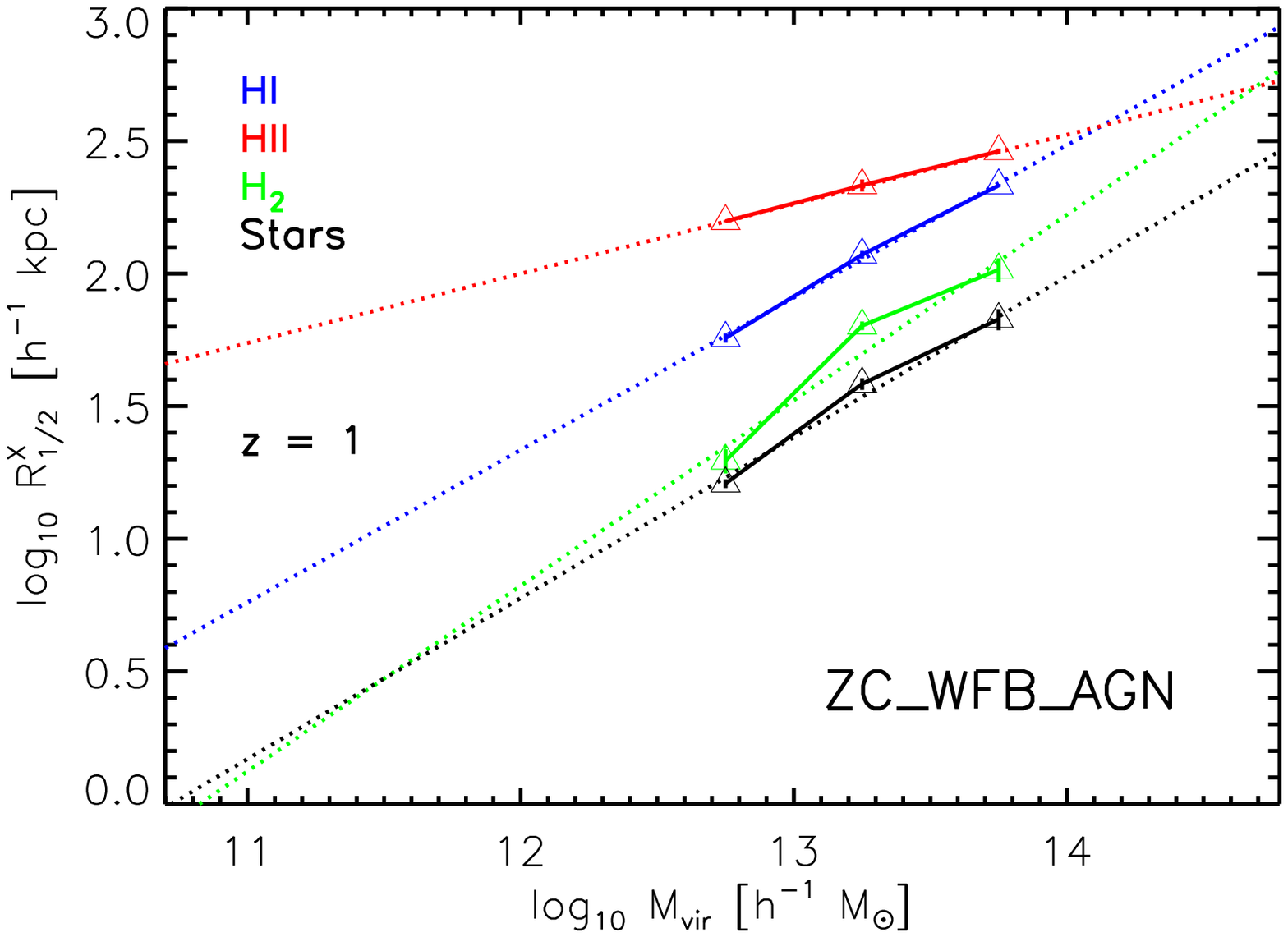, scale = 0.3} &
      \epsfig{figure=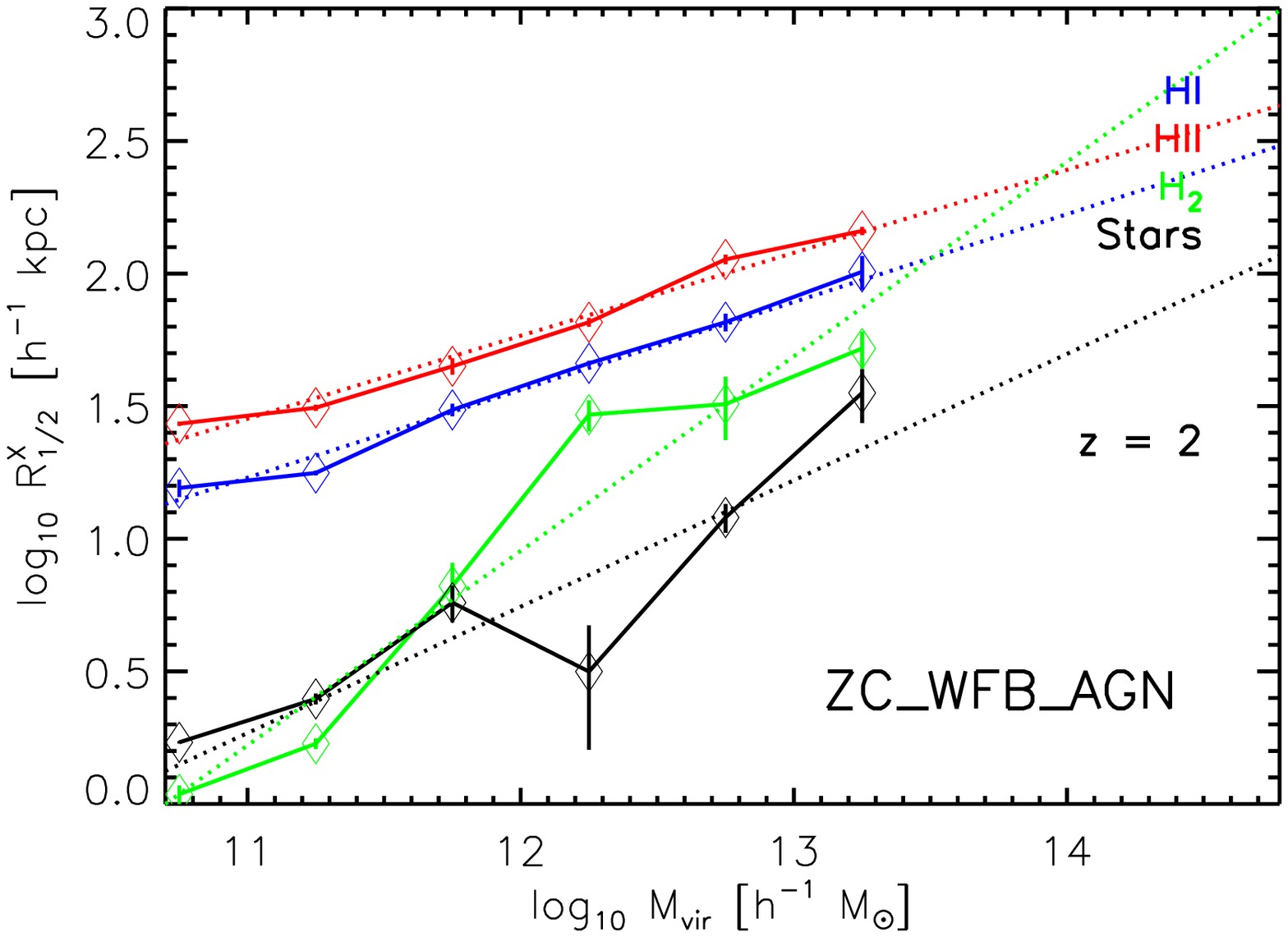, scale = 0.3} \\
      \end{tabular}
  \caption{The HI, HII, $\rm H_2 $ and stellar half-mass radii (in proper units) as a function of 
  	halo mass for $z=0$, $1$ and $2$, left to right respectively. 
    We consider runs with metal-line cooling and
    supernovae feedback, \emph{ZC\_WFB} in the top row and also with the addition of AGN in  \emph{ZC\_WFB\_AGN} in the bottom row.
      Power law fits to all radii-mass relations are shown, with best-fit values
      given in Table~\ref{tab:halfmass}.}
  \label{fig:halfmassfn_mhi_mfof}
  \end{center}
\end{figure*}

As well as the total mass of each species in a given halo, we can probe the spatial 
distribution of the gas and stars within the haloes, as shown in Fig.~\ref{fig:himap}. To quantify this Fig.~\ref{fig:halfmassfn_mhi_mfof} 
shows the half-mass radius for each species, $R^{\rm X}_{1/2}$, 
plotted as a function of halo mass for the different redshifts. Note that we only include particles that are gravitationally bound
to the halo. Again, results are shown for the \emph{ZC\_WFB} runs.

In haloes more massive than $\sim 10^{13} \hMsol$, the HI distribution rapidly becomes more extended
than both the stellar and molecular H$_2$ distributions, growing to nearly half the size of the
ionised diffuse component for haloes of mass $10^{14} \hMsol$. Since the molecular gas is only identified
with the high-density star-forming gas, it is not surprising that it has a similar half-mass radius as the stars
(large differences start to appear in low-redshift groups and clusters due to the presence of a diffuse stellar
component).

\begin{table*}
\begin{center}
\caption{Power-law fits to the half-mass radius versus halo mass relation for the various baryonic components
given by  $\log_{10} (R^{\rm X}_{1/2}/ \hkpc) = {\rm Norm} + {\rm Slope} \times \log_{10} (M_{\rm halo} / M_{\rm pivot})$
with a pivot mass of $2 \times 10^{12} \hMsol$ assumed. Quoted errors are 
68 per cent confidence limits, estimated from bootstrap resampling. 
The top series of values are for \emph{ZC\_WFB}
and the bottom set of three redshifts is for \emph{ZC\_WFB\_AGN}.}
\label{tab:halfmass}
\begin{tabular}{rrrrrrrrr}%{ccccccccc}%{|c|c|c|c|c|c|c|c|c|}
\hline
$z$ & \multicolumn{2}{c}{HI} & \multicolumn{2}{c}{HII} & \multicolumn{2}{c}{$\rm H_2$} & \multicolumn{2}{c}{Stellar} \\
 & Norm [$\hkpc$] & Ind & Norm [$\hkpc$] & Ind & Norm [$\hkpc$] & Ind & Norm [$\hkpc$] & Ind \\
 \hline
$0$ & $10.31\pm^{0.43}_{0.39}$ & $0.78\pm^{0.02}_{0.02}$  & $165.96\pm^{0.83}_{0.95}$ & $0.26\pm^{0.01}_{0.00}$  & $2.39\pm^{0.11}_{0.10}$ & $-0.06\pm^{0.04}_{0.10}$  & $3.79\pm^{0.07}_{0.07}$ & $0.25\pm^{0.01}_{0.01}$  \\
$1$ & $18.15\pm^{0.55}_{0.53}$ & $0.74\pm^{0.01}_{0.01}$  & $93.34\pm^{0.55}_{0.54}$ & $0.31\pm^{0.01}_{0.01}$  & $2.07\pm^{0.05}_{0.05}$ & $-0.01\pm^{0.11}_{0.02}$  & $2.03\pm^{0.05}_{0.04}$ & $0.16\pm^{0.07}_{0.03}$  \\
$2$ & $30.00\pm^{1.44}_{1.24}$ & $0.49\pm^{0.02}_{0.02}$  & $64.38\pm^{1.06}_{0.69}$ & $0.29\pm^{0.01}_{0.00}$  & $1.60\pm^{0.06}_{0.11}$ & $0.18\pm^{0.02}_{0.04}$  & $1.41\pm^{0.07}_{0.05}$ & $0.13\pm^{0.02}_{0.02}$  \\
\hline
$0$ & $5.95\pm^{0.94}_{0.60}$ & $1.01\pm^{0.03}_{0.05}$  & $232.16\pm^{4.06}_{3.48}$ & $0.21\pm^{0.00}_{0.01}$  & $2.78\pm^{0.37}_{0.33}$ & $1.08\pm^{0.04}_{0.04}$  & $14.93\pm^{0.59}_{1.16}$ & $0.52\pm^{0.04}_{0.01}$  \\
$1$ & $24.04\pm^{3.28}_{2.30}$ & $0.64\pm^{0.04}_{0.06}$  & $118.30\pm^{3.01}_{2.98}$ & $0.27\pm^{0.01}_{0.01}$  & $10.79\pm^{2.89}_{3.20}$ & $0.70\pm^{0.13}_{0.09}$  & $9.10\pm^{0.91}_{0.76}$ & $0.61\pm^{0.04}_{0.04}$  \\
$2$ & $41.73\pm^{2.95}_{2.14}$ & $0.38\pm^{0.04}_{0.02}$  & $71.62\pm^{2.83}_{1.77}$ & $0.32\pm^{0.02}_{0.01}$  & $14.97\pm^{1.42}_{1.65}$ & $0.73\pm^{0.03}_{0.04}$  & $7.72\pm^{2.17}_{1.78}$ & $0.48\pm^{0.05}_{0.06}$  \\
\hline
\end{tabular}
\end{center}
\end{table*}

The power-law dependence of size on halo mass for the ionised and stellar components 
is a generic prediction given by~\citet{MMW:98}, in which they also predicted that haloes 
of a given mass will have smaller half-mass radii at higher redshift (due to the higher density
and hence smaller virial radius). We present simple power-law fits to our simulated data in
Table~\ref{tab:halfmass}. The half-mass radius of the ionised hydrogen has a cube-root dependency on the halo mass 
which is a generic prediction for a component with a smoothly distributed isothermal profile.
The molecular hydrogen and stellar components are more centrally concentrated than the HI and HII
at all masses and redshifts. They are sensitive to the implementation of the AGN feedback which
dramatically increases the typical half-mass radius by nearly an order of magnitude. However, at $z=0$
the effects of the AGN feedback in increasing the half-mass radius of the stars and H$_2$ is lessened,
although the stellar component is, roughly, a factor 5 more extended due to the suppression of cooling
flows which would lead to a highly concentrated, massive central galaxy (a similar effect was noted 
for galaxies at $z=2$ by~\citealt{Sales:10} using the AGN feedback scheme discussed here).

\section{Neutral Gas Mass functions}\label{sec:HImassfn}
A key observable for each state of hydrogen is its mass function. In this section we focus
on the HI mass function (Section~\ref{sec:HImassfnres}) and the molecular and cold
gas mass functions (Section~\ref{sec:H2massfn}), for the \emph{ZC\_WFB} simulations as
they span the largest range in mass. Results are presented at  $z=0$, $1$ and $2$  
for the three different self-shielding methods. The effects of feedback and metal-line cooling 
will be investigated in the following section.

\begin{figure*}
  \begin{center}
    \begin{tabular}{ccc}
    $\SSNO$ & $\SSALF$ & $\SSDLA$ \\

    \epsfig{figure=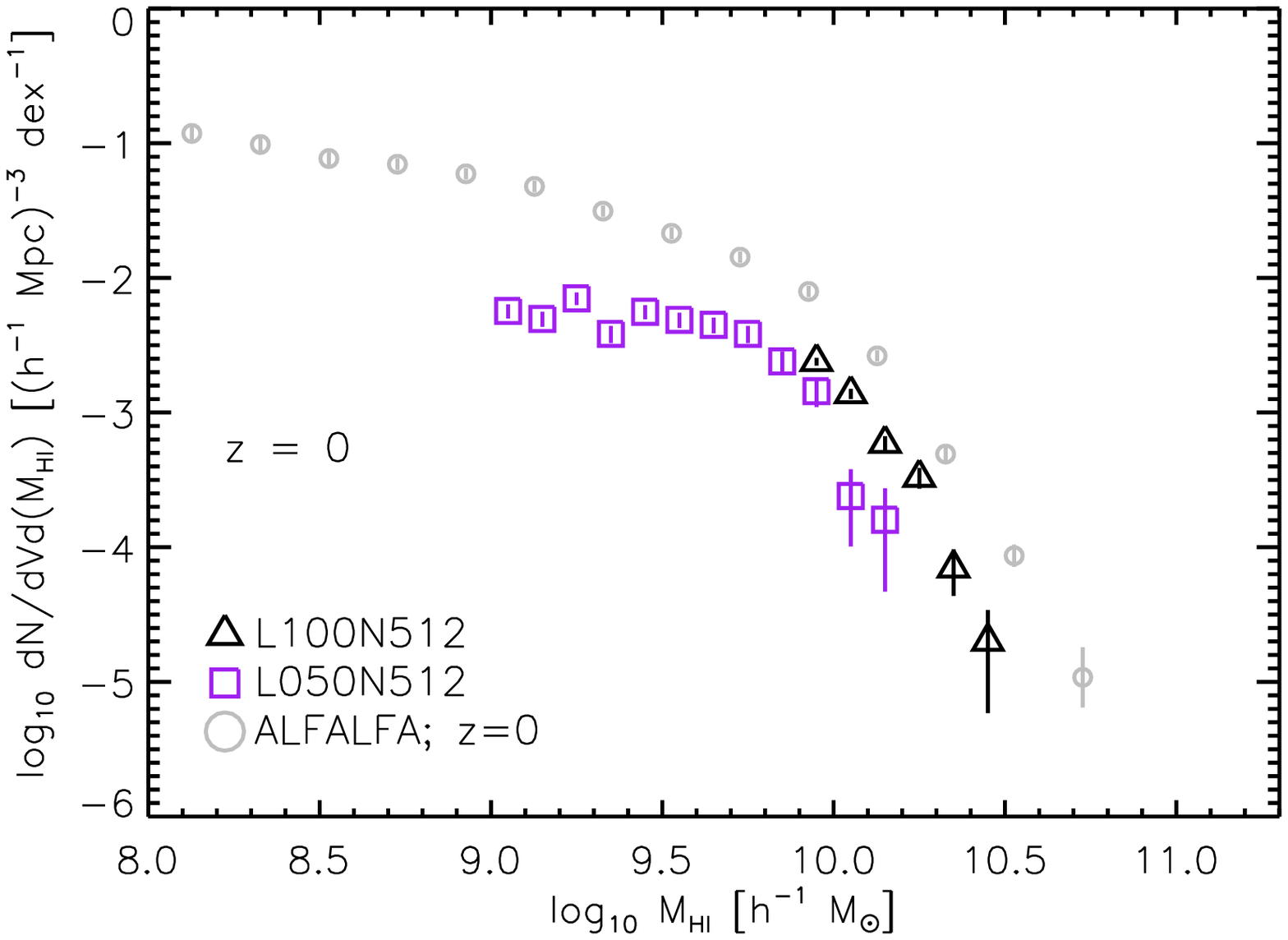, scale=0.32} &
    \epsfig{figure=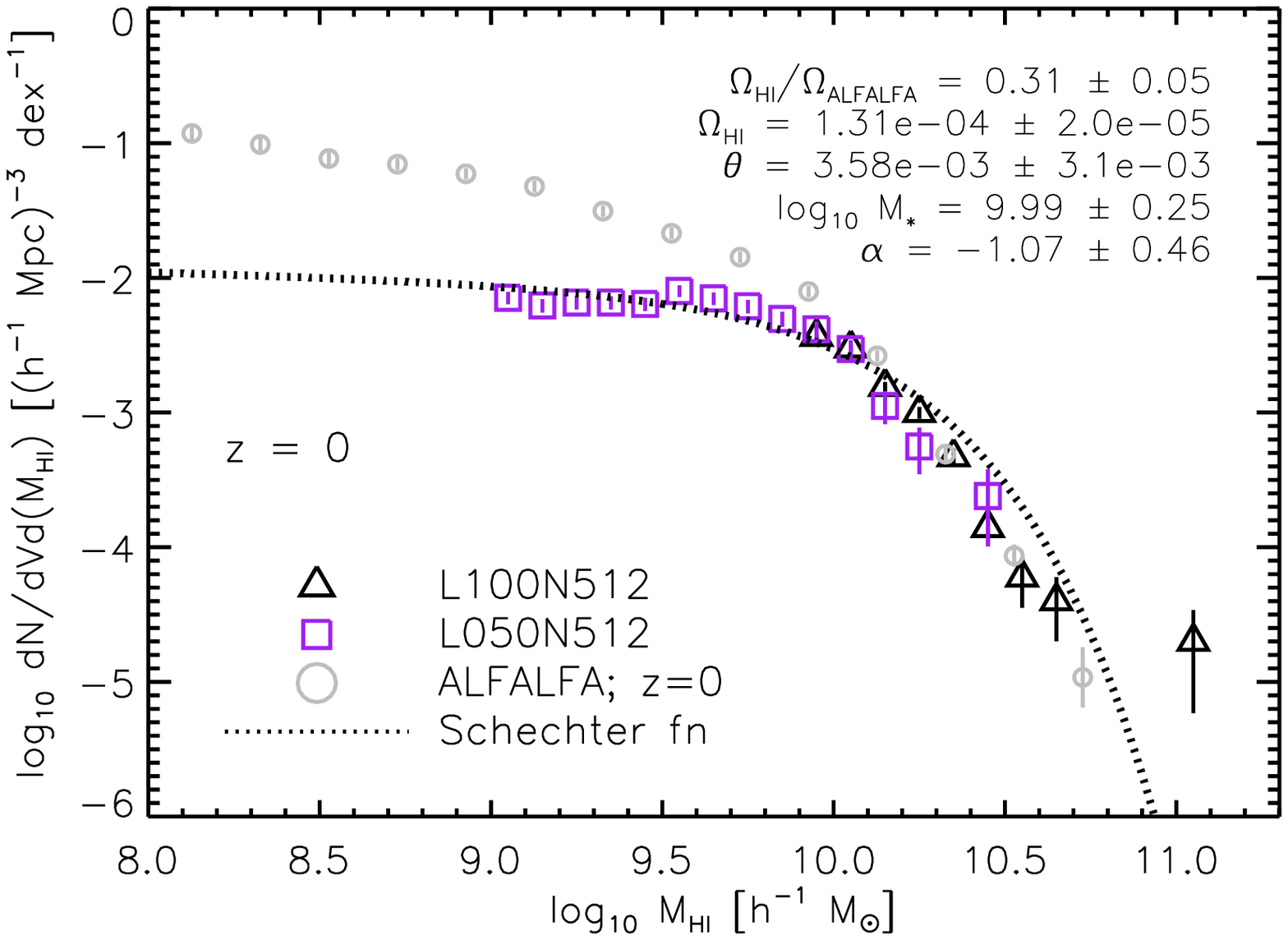, scale=0.32} & 
    \epsfig{figure=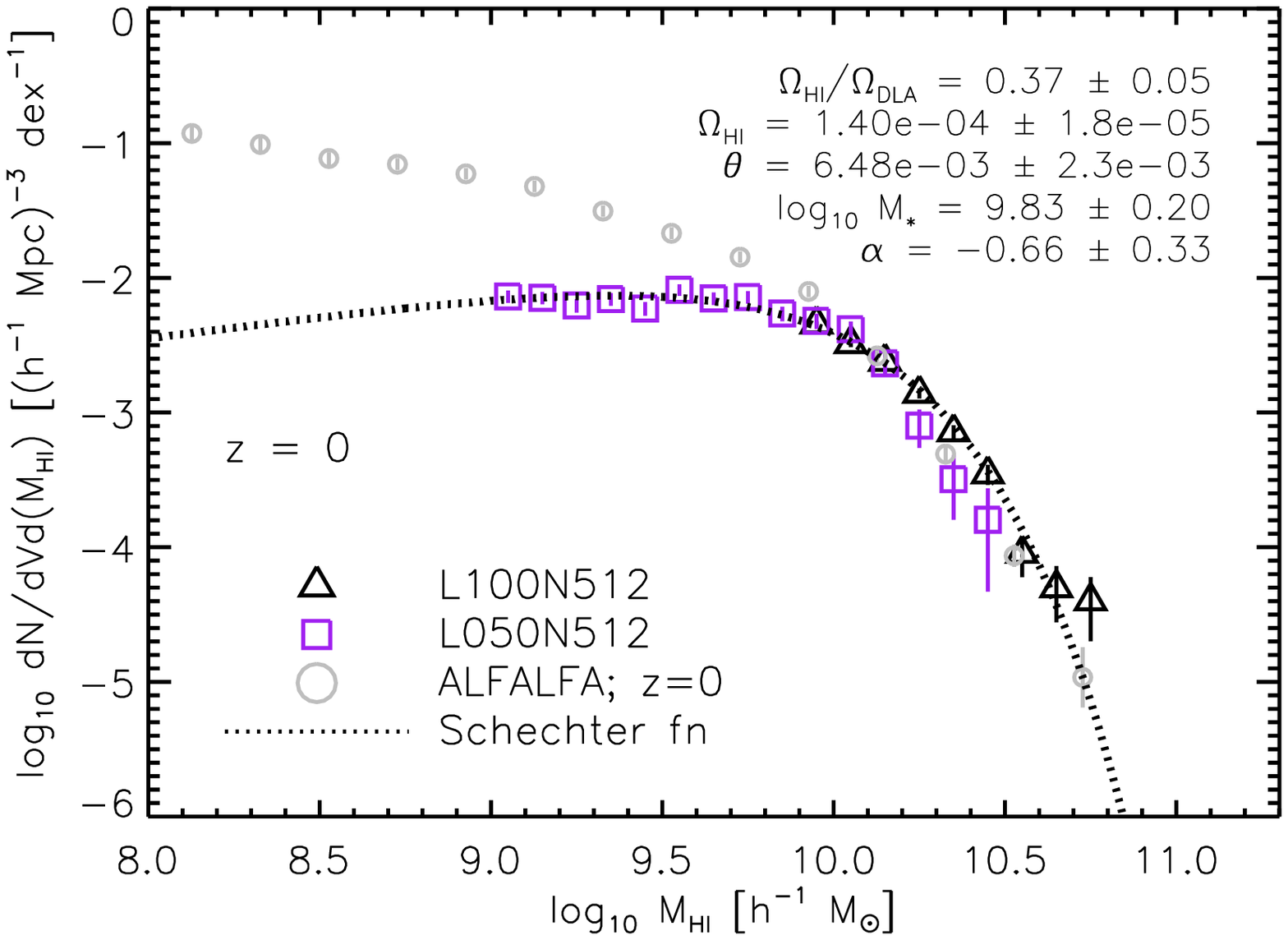, scale=0.32} \\

    \epsfig{figure=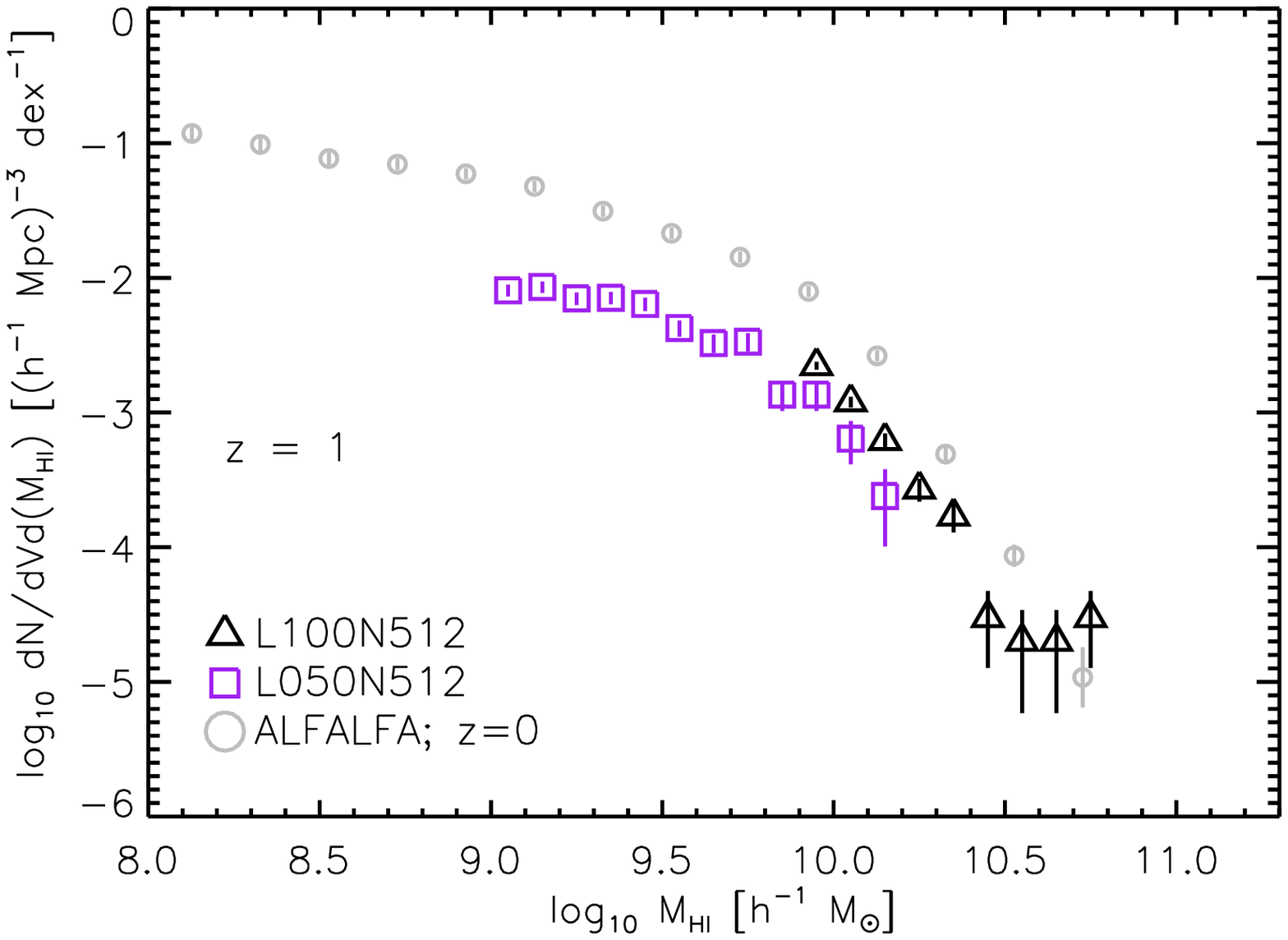, scale=0.32} &
    \epsfig{figure=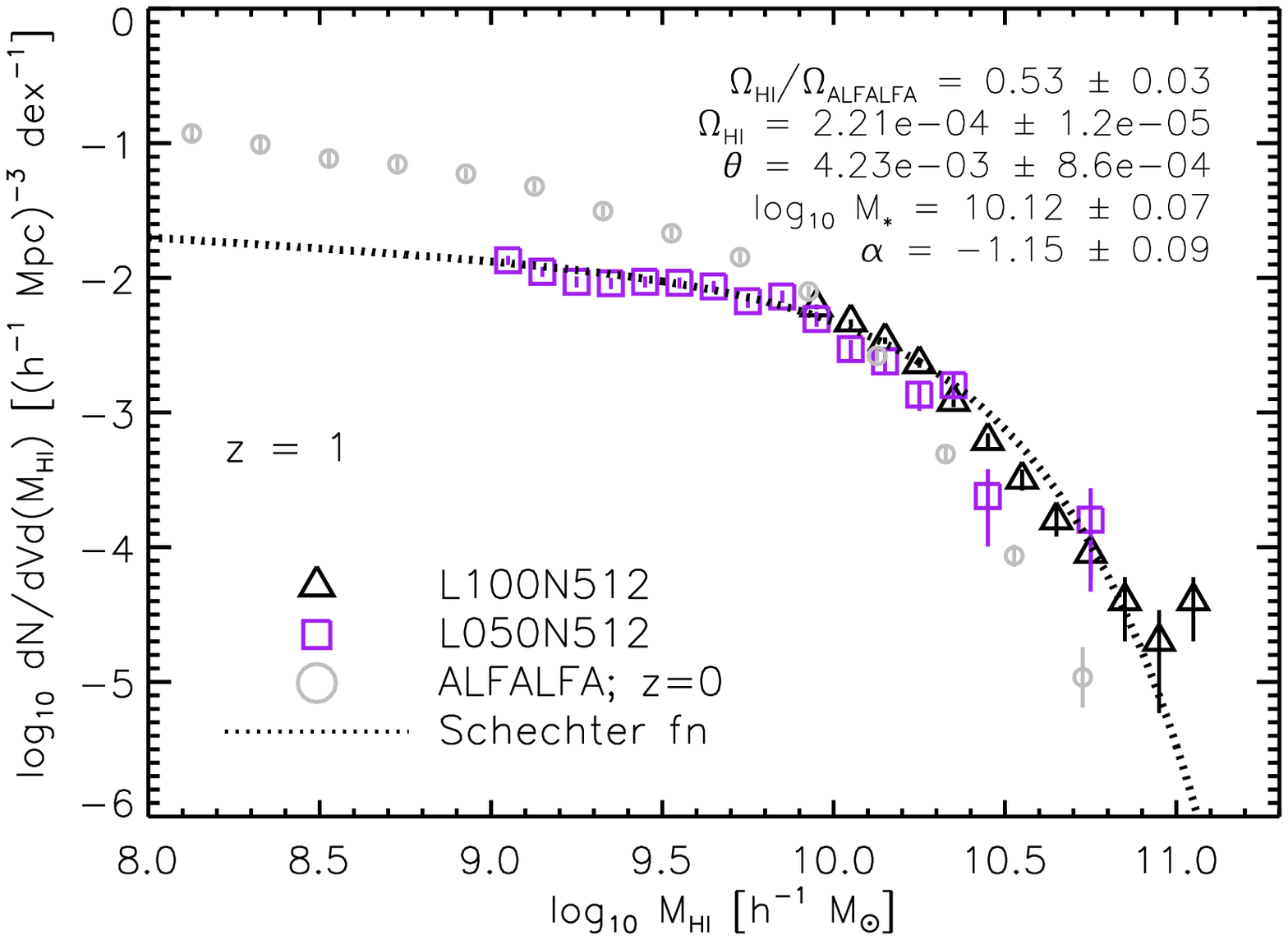, scale=0.32} & 
    \epsfig{figure=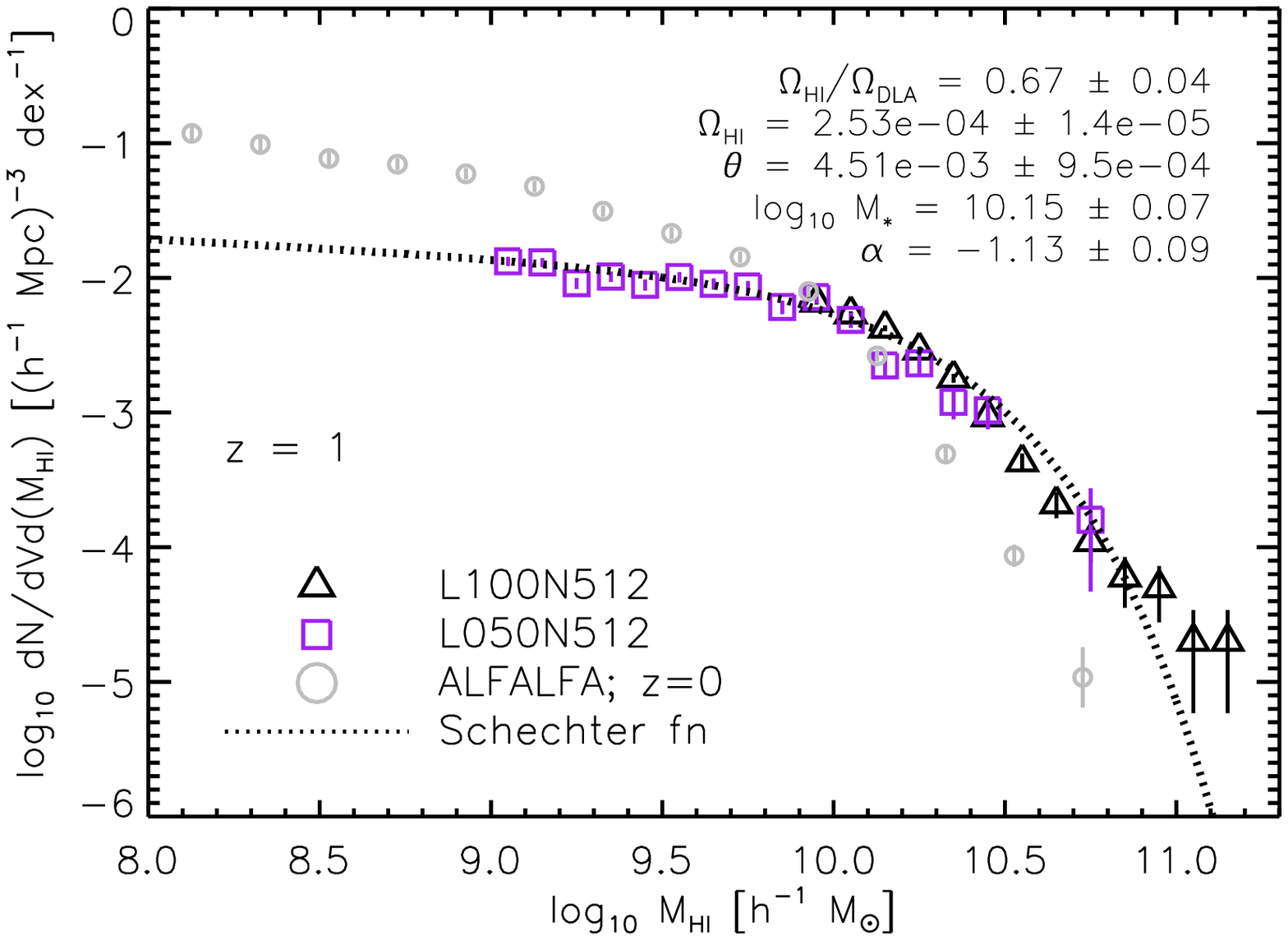, scale=0.32} \\

    \epsfig{figure=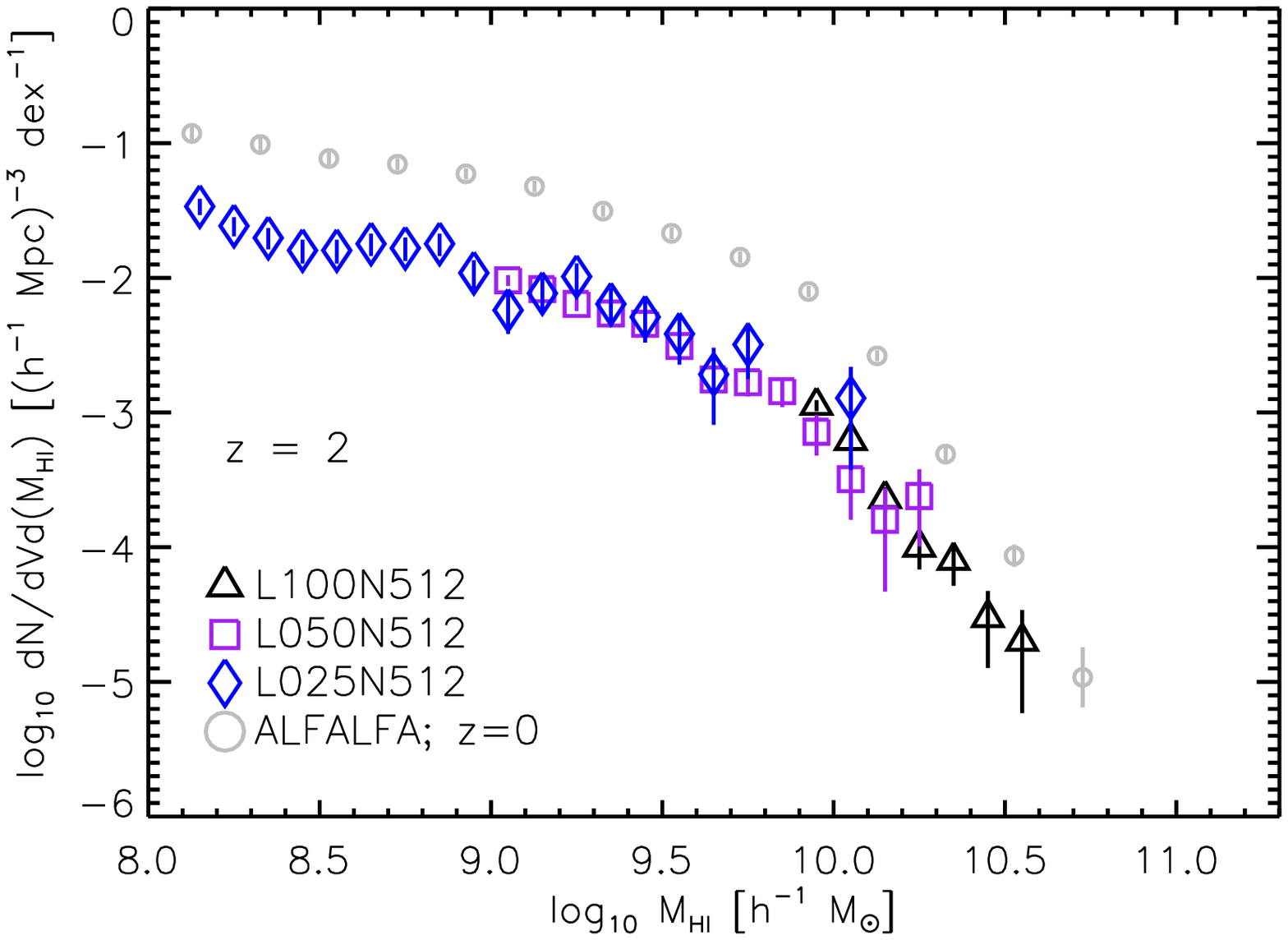, scale=0.32} &
    \epsfig{figure=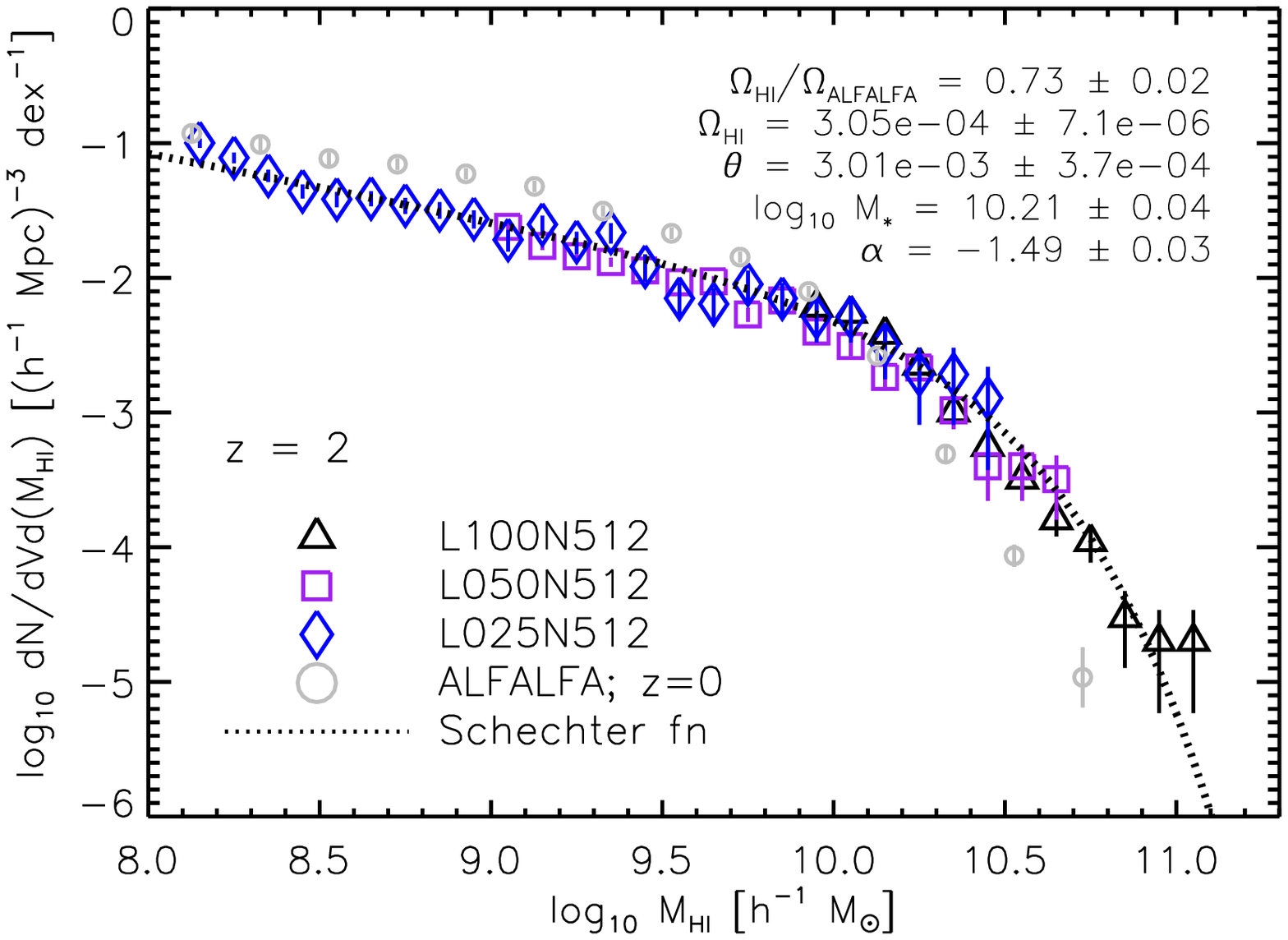, scale=0.32} & 
    \epsfig{figure=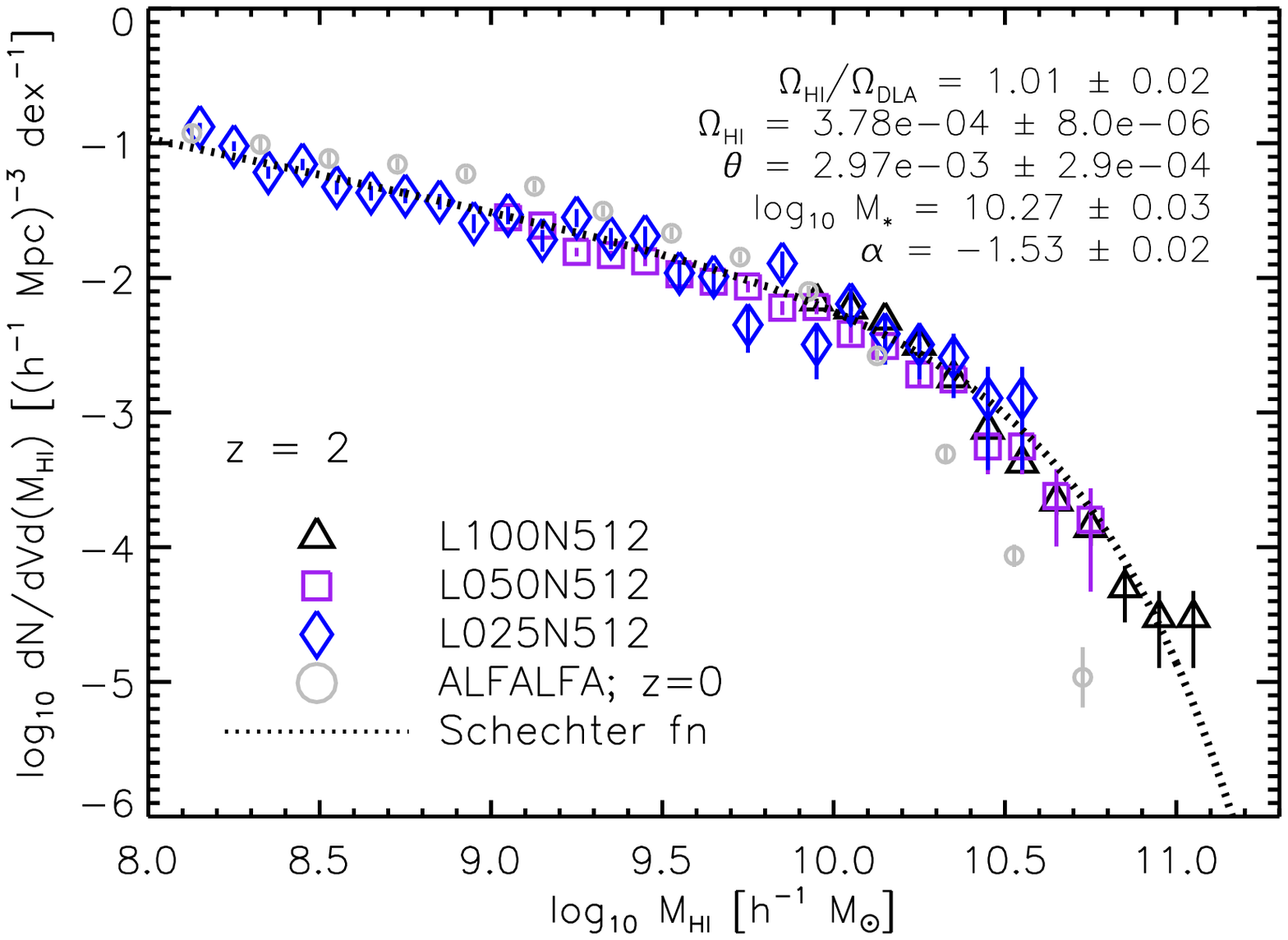, scale=0.32} \\
    \end{tabular}
    \caption{The HI mass function from \emph{ZC\_WFB} at $z=$ 0, 1 and 2; top, middle and bottom 
respectively, with the $z=0$ ALFALFA HI mass function shown for comparison. Each column displays
results for the different self-shielding methods: the left column is method $\SSNO$ (no self-shielding); the 
middle column is method $\SSALF$ (pressure threshold tuned to match the ALFALFA data in the range 
$M_{\rm HI} = 10^{10} - 10^{11} \hMsol$ ) and the right
column is method $\SSDLA$ (pressure threshold tuned to match the $z=2$,  $\Omega_{\rm DLA}$ 
results of~\citealt{Prochaska:09}). Typically, the pressure value found to match the $z=2$ 
DLA results is higher than that used for $z=0$, allowing us to bracket the available 
range of possible HI mass functions for the \emph{ZC\_WFB} simulations. Poisson errors are
assumed for individual simulation data points. We also show a Schechter function fit to the 
simulated data (for methods $\SSALF$ and $\SSDLA$) and present best-fit values for $\alpha$, $\theta$ and $M_{*}$ 
(and their errors, from bootstrap resampling) in the legend.
Values for $\Omega_{\rm HI}$, calculated from the integrated Schechter function, are also shown. Clearly,
$\Omega_{\rm HI} \simeq \Omega_{\rm DLA}$ for method $\SSDLA$ at $z=2$ but $\Omega_{\rm HI} <  \Omega_{\rm ALFALFA}$
in method $\SSALF$ (i.e. this model predicts less HI at high-redshift). }
    \label{fig:himassfn_hicut}
  \end{center}
\end{figure*}

\subsection{HI Mass Functions}
\label{sec:HImassfnres}

The HI mass functions for \emph{ZC\_WFB} are shown in Fig.~\ref{fig:himassfn_hicut}. 
As discussed earlier, we use individual subhaloes to represent the galaxies and 
we determine the minimum HI mass for each simulation according to the
values given in Equation~\ref{eqn:hires}. In all cases, the number density of objects decreases as a function of 
increasing HI mass, with a strongly suppressed high-mass end and a more gradual
slope at the faint end (where simulation data is available). The good agreement in abundance at the
overlap between the different simulation volumes, increases our confidence in the numerical 
resolution cuts we have made. 

We consider the no-shielding case first (left column of Fig.~\ref{fig:himassfn_hicut}) 
and can immediately see that at $z=0$, top panel, the simulation points lie below
the ALFALFA results. This justifies our assumption that the self-shielding correction
(which increases the amount of HI at moderate densities) must be implemented 
to match observations. In the absence of self-shielding there appears to be little 
evolution in the HI mass function between $z=0$ and $z=2$.

We can tune the self-shielding pressure threshold to match the `knee' of the 
$z=0$ ALFALFA mass function and utilise this same pressure cut at higher redshifts, 
as shown in the middle column of Fig.~\ref{fig:himassfn_hicut}. Interestingly, the simulations predict
relatively little change in the abundance of haloes in this mass range with redshift. The high-mass
end matches the ALFALFA data very well.
At the low-mass end, the simulation under-predicts the number of objects; this {\it flattening} is also seen in the
cold gas mass function (see Section~\ref{sec:H2massfn}) and thus appears to be related to the 
galaxy formation modelling in the simulation. 

The HI mass functions are well parameterised by a Schechter function (particularly at $z=2$ where we have the
highest dynamic range) given by
\begin{eqnarray}\label{eqn:schechterfn}
\frac{ {\rm d}n}{{\rm d}M} = \frac{\theta}{M_{*}} \left( \frac{M}{M_{*}} \right)^{\alpha} {\rm e}^{-\frac{M}{M_{*}}} \,,
\end{eqnarray}
with normalisation $\theta$, low-mass slope $\alpha$ and 
characteristic mass scale $M_{*}$, respectively. 
This can be integrated to give the mass density 
\begin{eqnarray}\label{eqn:int_schechterfn}
\rho &=&\int^{\infty}_{0} M \, \frac{{\rm d}n}{{\rm d}M} \, {\rm d}M \,,\nonumber \\
  &=& \theta\int^{\infty}_{0} \left( \frac{M}{M_{*}} \right)^{\alpha+1} {\rm e}^{-\frac{M}{M_{*}}} \, {\rm d}M. 
\end{eqnarray}
The global HI density (assuming the above masses are HI masses) in units of the critical density, $\rho_{\rm crit}$,  is then
\begin{eqnarray}\label{eqn:ohi_int_schechterfn}
\Omega_{\rm HI} = \theta \Gamma(\alpha +2) M_{*}/\rho_{\rm crit}\,,
\end{eqnarray}
where $\Gamma(x)$ is the standard gamma function. Comparing our
results at $z=2$ with the low-redshift data, we predict mild evolution of the HI
mass function with the global HI density decreasing from $0.7$ to $0.3$ at $z=0$. 
This has important ramifications for upcoming
deep HI surveys such as those to be performed with the Square Kilometre Array
which will detect more HI sources than an extrapolation from the $z=0$ mass function
would suggest.

Our third method uses a pressure threshold set by matching the integrated 
mass function of the simulations at $z=2$ to the DLA inferred cosmic density of HI at 
$z=2.2$ from~\citet{Prochaska:09}. The right panels in 
Fig.~\ref{fig:himassfn_hicut} show that this leads to very similar results as method 
$\SSALF$. The higher global HI density at $z=2$ leads to a slightly larger value for $M_*$
and a slightly steeper low-mass slope, $\alpha$. Again, comparing results with the 
ALFALFA data, there is only mild evolution in the global HI density with redshift.

For all models there is however strong evolution seen in the low-mass tail, from a 
steep index of $-1.5$ at $z=2$ to $\ge -1$ at $z=0$. This evolution is also seen in the cold
gas mass function in Fig.~\ref{fig:htwomassfn_hicut} which indicates that it is not subject to
the self-shielding methodology. This result was also seen in semi-analytic work by~\citet{Power:10}
who found that cold gas was being turned into stars in these low-mass systems. 

We summarise the HI evolution in Fig.~\ref{fig:cosmichi} where we integrate the Schechter function described in Fig.~\ref{fig:himassfn_hicut}
even though we can clearly see in those figures that the low-mass tail undergoes significant evolution in model \emph{ZC\_WFB} and is 
shallower than measured locally. The high-mass tail appears to show much more gradual evolution over this period. 
Observers will be constrained to observing only the largest systems so to, crudely, mimc this selection we limit the fit of the Schechter function 
to HI masses above $10^{10} \hMsol$ and fix the faint end to $-1.5$ (the value we find at $z=2$). In this case we well agree with observations 
at low redshift and predict little evolution will be measured to high redshift. We show a 
number of observational measurements in Fig.~\ref{fig:cosmichi} and, while we did not attempt to make an exhaustive compilation, we consider this a 
reasonable summary of the current understanding of the cosmic HI density evolution. The results range from direct, local HI detections 
by~\citet{HIPASS} and~\citet{Martin:10} to a stacking of the HI signal at a moderate redshift by~\citet{Lah:07}. Then there are inferred
cosmic HI densities through the use of HST observations of low-redshift DLA systems traced by MgII absorbers~\citep{Rao:06,Meiring:11} and then at 
higher redshift by~\citet{Prochaska:09} and~\citet{Noterdaeme:09} using SDSS quasar spectra. There is disagreement between these last two results 
due, primarily, to a more aggressive treatment for the bias affecting strong absorbers in the latter work.

\begin{figure}
   \epsfig{figure=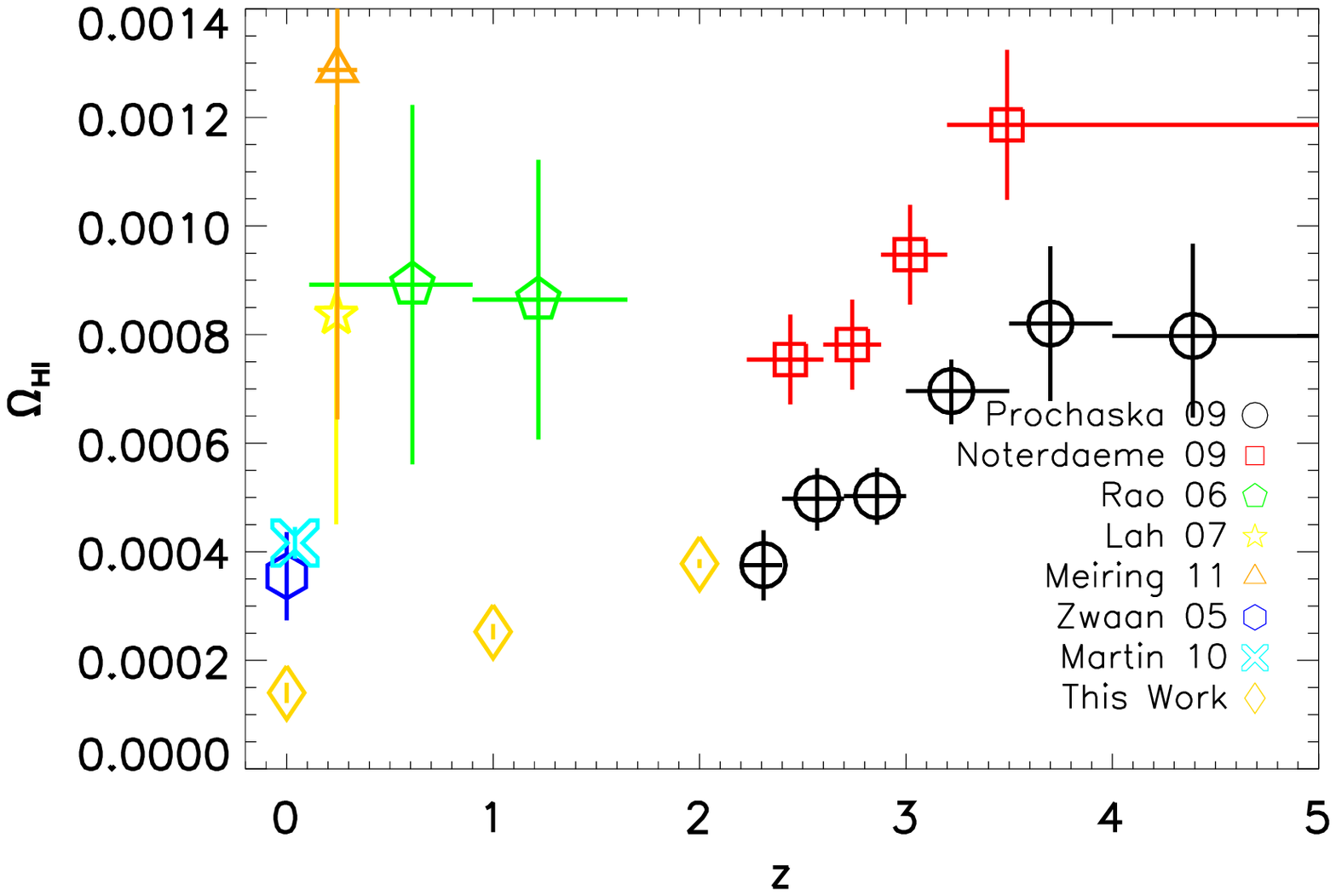, scale=0.45} 
    \caption[HI evolution]
            {A compilation of some recent measurements of the cosmic HI densities at various epochs. We have included the integrated values for
            the simulation \emph{ZC\_WFB} (with self-shielding method $\SSDLA$) 
            to demonstrate the difficulty in reproducing the observed HI density. Even though at $z=0$ we can reproduce the mass function about 
            the `knee' of the Schechter function, we do not have enough low-mass objects to recover the observed $\Omega_{\rm HI}$. However, if
            we fix the faint end slope to the value at $z=2$ ($\alpha = -1.5$) and limit the fit to haloes with $M_{\rm HI} > 10^{10} \hMsol$ then we recover
            the blue diamond points which are in better agreement with the low redshift results. They also demonstrate that, in the simulations at least,
            the evolution in HI is constrained to the faint end.
            Note that we have tuned the $z=2$ case to agree with~\citet{Prochaska:09}
            rather than the results of~\citet{Noterdaeme:09}, see text. In our models the evolution of $\Omega_{\rm HI}$ is driven by the low-mass tail, 
            whereas most 21cm observations will be unable to detect this population.}
    \label{fig:cosmichi}
\end{figure}

\subsection{Molecular and Cold Gas Mass Functions}
\label{sec:H2massfn}

We now consider the molecular mass function, shown in the left column of 
Fig.~\ref{fig:htwomassfn_hicut}, and the more general cold gas mass function,
shown in the right column. Cold gas is defined as all gas that is gravitationally 
bound to the halo and that has temperature $T<10^{4.5}$K, as well as all star-forming gas.
Note that neither the cold or molecular gas mass depends on the self-shielding measure used.

\begin{figure*}
  \begin{center}
   \begin{tabular}{cc}
    \epsfig{figure=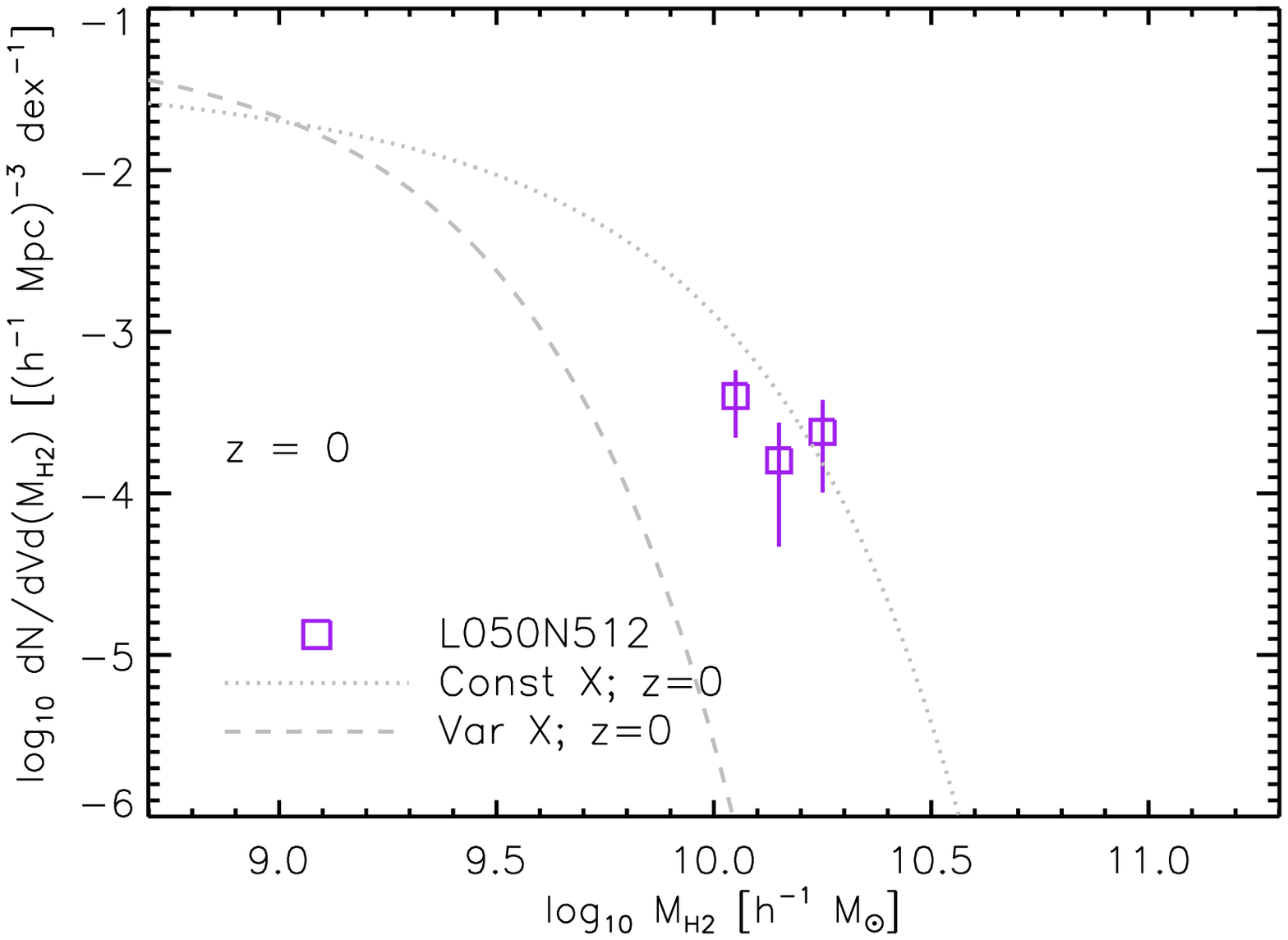, scale=0.35} &
    \epsfig{figure=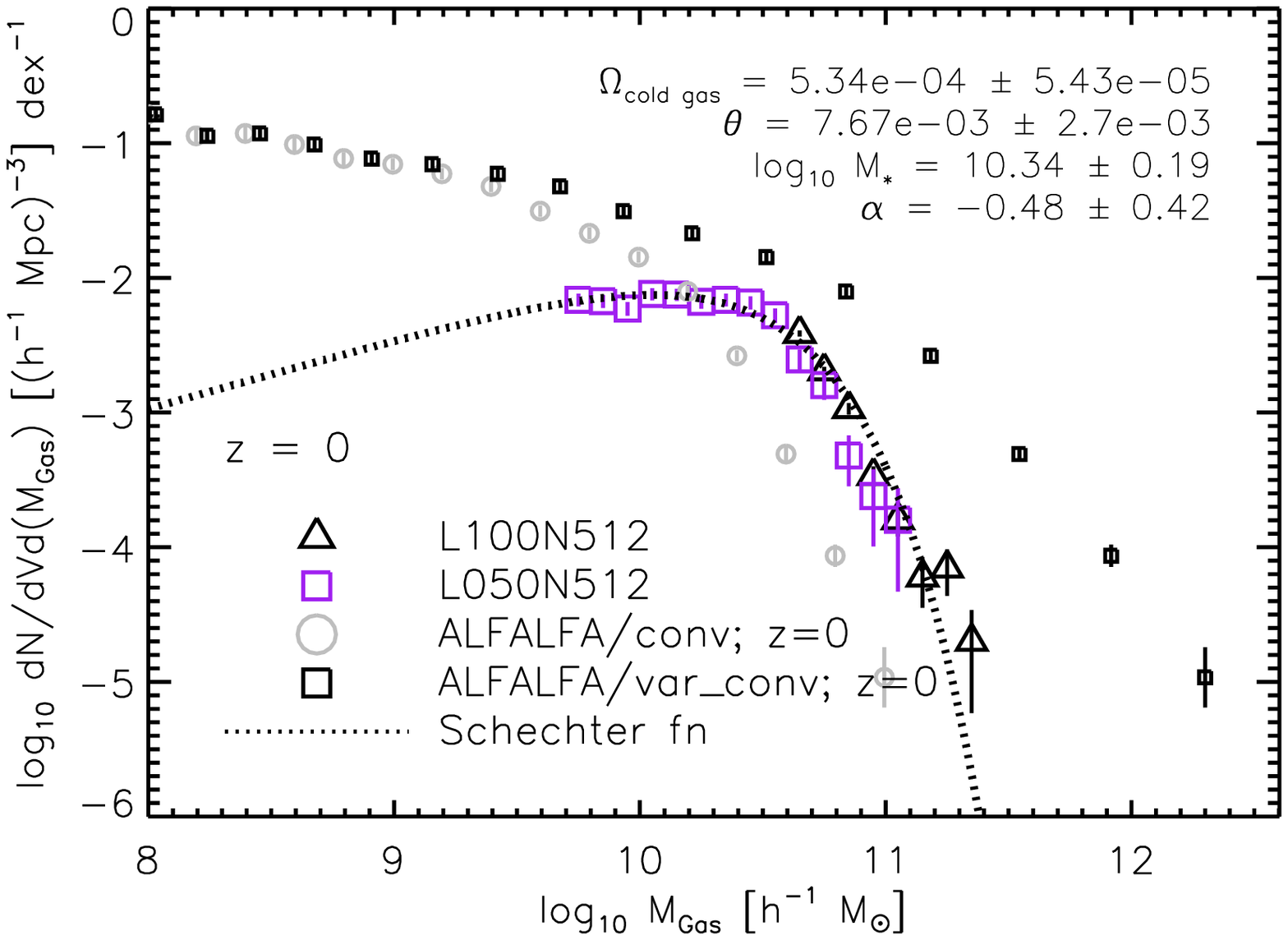, scale=0.35} \\
    \epsfig{figure=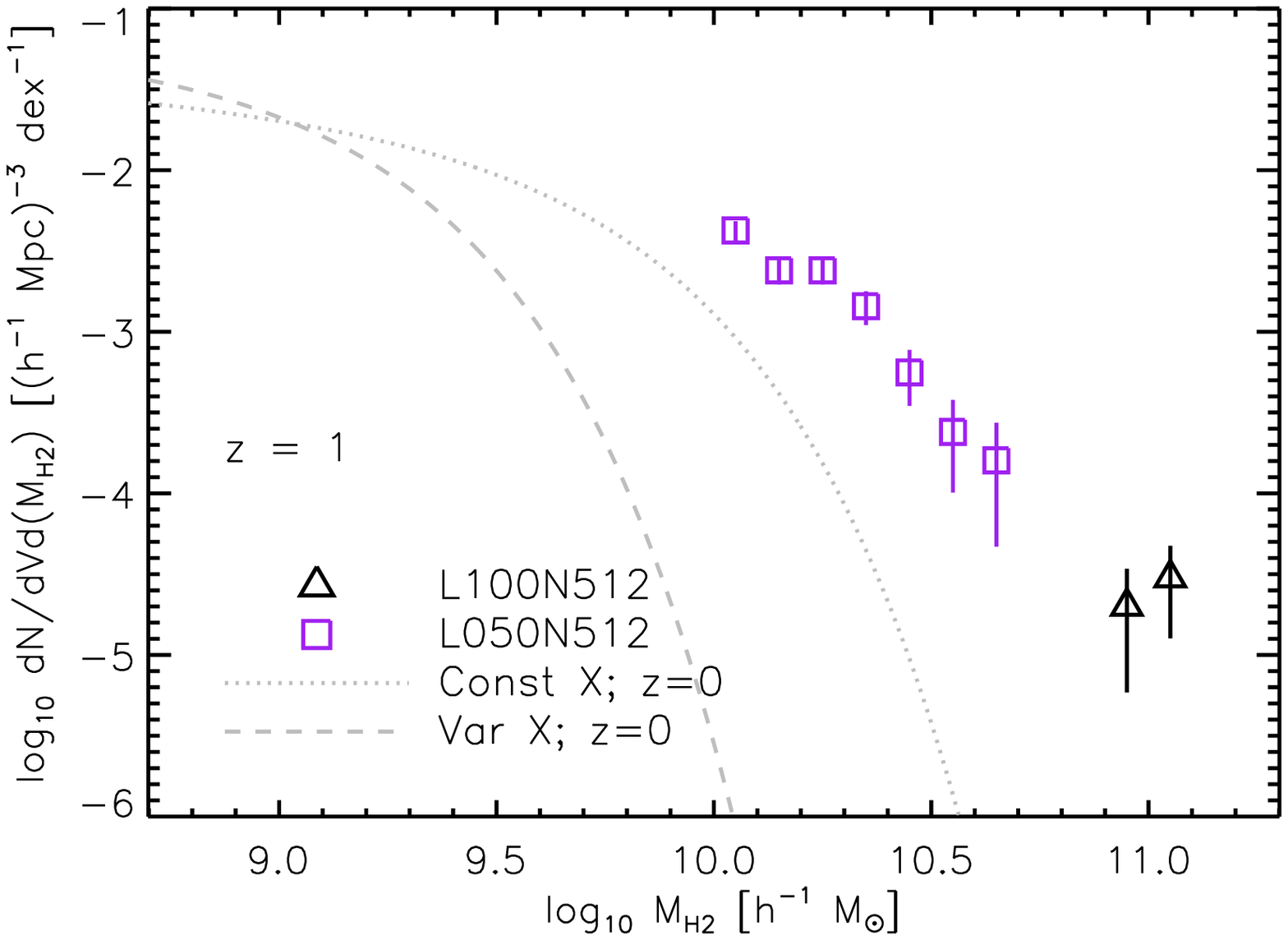, scale=0.35} &
    \epsfig{figure=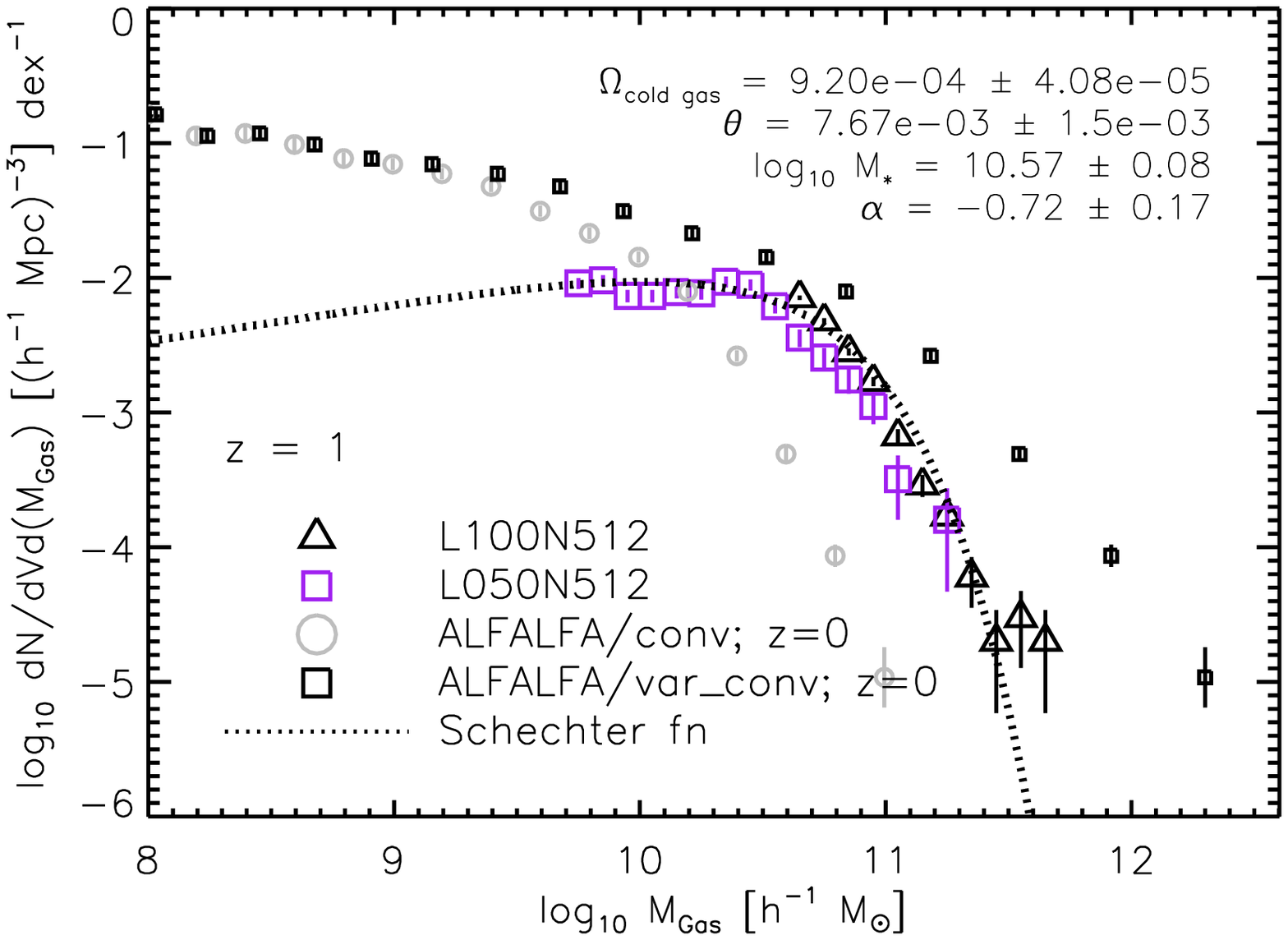, scale=0.35} \\
    \epsfig{figure=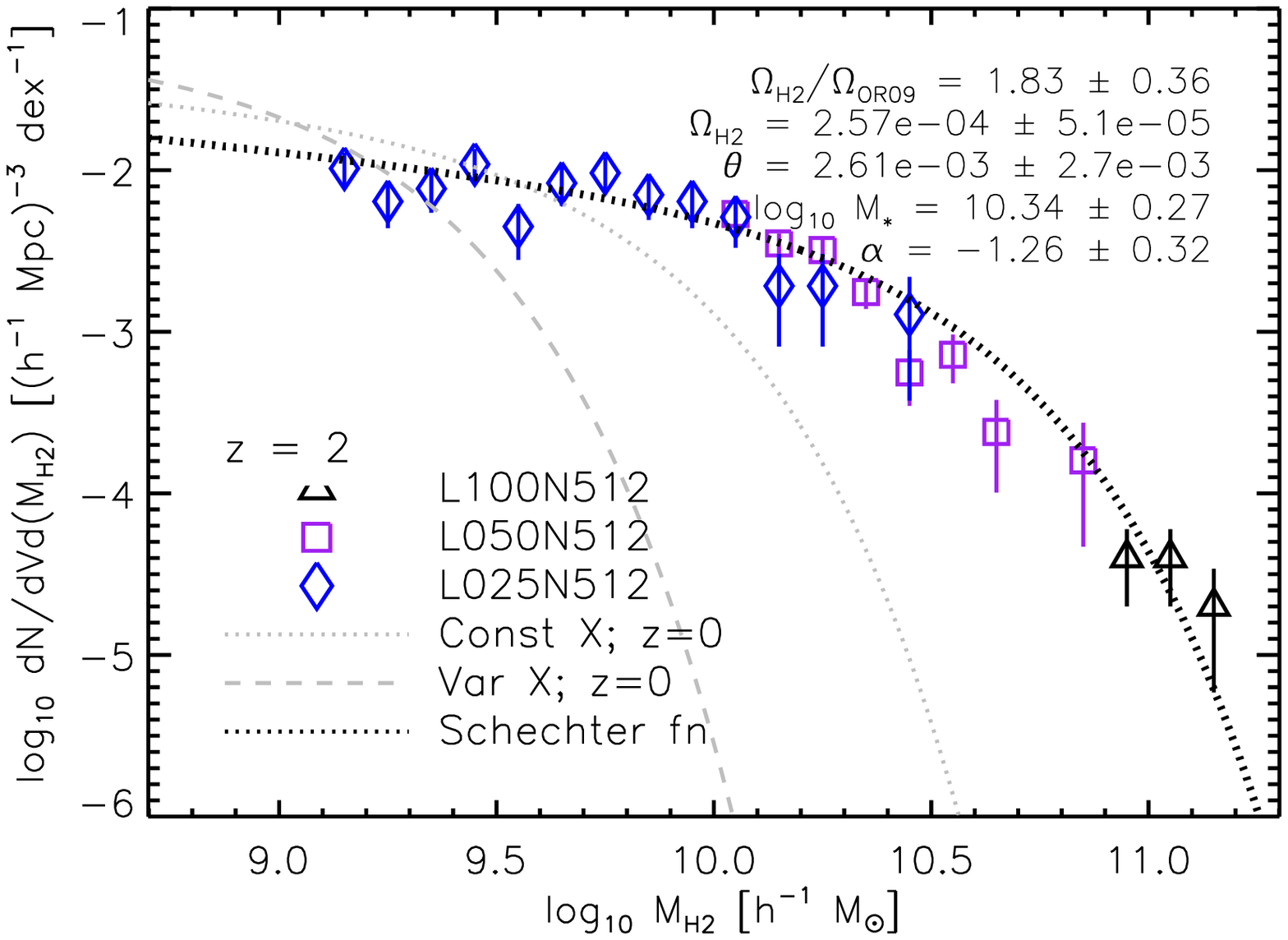, scale=0.35} &
    \epsfig{figure=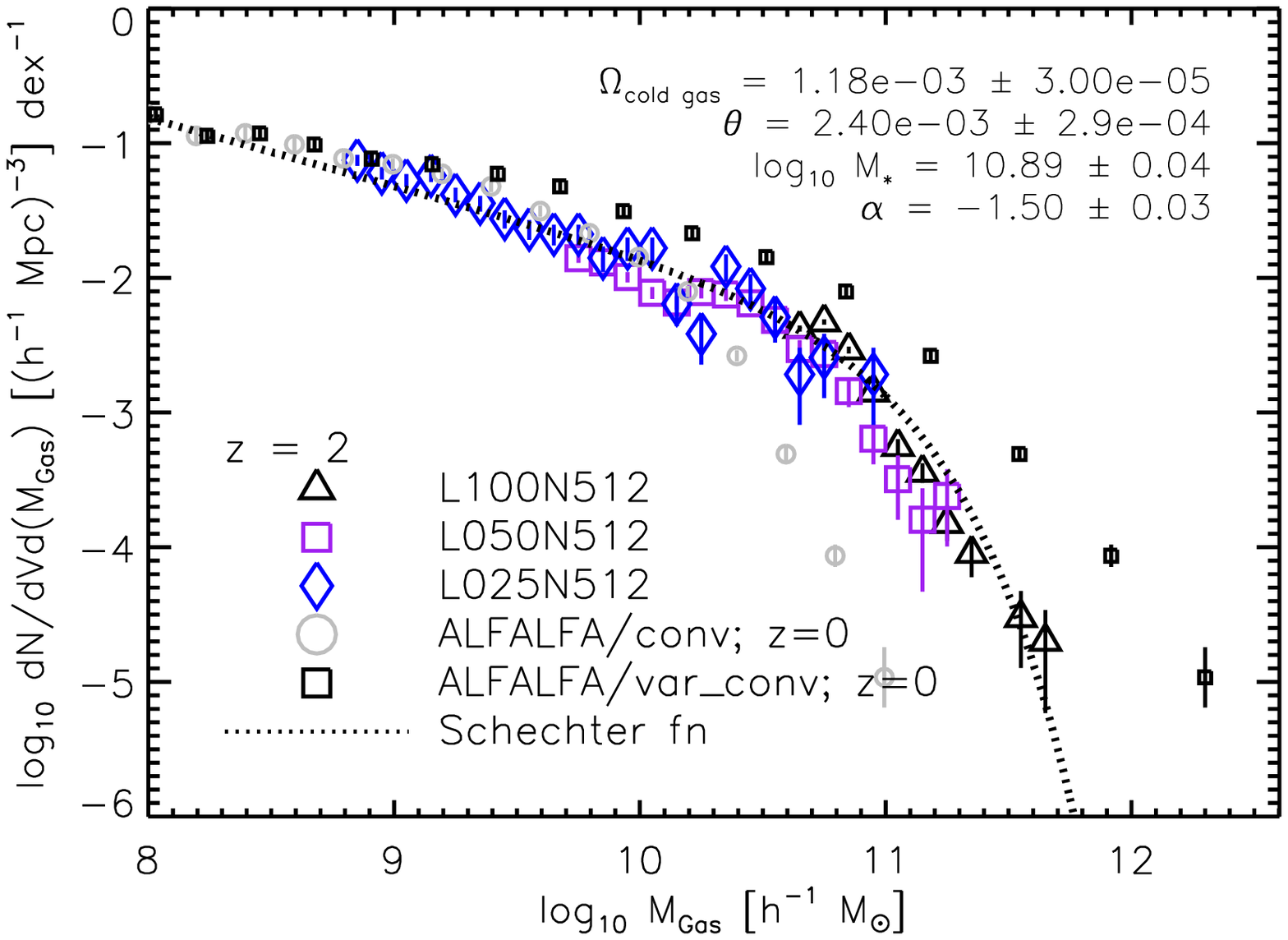, scale=0.35} \\
    \end{tabular}
    \caption{The simulated $\rm H_{2}$ (cold gas) mass function is shown in the left (right)
column at $z=$ 0 (top), 1 (middle) and 2 (bottom), with the $z=0$ $\rm H_{2}$ mass function results from the CO 
mass function of~\citet{Keres:03} when using a constant ($X$-factor) 
conversion or the varying scheme proposed by~\citet{Obreschkow:09a} (denoted with the dotted and dashed lines, respectively). 
The cold gas mass function is estimated by converting the ALFALFA detections through either a constant
assumed atomic-molecular ratio or a variable scheme that depends on the disc gas and 
stellar mass, with both values taken from~\citet{Power:10}.
The errors on the points are Poissonian.
We have made a resolution cut in $M_{\rm H2}$ ($M_{\rm Gas}$) using Equation~\ref{eqn:hires}. Note that the molecular
mass function evolves more strongly than the cold gas mass as the high mass end.}
    \label{fig:htwomassfn_hicut}
  \end{center}
\end{figure*}

For the $z=0$ molecular hydrogen mass function, top left panel, we have only a limited
mass range ($M_{{\rm H}_2} \simeq 1-2 \times 10^{10}\hMsol$), but the 
abundance of objects is in reasonably good agreement with the data of \citet{Keres:03},
using a constant (X-factor) conversion between CO and H$_2$ mass. 
(We implicitly used this conversion rather than the variable measure
advocated by~\citealt{Obreschkow:09a}, when we adopted the~\citealt{Leroy:08} molecular-atomic
abundance normalisation.)
It is clear that the high-mass end of the Schechter function evolves strongly as a function of redshift, with many more high
molecular mass systems at higher redshift.
The overall cosmic molecular mass density at $z=2$ is predicted to be a factor two larger than observed
locally. Thus, the large increase in stellar density between $z=2$
and the present day is reflected in the molecular mass function rather than the
HI mass function, which appears to be demonstrate mild evolution. This overall effect is also seen 
for a range of other physics schemes probed in Fig.~\ref{fig:htwomassfn_multi_hicut}.

The cold gas mass function can be estimated observationally by converting the HI mass function from ALFALFA, corrected for the He fraction
assuming a molecular - atomic ratio based on the ratio of the global densities of each. 
As advocated by~\citet{Power:10}, for a fixed molecular-atomic ratio this conversion is akin to dividing the HI mass by $0.54$. 
For a conversion factor that depends on galactic scale properties, as considered in~\citet{Obreschkow:09a} the HI mass - cold mass conversion 
factor is given in Equation 4. in~\citet{Power:10} as 
\begin{equation}
M_{\rm HI} = 0.76 M_{\rm cold} / (1+R_{\rm surf})\,,
\end{equation}
where we previously defined $R_{\rm surf}$ in Equation~\ref{eqn:rsurf}. 
The exact conversion of HI mass to cold gas mass is especially important at the high-mass end. If we use a variable conversion ratio,
then our simulated gas masses are too low at the high-mass end. The situation is reversed in the case of a constant conversion factor. For the
case of the constant conversion factor,~\citet{Power:10} find that the semi-analytic prescriptions also typically overproduce the number of 
gas-rich systems. There appears to be a generic disagreement between both hydro and semi-analytic simulations with the data if a 
constant HI/H$_2$ is assumed. This lends increasing support to the idea that the ratio depends on 
the global properties of the system in which it is measured, e.g. as considered in~\citet{Obreschkow:09a}.

\section{Effects of feedback and metal cooling on HI mass function}\label{sec:HImassfn_multi}
As we have demonstrated in previous work~\citep{Duffy:10},
the gas distribution in a halo can vary significantly with the 
physics implementation. 
This is especially true as the \emph{ZC\_WFB} model is unable to remove gas from the 
largest galaxies, and hence is unable to form red ellipticals by suppressing star
formation in the most massive systems.
We attempt to analyse the sensitivity of our results to the wide array of possible feedback types, as well as cooling
schemes, used in this previous work.
We have created the HI mass function at $z=0$ and $2$ for the different
physics prescriptions and present these in Fig.~\ref{fig:himassfn_multi_hicut}, again applying
the HI mass cut from Equation~\ref{eqn:hires} and tuning each physics simulation 
independently in the same manner as Section~\ref{sec:noneosgas} with pressure
values given in Table~\ref{tab:pressure}. Equivalent plots to Fig.~\ref{fig:htwomassfn_hicut},
for the H$_2$ and cold gas mass functions, are shown in Fig.~\ref{fig:htwomassfn_multi_hicut}.

\begin{figure*}
  \begin{center}
   \begin{tabular}{ccc}
    $\SSNO$ & $\SSALF$ & $\SSDLA$ \\
    \epsfig{figure=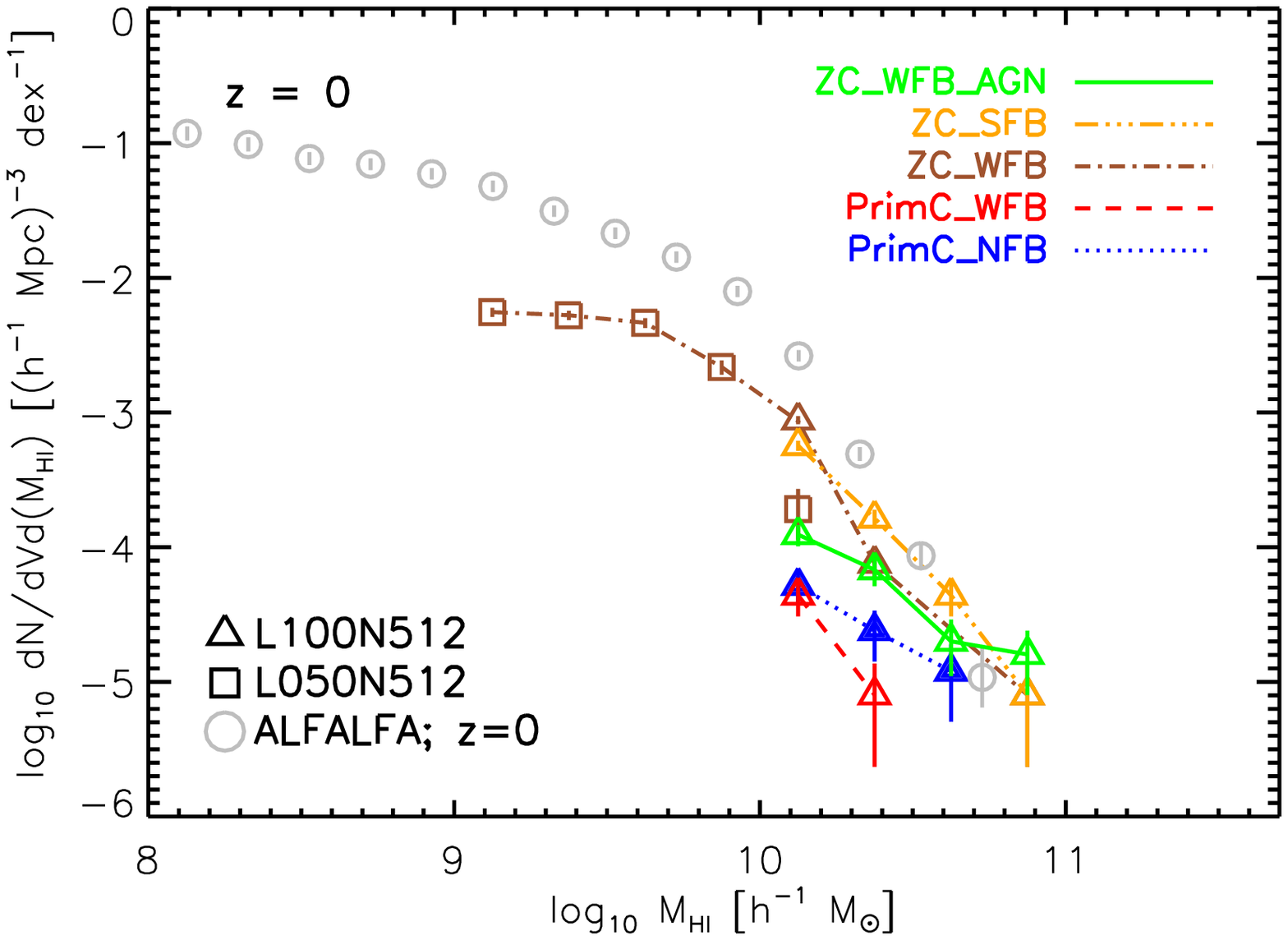, scale=0.32} &
    \epsfig{figure=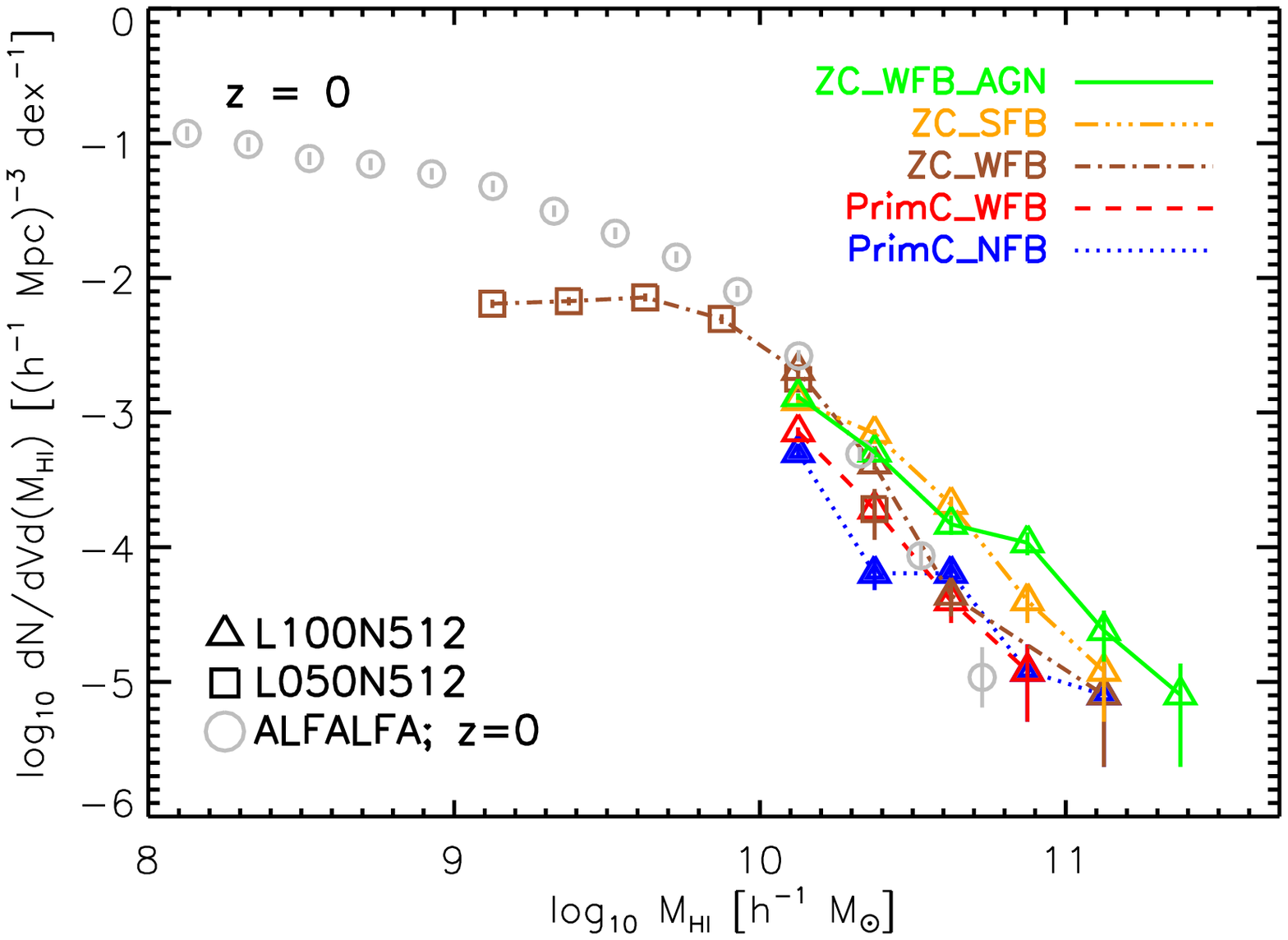, scale=0.32} &
    \epsfig{figure=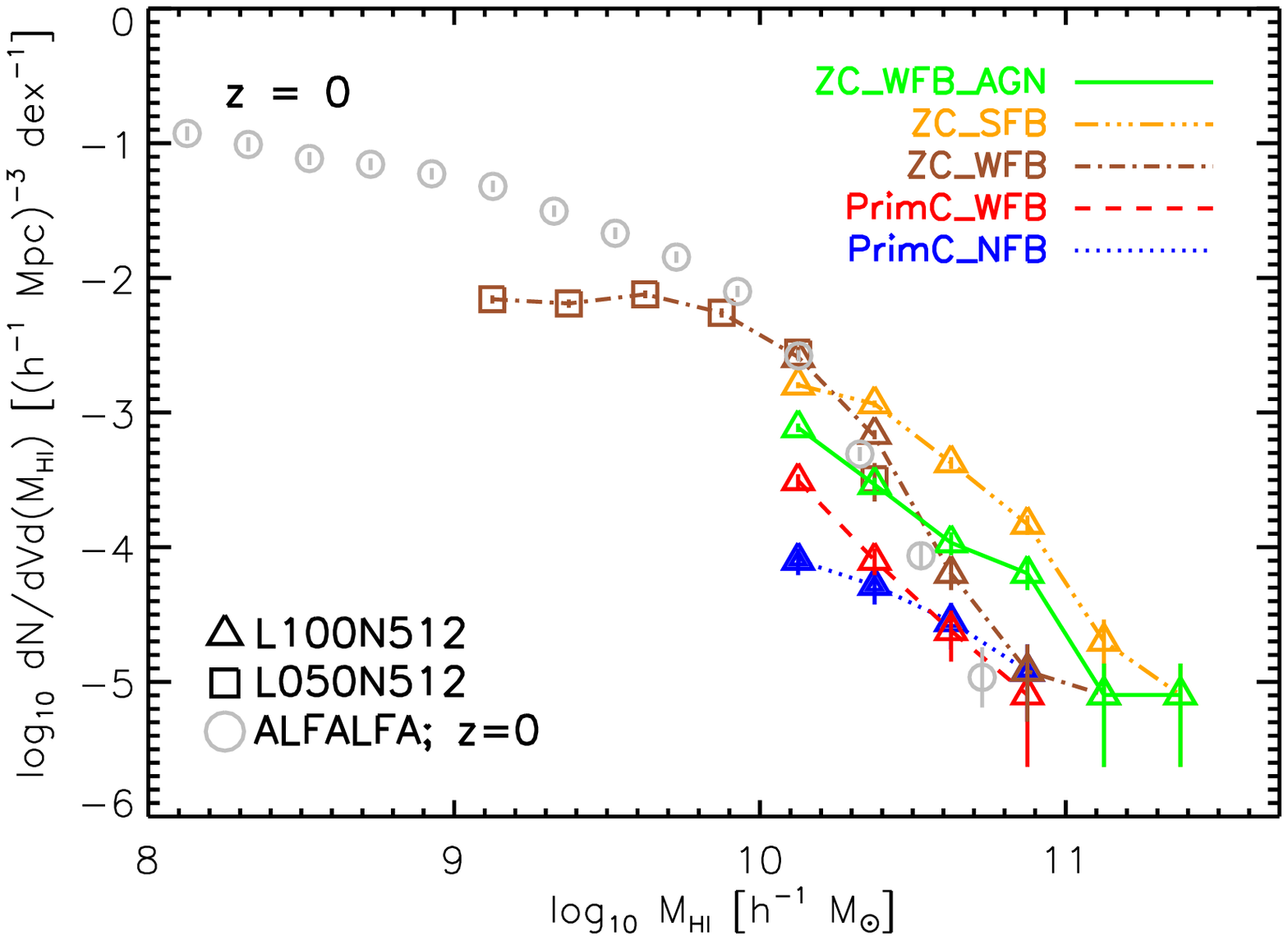, scale=0.32} \\
    \epsfig{figure=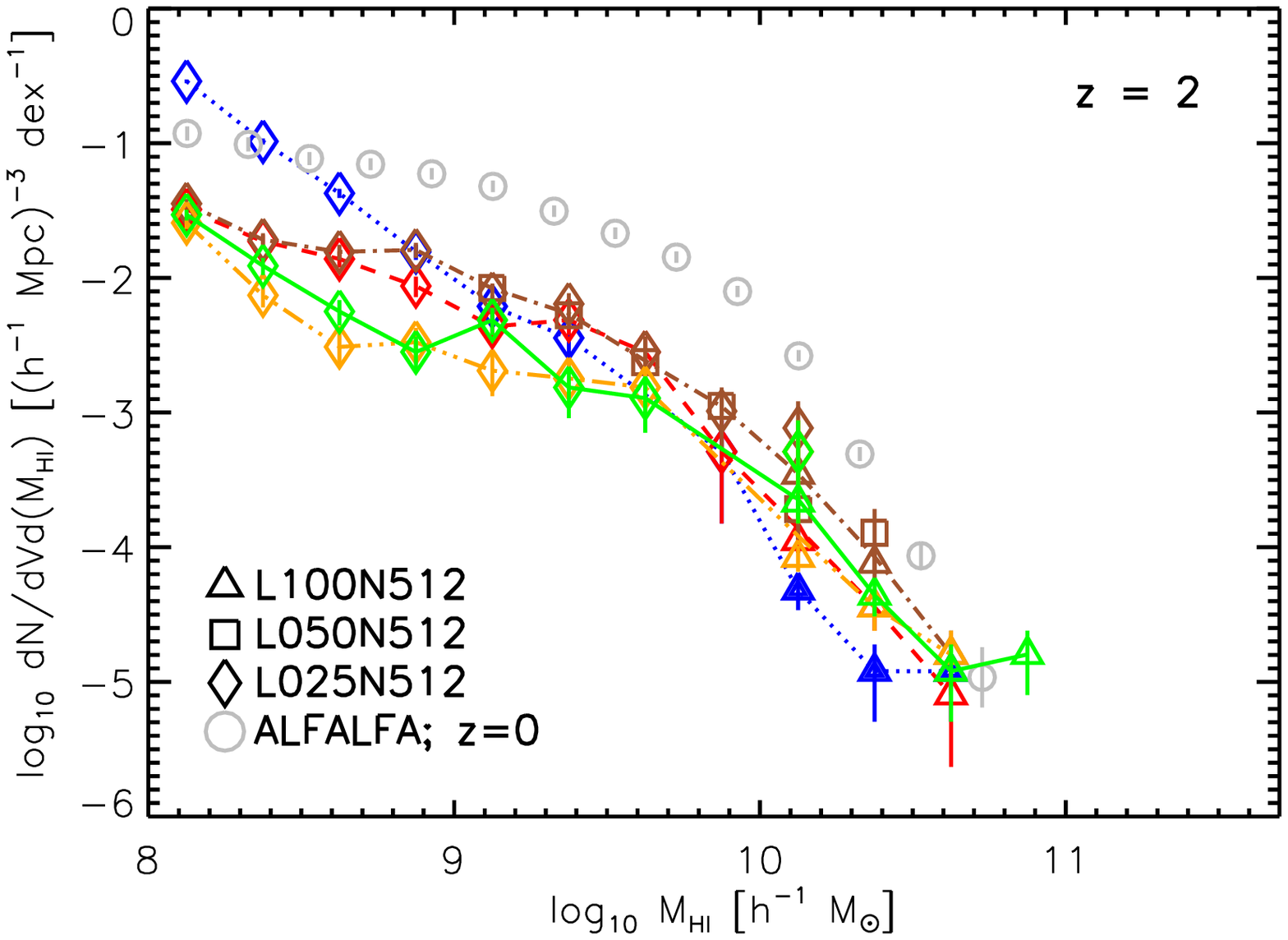, scale=0.32} &
    \epsfig{figure=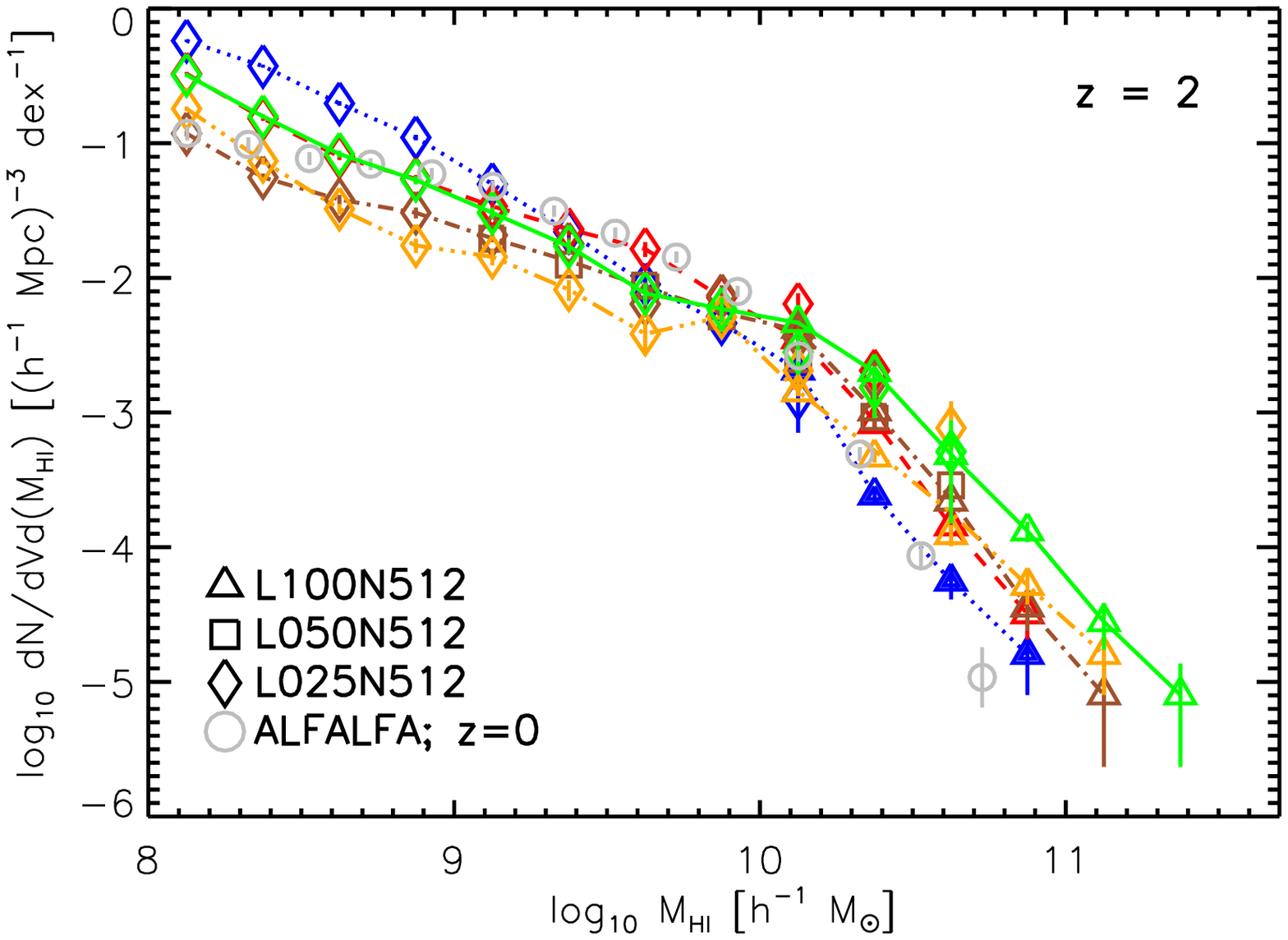, scale=0.32} &
    \epsfig{figure=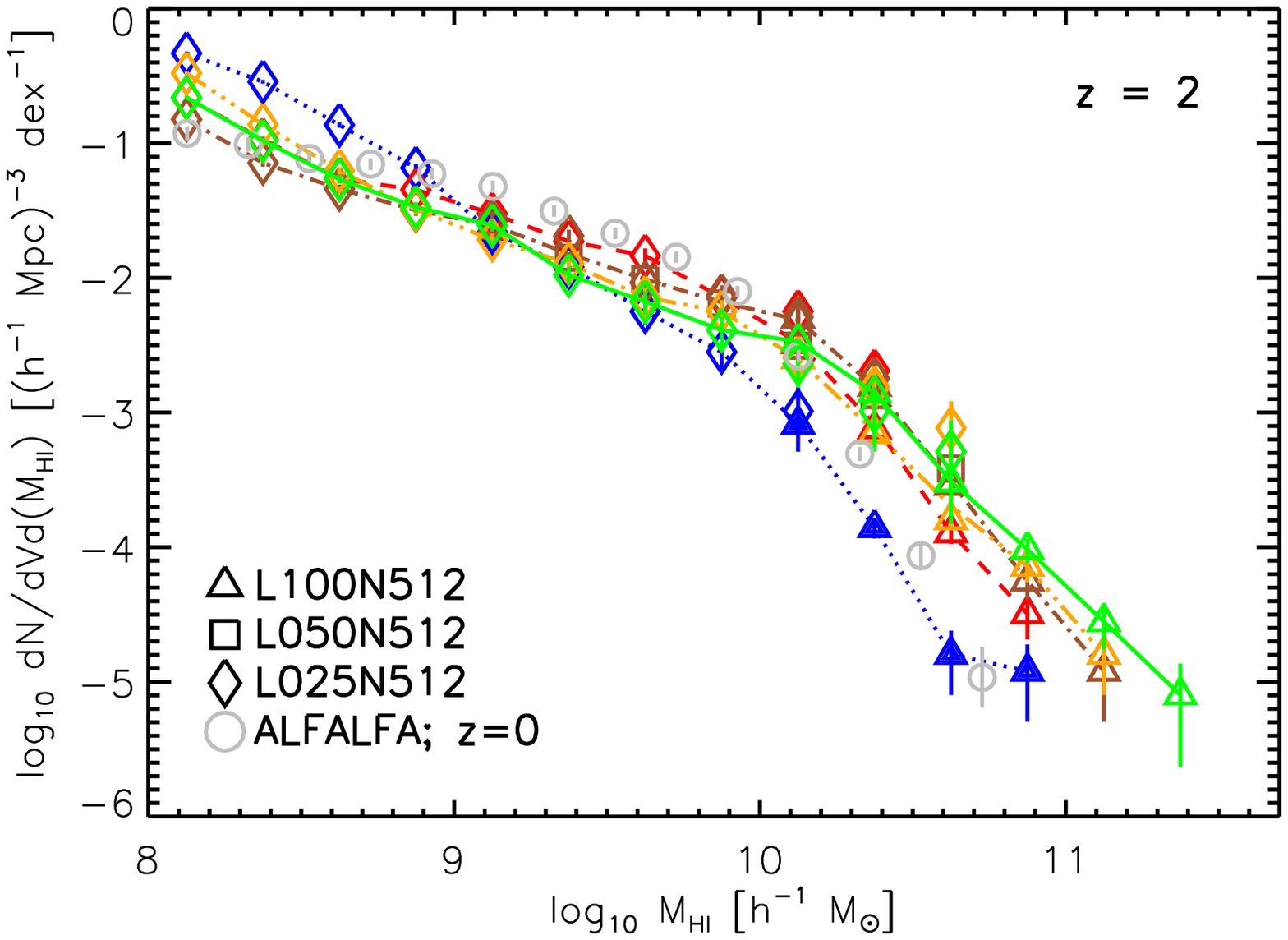, scale=0.32} \\
   \end{tabular}
    \caption{As Fig.~\ref{fig:himassfn_hicut} but now showing a comparison of HI mass functions for runs with different physics prescriptions, 
    at $z=0$ (top panels) and $z=2$ (bottom panels). Again, the columns show results for self-shielding 
    methods $\SSNO$, $\SSALF$, $\SSDLA$, left to right respectively. The low-redshift result from ALFALFA is also shown for comparison.
    Note that certain simulations, e.g. \emph{PrimC\_WFB} and \emph{PrimC\_NFB}, have too little cold gas at low redshift that even when setting
    the $P_{\rm shield} = 0$, converting all cold gas in the galaxy to be neutral. This is seen most clearly in the top-middle panel in which the 
    red dashed and blue dotted lines lie below the other simulations at all mass bins.}
    \label{fig:himassfn_multi_hicut}
  \end{center}
\end{figure*}

\begin{figure*}
  \begin{center}
    \begin{tabular}{cc}
    \epsfig{figure=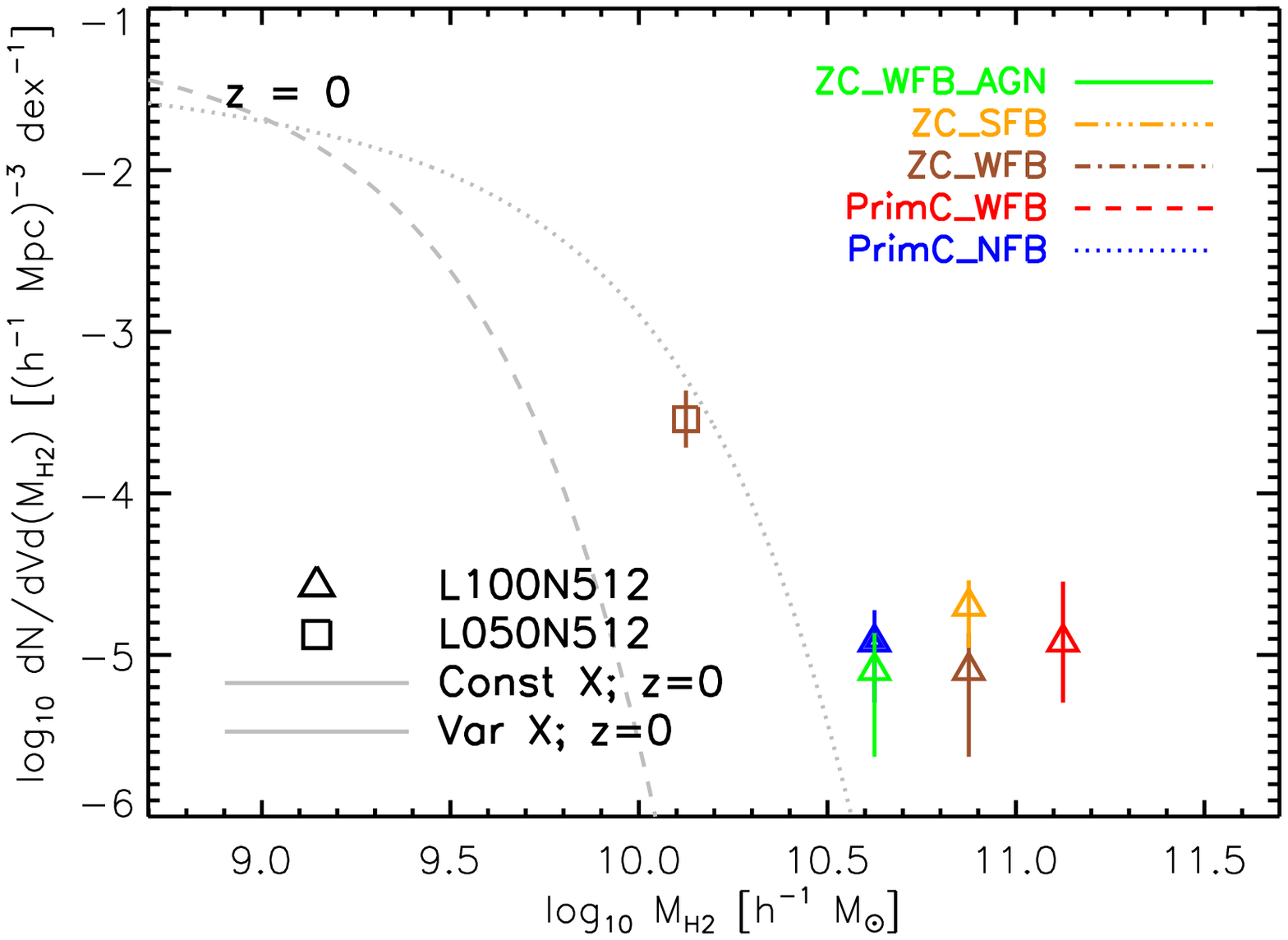, scale=0.35} &
    \epsfig{figure=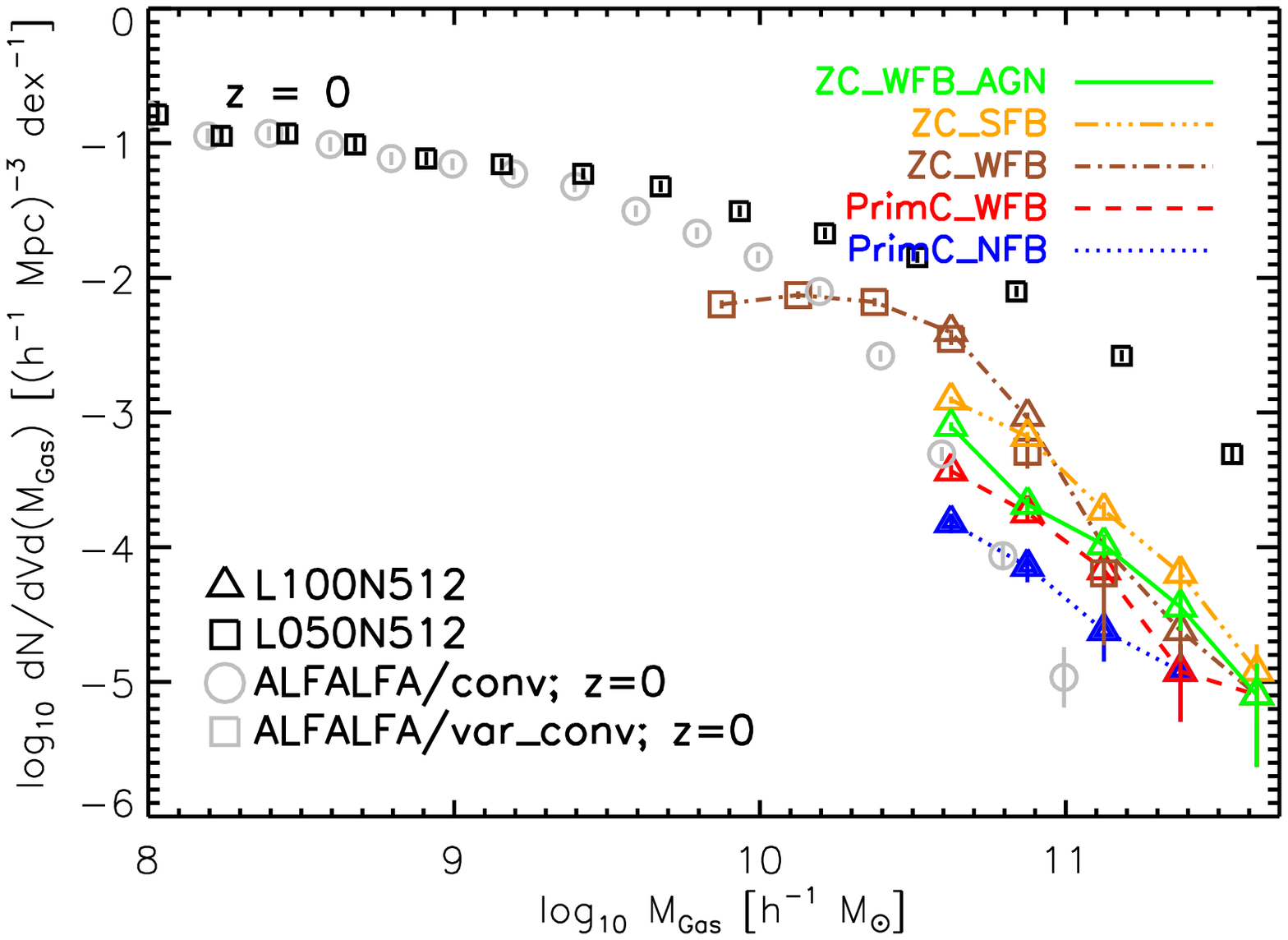, scale=0.35} \\
    \epsfig{figure=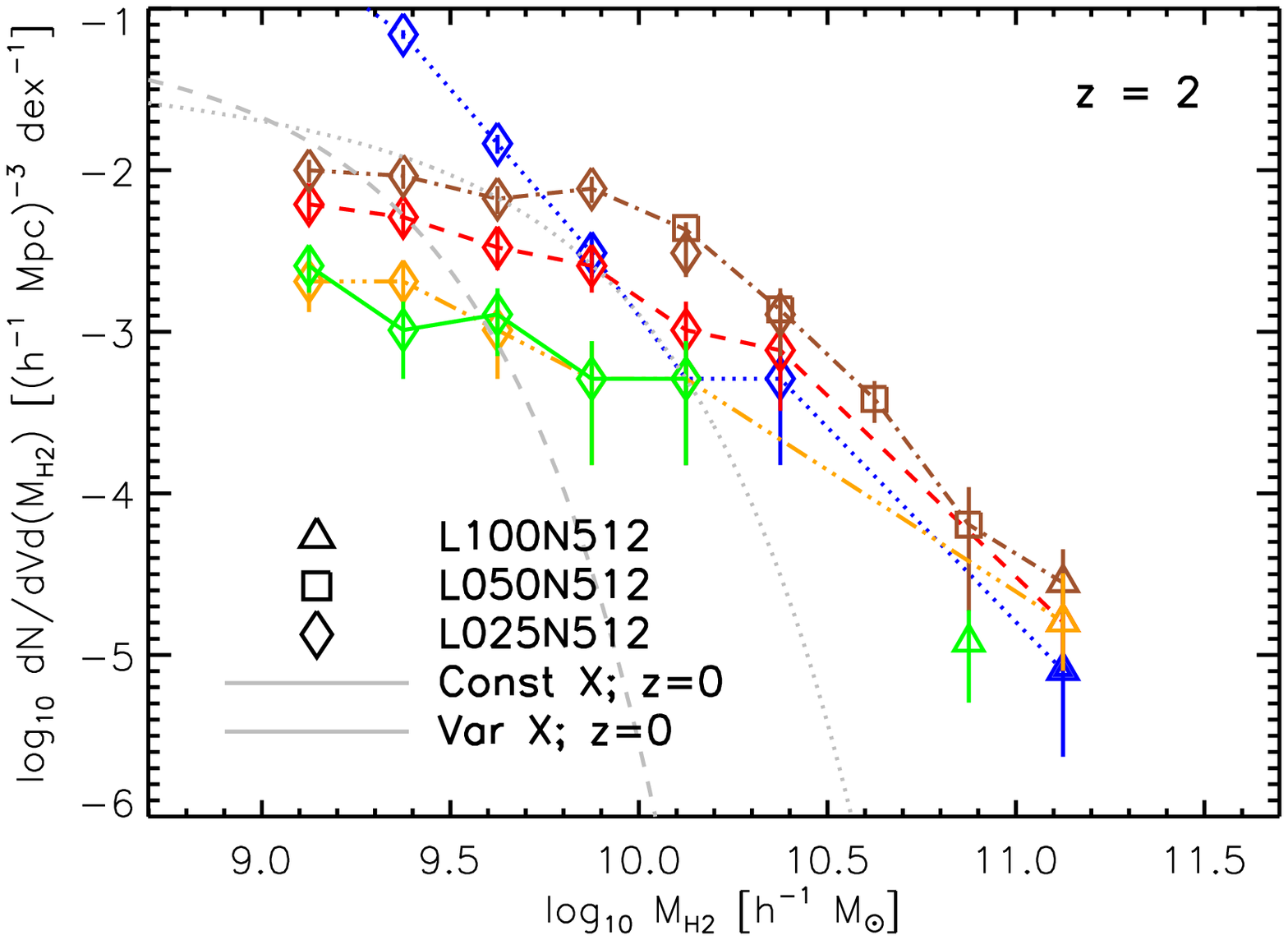, scale=0.35} &
    \epsfig{figure=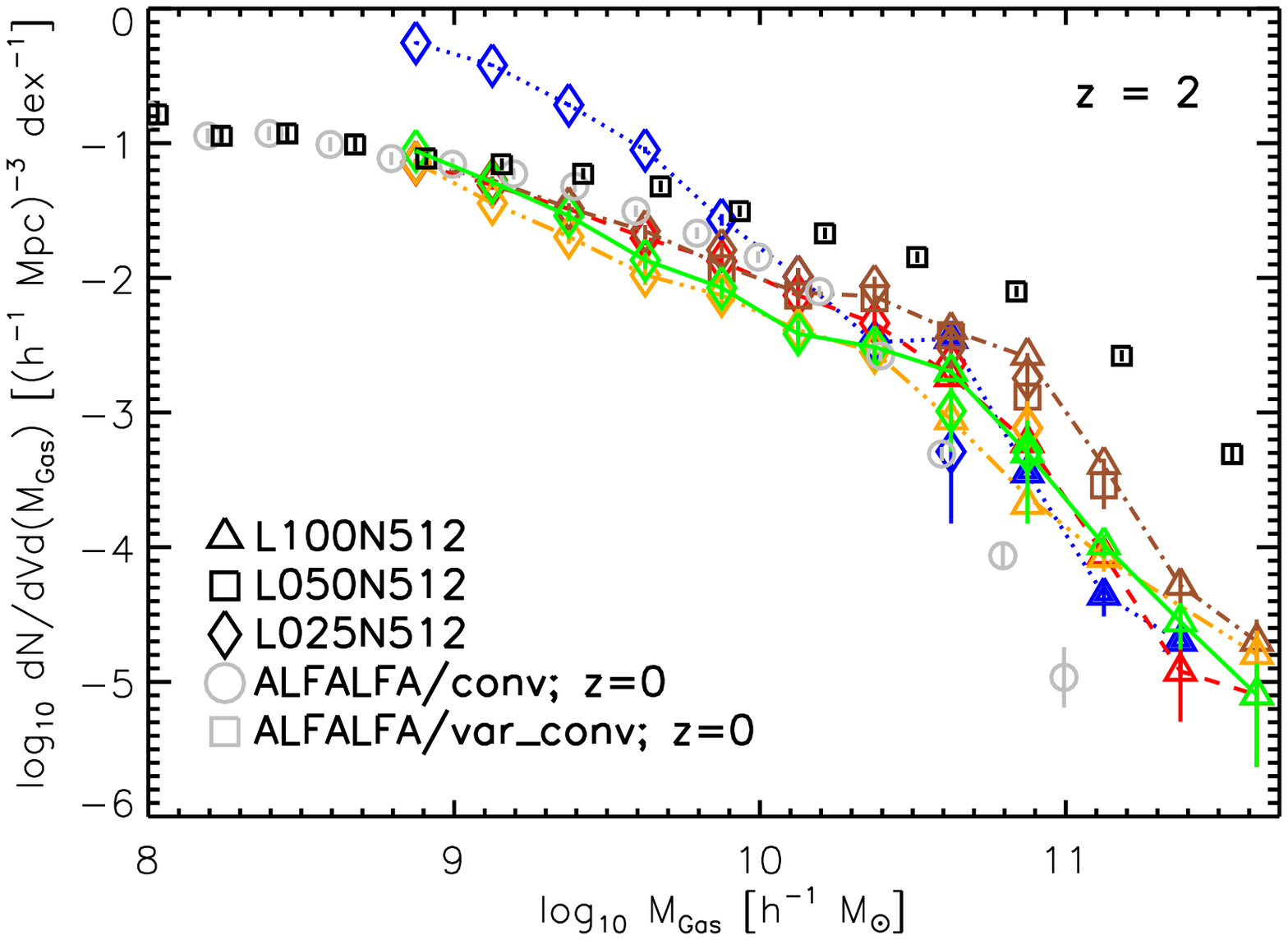, scale=0.35} \\
   \end{tabular}
    \caption{As Fig.~\ref{fig:himassfn_multi_hicut} but now showing a comparison of H$_2$ (left) and cold gas (right)
    mass functions for the different physics prescriptions. The self-shielding method is irrelevant for these results.  
    The grey curves are the $z=0$ $\rm H_{2}$ mass function results from the CO mass function of~\citet{Keres:03} 
    when using a constant ($X$-factor) conversion or the varying scheme proposed by~\citet{Obreschkow:09a} 
    (denoted with the dotted and dashed lines, respectively).}
    \label{fig:htwomassfn_multi_hicut}
  \end{center}
\end{figure*}

Without any self-shielding implemented at $z=0$ (top-left panel of Fig.~\ref{fig:himassfn_multi_hicut})  
the simulations with strong feedback (\emph{ZC\_WFB\_AGN} and \emph{ZC\_SFB}) 
result in a greater number of massive HI systems than the schemes without effective feedback. This is a result of more cold
gas being prevented from forming stars, as seen in the top right panel of Fig.~\ref{fig:htwomassfn_multi_hicut}.
When we tune the self-shielding prescription to match the ALFALFA HI cosmic density (over the HI
mass range $10^{10} - 10^{11} \hMsol$; method $\SSALF$)
 the difference between the various physics schemes at $z=0$, top middle panel, is reduced as expected.
 
At $z=2$ we can tune to the DLA result (method $\SSDLA$), shown in the bottom right panel, which also shows very 
little difference between the schemes, although the run with no feedback (\emph{PrimC\_NFB}) shows a lack
of massive HI systems in comparison to the other runs. We see from the molecular and cold gas
mass functions at $z=2$ (bottom panels of Fig.~\ref{fig:htwomassfn_multi_hicut}) that more molecular hydrogen is
present at high-redshift, due to both the lack of feedback and time to use up the gas and form stars.

It appears that the HI mass function is at least as sensitive, if not more so, to the modelling of 
self-shielding as to the sub-grid baryonic physics.
This could be seen as an unexpected result and highlights
the need for a more careful modelling of the optically thick gas in and around galaxies.

There is significantly more variation in the molecular hydrogen mass function when the baryonic physics
is varied, as shown in the left column of Fig.~\ref{fig:htwomassfn_multi_hicut}.
The situation at $z=2$, bottom left panel, is complex but apparently the strong feedback schemes
remove large amounts of the densest gas relative to the simulations without feedback.
The cold gas at $M_{\rm Gas} \ge 10^{9.5} \hMsol$ is, on the other hand, completely insensitive to the sub-grid physics scheme
used provided there is some feedback to suppress the rapid overcooling demonstrated by the \emph{PrimC\_NFB}
model in the bottom right panel. The potential paradox presented here between the greater sensitivity of the H$_2$ to the feedback schemes
than the HI when the cold gas also appears insensitive is a consequence of the cold gas having no dependancy
on the self-shielding prescription used to calculate the HI. As a result the HI fraction of the cold gas can vary to match
the observations, giving a greater freedom in the HI results compared to the H$_2$ which have no such variability. Thus
the cold gas mass across simulations is similar but the distribution of this gas as a function of pressure is not, which is then
most easily visible in the H$_2$ mass functions shown in Fig.~\ref{fig:htwomassfn_multi_hicut}.

\section{Conclusions}
\label{sec:HI_conclusion}

We have used hydrodynamical simulations from the OWLS project to investigate 
the distribution of neutral (atomic HI and molecular H$_2$) hydrogen in cosmological volumes across the redshift range $z=0-2$. The large
suite of simulations allowed us to probe both the effects of numerical resolution
and the sub-grid baryonic physics implemented (especially the effects of metal enrichment on
the gas cooling rate and the different methods and sources of feedback). 
In addition, we considered several methods to correct for the lack of self-shielding
in our simulations in moderate-high-density regions, where the assumption that the gas is optically 
thin to a uniform UV/X-ray  background breaks down.
With this in mind, we have found that, together with an empirical pressure-based law to
discern the molecular fraction in galaxies, the distribution of HI can be successfully simulated in our 
cosmological volumes. We have succeeded in recreating the observed local HI mass function for
objects with HI mass greater than $10^{10} \hMsol$ but with a significantly shallower faint-end slope.
We consider evolution across the redshift range $z=0 - 2$ and detect significant increases in low-mass
HI objects which corresponds into a small, yet systematic, increase in the HI cosmic density with increasing redshift.
If we mimc an observational magnitude limited survey, namely objects with HI masses above $10^{10} \hMsol$, and 
fix the faint end slope to the $z=2$ value then we find that the inferred cosmic HI density is in better agreement
with observations at $z=0$ and that observers will infer little evolution in the period $z=0-2$. During this time 
the molecular hydrogen component evolves strongly, with
many more massive systems in the distant past than now, in qualitative agreement
with~\citet{Obreschkow:09b}.

When comparing the predictions of the simulations with varying physics prescriptions,
including an AGN feedback model, we found that the HI mass functions were insensitive to these 
treatments. Provided one can tune the pressure based self-shielding threshold, we can recover the overall HI distribution 
to within a factor of two using just a single threshold value.

However, the implementation of the self-shielding threshold has at least as large an effect on the HI mass function
than the differences between the sub-grid cooling and feedback models we consider here, these models
essentially bracket the extrema of sub-grid implementations and hence to improve on this level of accuracy in the future
we will utilise radiative transfer schemes such as in~\citet{Altay:10} and~\citet{traphic}.

For the case of the simulation with supernovae feedback and realistic cooling we found a 
peak HI - halo mass fraction at a given halo mass across our redshift range, $z=0-2$, which we attribute to two effects. 
The first is that in more massive haloes the neutral component is dominated by denser $\rm H_{2}$ while at lower masses the overall
gas fraction of the halo has been reduced due to feedback. 
This peak is also found to be the `knee' of the HI mass function, which is well
reproduced in our simulation with supernova feedback. 

The typical sizes, denoted by the half-mass radius, of the HI, H$_2$ and stars in this work
were found to vary as a function of total halo mass and redshift. We found, however that the HII radius increased
with the cube root of the halo mass, characteristic of a smoothly distributed isothermal profile. The typical sizes of the HI disk increased with redshift
but that the steepness of this relation with halo mass decreased from a linear dependency to approximately a cube root (tracing of the 
more diffuse, ionised component).
The stars and molecular hydrogen closely traced each other throughout redshift, with an identical half mass radius at galaxy scales, i.e. the pivot mass of 
$2 \times 10^{12} \hMsol$, but differed significantly at $z=0$ for galaxy group and cluster-sizes, with the stellar radius a factor 5 more extended the 
molecular component. We also tested the feedback from an AGN and found that the molecular and stellar components were more extended
than their counterparts in the simulation without an AGN, at all redshifts. This is in agreement with~\citet{Sales:10} who studied galaxies at $z=2$
and found that such strong feedback resulted in the formation of large, extended galaxies.

Finally, we have found that the ratio of HI (H$_2$) to stellar mass of our objects is negatively (positively) correlated with the stellar mass. 
The normalisation of the model with SNe feedback was found to be in particularly good 
agreement with observations.
Furthermore, we predict that the ratio of HI (H$_2$) to stellar mass at a given stellar mass increases significantly with redshift. 
This means that the hydrogen signal from small systems will be easier to find than an extrapolation of 
from larger optical counterparts would suggest. This improves the chances that 
optically faint galaxies may be detectable in upcoming ultra-deep 21 cm and molecular hydrogen observations. 

\section*{Acknowledgements}
ARD was supported under an RAS Centre of Excellence grant and further
acknowledges this work was begun during an STFC PhD studentship. The authors wish to thank Martin 
Meyer, Rob Crain and Max Pettini for helpful and stimulating discussions. Further thanks to 
Kirsten Gottschalk, Serena Bertone and Freeke van de Voort for extensive technical help.
Additionally ARD would like to thank Attila Popping and Ann Martin for making available comparison data and useful 
discussions of results.
The simulations presented here were run on Stella, the LOFAR 
BlueGene/L system in Groningen and on the Cosmology Machine at the
Institute for Computational Cosmology in Durham as part of the Virgo
Consortium research programme. This work
was sponsored by the National Computing Facilities Foundation (NCF) for
the use of supercomputer facilities, with financial support from the
Netherlands Organization for Scientific Research (NWO). 
This work was supported by a Marie Curie Initial training Network
CosmoComp (PITN-GA-2009-238536).

%\bibliographystyle{mn2e}
%\bibliography{full_references}

\appendix
\section{Resolution Tests}
\label{appendix:restest}

Here we investigate how numerical resolution and simulation box size 
affect our main results. In particular, we study how increasing
the number of particles (and decreasing the gravitational 
softening length) affects the properties of the same sample of haloes. 
We also investigate the effects of increasing the simulation
volume while keeping the resolution constant. As a result, we identify
a simple scaling between the minimum required halo mass (that depends
on the particular baryonic species under consideration) and the particle mass.
We also show that our results cannot be significantly affected by the lack of
structure on scales larger than our available box sizes. Although we consider
the case of the neutral phases of hydrogen, similar tests have been performed
for the stars and ionised hydrogen to create the mass limits quoted in Equation~\ref{eqn:hires}.

\subsection{HI mass range}
\label{appendix:restest_hi}

\begin{figure}
  \epsfysize=2in \epsfxsize=4in
  \epsfig{figure=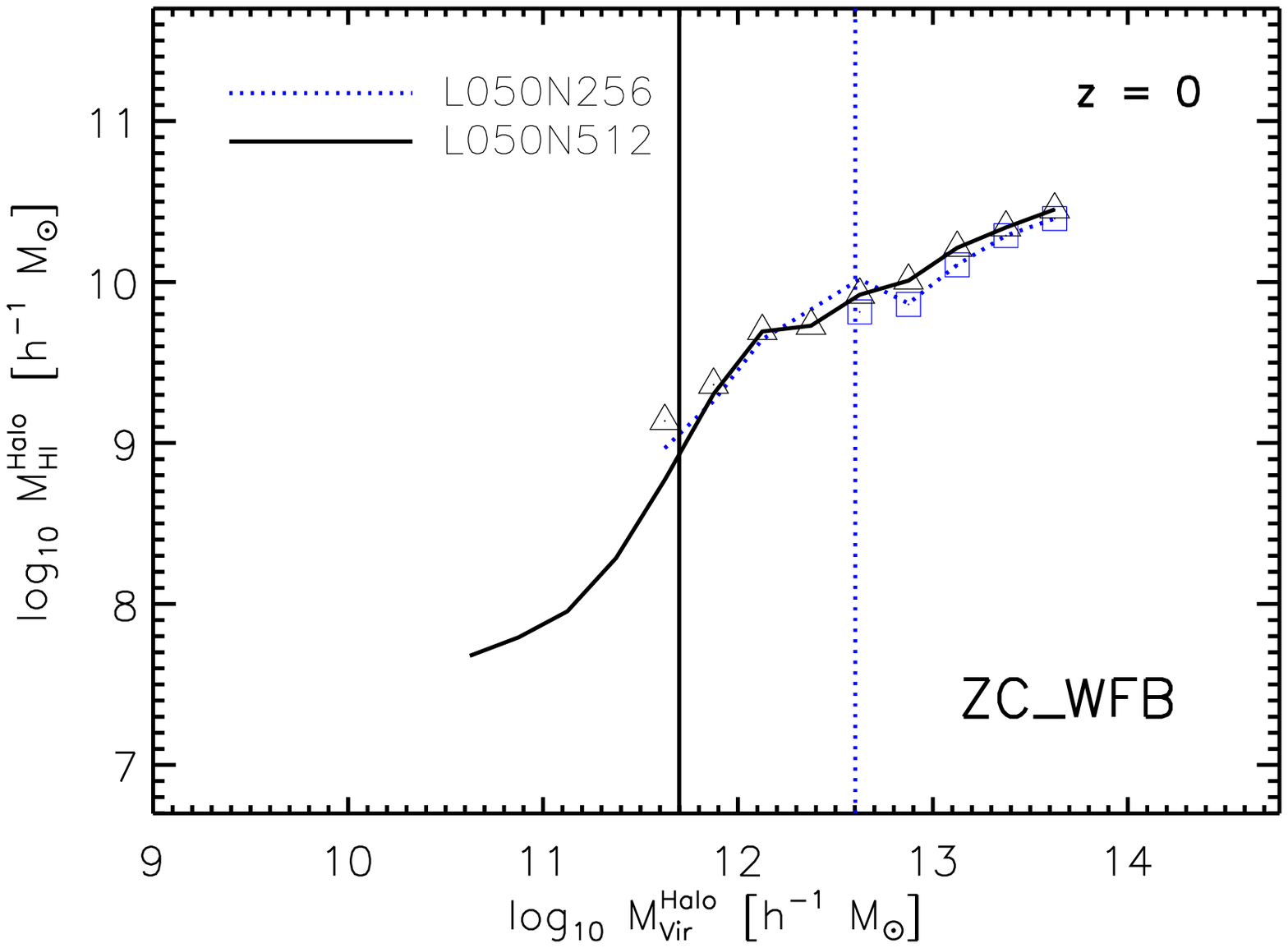, scale =0.45} 
  \epsfig{figure=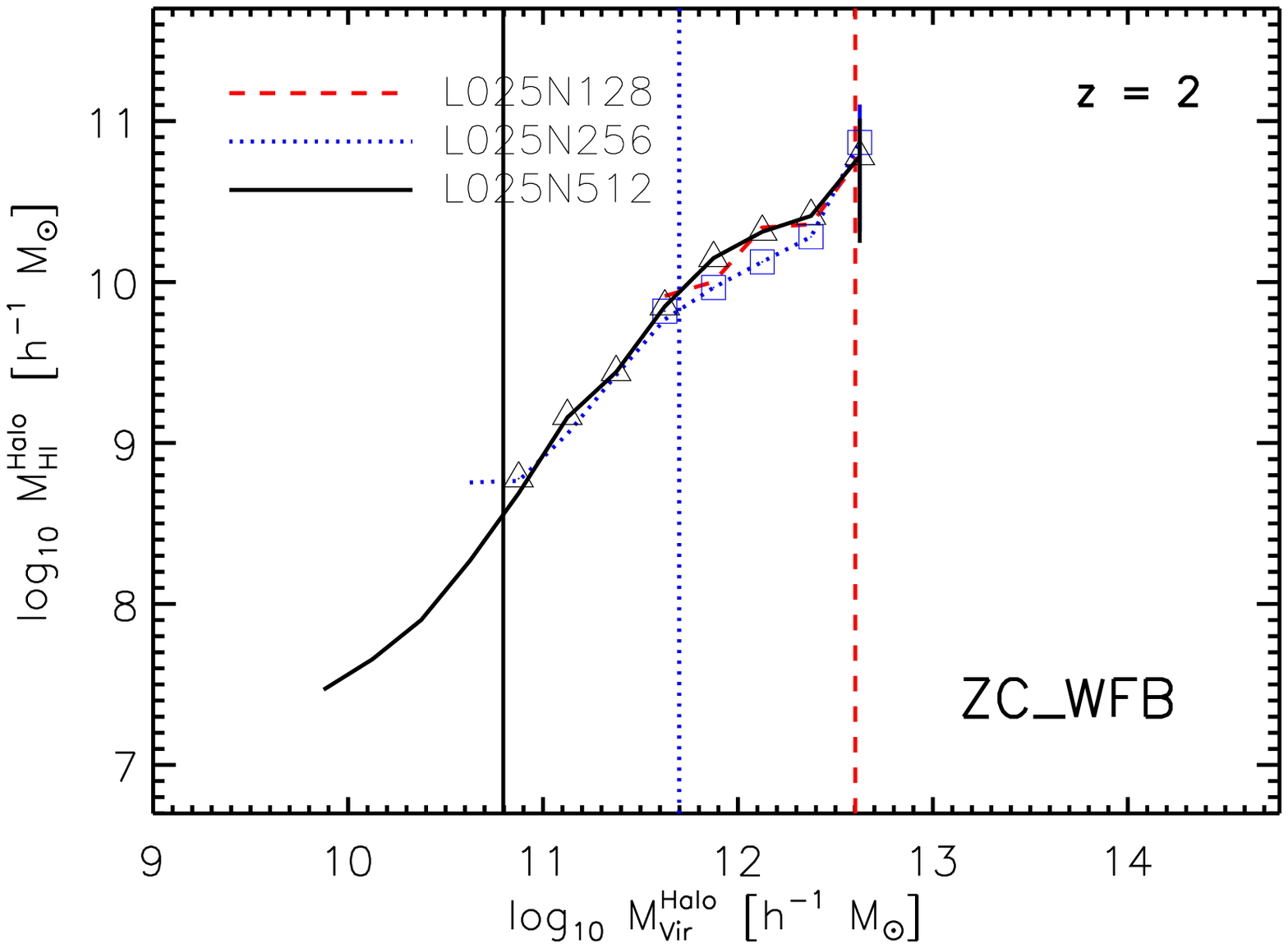, scale=0.45} 
  \caption{Resolution tests for the HI mass 
    in the FoF group as a function of ${M^{\rm Halo}_{\rm vir}}$ in \emph{ZC\_WFB}. The top panel 
    shows results for the  $50\, h^{-1}\,{\rm Mpc}$ box at $z=0$, while the bottom panel is for
    the 25 $\hMpc$ box at $z=2$. At $z=2 \, (0)$ we show results for 3 (2) simulations of the same box,
    spanning a factor of 64 (8) in particle number/mass. 
    The vertical lines indicate the mass limit given in Equation~\ref{eqn:hires} (roughly speaking this is equivalent
    to 7,000 dark matter particles) for each run. The curves correspond to haloes an order of magnitude in mass below
    this limit; the results appear converged, therefore we can be confident with our conservative limit in Equation~\ref{eqn:hires}. 
    We bin the haloes that have masses greater than the vertical mass limits and denote their median values with symbols
   (triangles for the solid curve, squares for the dotted).
   Errors are $1\sigma$ values, estimated by bootstrap analysis and added 
    in quadrature with Poissonian error estimates.}
  \label{fig:restest_box_m_mfof_z2}
\end{figure}

In order to determine the minimum halo size, and the corresponding HI mass 
resolution, we select FoF groups from the `\emph{ZC\_WFB}' simulation
according to their total mass, $M^{\rm Halo}_{\rm vir}$. 
After grouping the objects in fixed mass bins we calculate 
the median HI mass of the bin. The variation in median 
HI mass with halo mass is shown for two simulation sets in 
Fig.~\ref{fig:restest_box_m_mfof_z2}; with increasing resolution in
the $50 \hMpc$ box at $z=0$ (top panel);  and those with increasing
resolution in the $25 \hMpc$ box at $z=2$ (bottom panel). 

As one would expect, the curves demonstrate a positive correlation between
HI and halo mass. Those haloes which lie on the same relation between simulations typically
have at least $\sim 7,000$ dark matter particles. We adopt a conservative mass limit, shown as a vertical line for each simulation,
and identify the HI mass corresponding to this halo mass for each simulation. As mentioned in Section~\ref{sec:reseffects}, we found that we
could summarise this limit for all species as a simple power-law scaling
relation of the simulation mass resolution, as given
in Equation.~\ref{eqn:hires} in the main body of the text.
In our study, we 
have applied this limit to the baryonic masses, resulting in an even more conservative 
lower HI mass limit.

\begin{figure}
  \epsfysize=2in \epsfxsize=4in
  \epsfig{figure=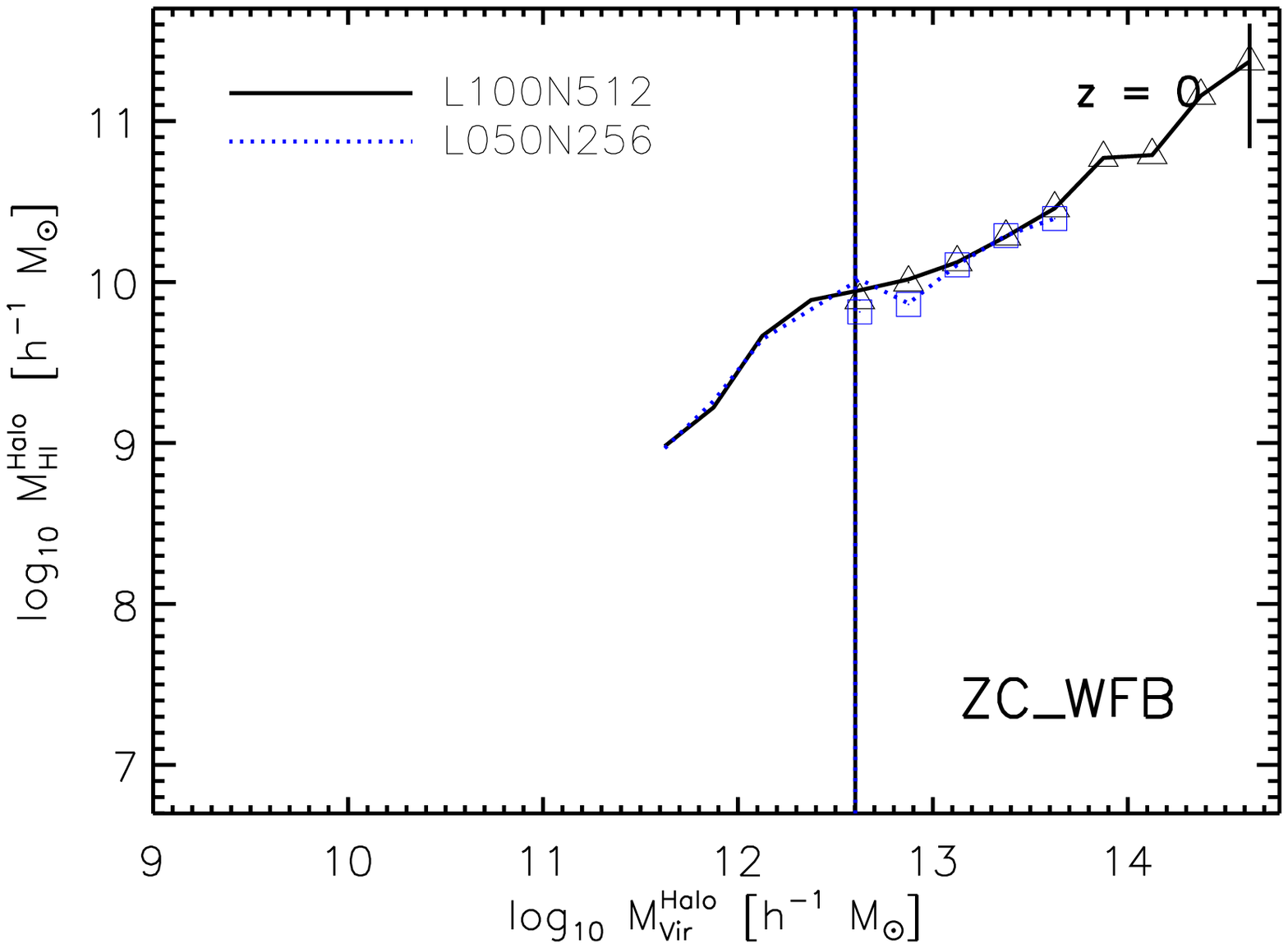, scale=0.45}
  \epsfig{figure=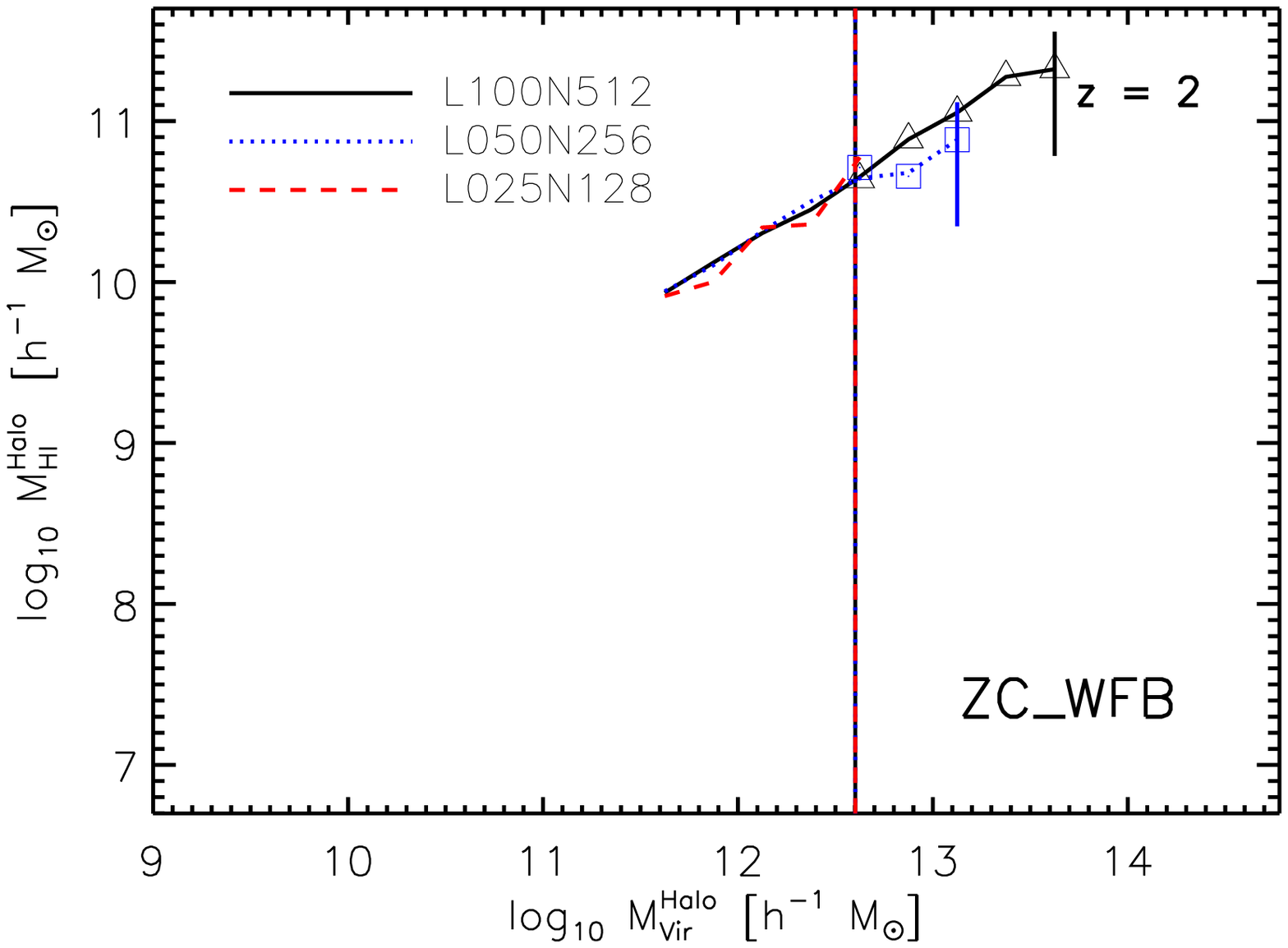, scale=0.45}
  \caption{
  As Fig.~\ref{fig:restest_box_m_mfof_z2} but comparing different box sizes (25, 50 \& 100 $\hMpc$,
where available)  at fixed resolution (where available) at $z=0$ and 2.} 
  \label{fig:sameres_np_m_mfof}
\end{figure}

We also compare the same relation for different 
simulation volumes at the same resolution in Fig.~\ref{fig:sameres_np_m_mfof}. 
These demonstrate that the results are insensitive to the total volume 
simulated.

\subsection{HI mass function}
\label{appendix:restest_himassfn}

Given our established lower HI mass limits, we now investigate the range of masses
over which we can trust the HI mass function.
To that end we show the HI mass functions for the simulations (going down an order
of magnitude in mass) in Fig.~\ref{fig:restest_z0_hi}, illustrating our 
minimum HI mass cut-off as a vertical line.
For comparison we also overplot the datapoints from P09; the resolution of their
simulation (with $256^3$ dark matter particles in a $32 \hMpc$ box) lies in between
our $100\hMpc$ and $50\hMpc$ simulations with $512^3$ dark matter particles.

\begin{figure}
  \epsfysize=2in \epsfxsize=4in
   \epsfig{figure=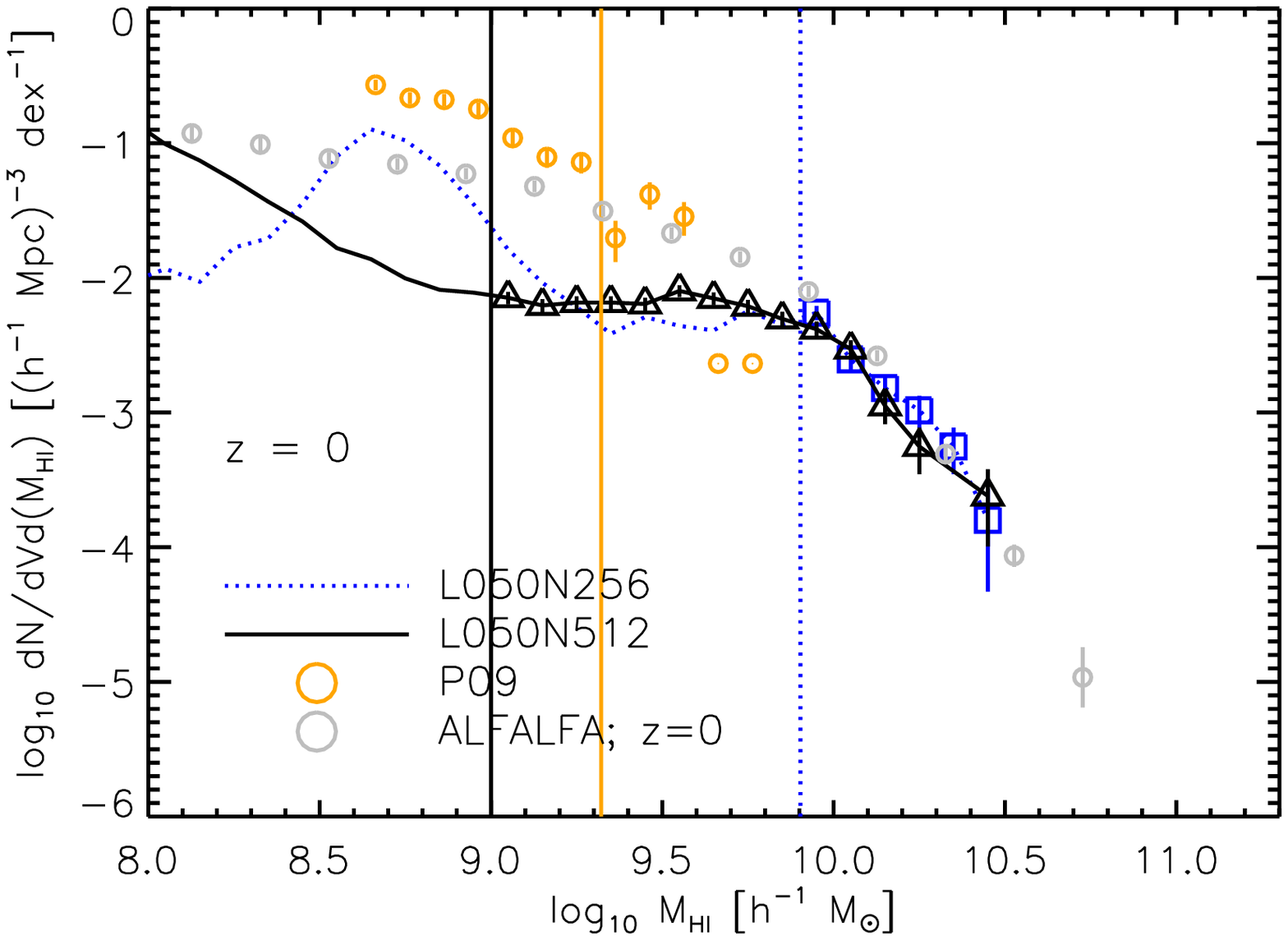, scale=0.45}
  \epsfig{figure=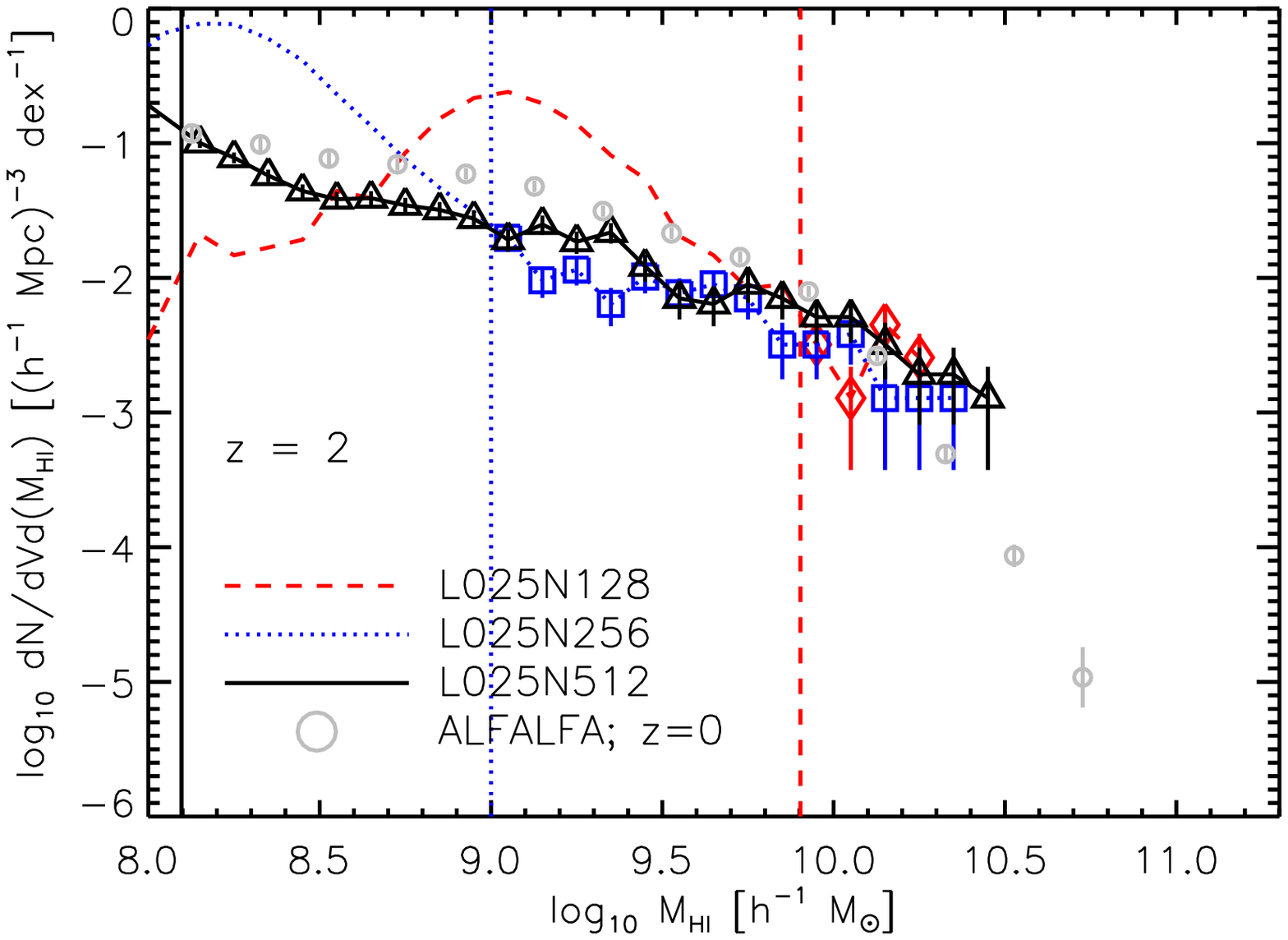, scale=0.45} 
  \caption{Effects of resolution on the HI mass function for the 50 $\hMpc$ boxes at
  $z=0$ and the 25 $\hMpc$ boxes at $z=2$ (top and bottom rows respectively). These results span 
  a factor of 64 in particle mass between the lowest
  and highest resolutions. Lower HI mass limits are shown as vertical lines, above which the
  abundance of haloes is reasonably well converged (Poisson errors in abundance
  are assumed). The mass function starts to rises more steeply below the resolution limit, suggesting
  a numerical artefact. We have also included the $z=0$ data-points from~\citet{Popping:09}
  in the top panel. We note that our HI mass limit applied to their simulation (orange line) suggests 
  that their results may suffer from numerical effects.}
  \label{fig:restest_z0_hi}
\end{figure}

It is clear that our mass function is reasonably well converged above our HI mass limits;
below these limits there is an artificial rise in the number of haloes, an effect that may be present
in the results of P09 (also shown in the figure) who study the mass function down to $3.5 \times 10^{8} \hMsol$;
we would estimate their resolution limit to be cut at masses 40\% higher, close to where their simulations exhibit
the characteristic rapid rise in the HI mass function.
We have also checked the effects of increasing volume at fixed resolution and find excellent convergence between the
simulations.

\subsection{Molecular Hydrogen}
\label{appendix:restest_htwomassfn}

The molecular hydrogen in the simulations comes from higher density regions in galaxies than
the HI and therefore may be expected to be a more demanding quantity to resolve. Again, by comparing the
molecular mass from simulations with different resolution but of the same volume,
we can determine a resolution-dependent mass limit. As expected, our H$_2$ mass limit is somewhat higher
(by around an order of magnitude) than the HI mass limit, as given in Equation~\ref{eqn:hires}. 

In Fig.~\ref{fig:restest_z2_htwo} we show the molecular H$_2$ mass functions for simulations
with increasing resolution, in the $50\hMpc$ box at $z=0$ (top panel) and the $25\hMpc$ box at $z=2$
(bottom panel). Vertical lines denote the resolution limits for the simulations. The overlapping curves 
demonstrate a reasonably well converged result, although it is clear that the mass resolution is 
extremely demanding. The $z=0$ results from P09 are also shown (we estimate their limit is $5 \times 10^{9} \hMsol$,
they choose a limit 75 times less than this).

\begin{figure}
  \epsfysize=2in \epsfxsize=4in
  \epsfig{figure=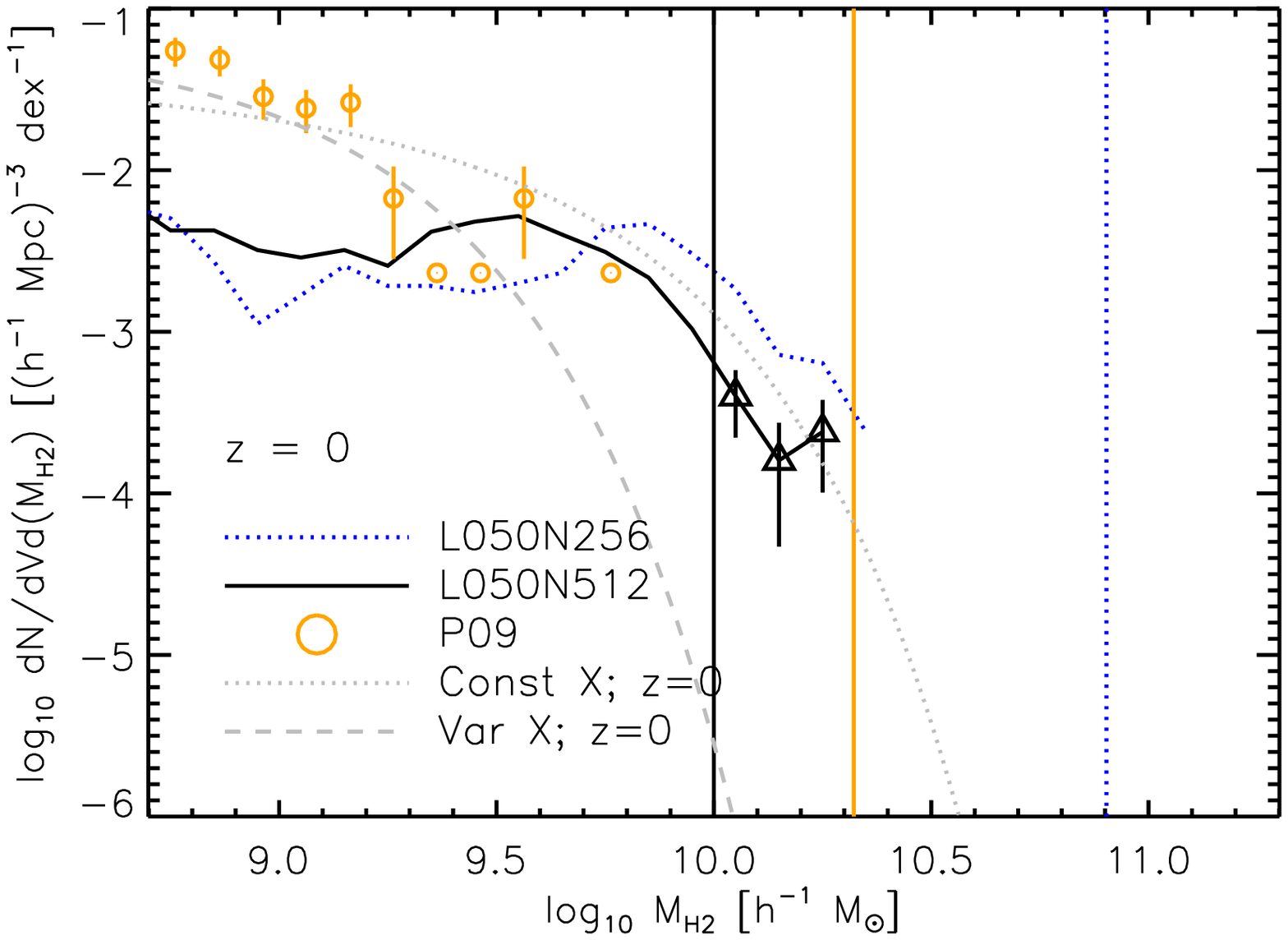, scale=0.45}
  \epsfig{figure=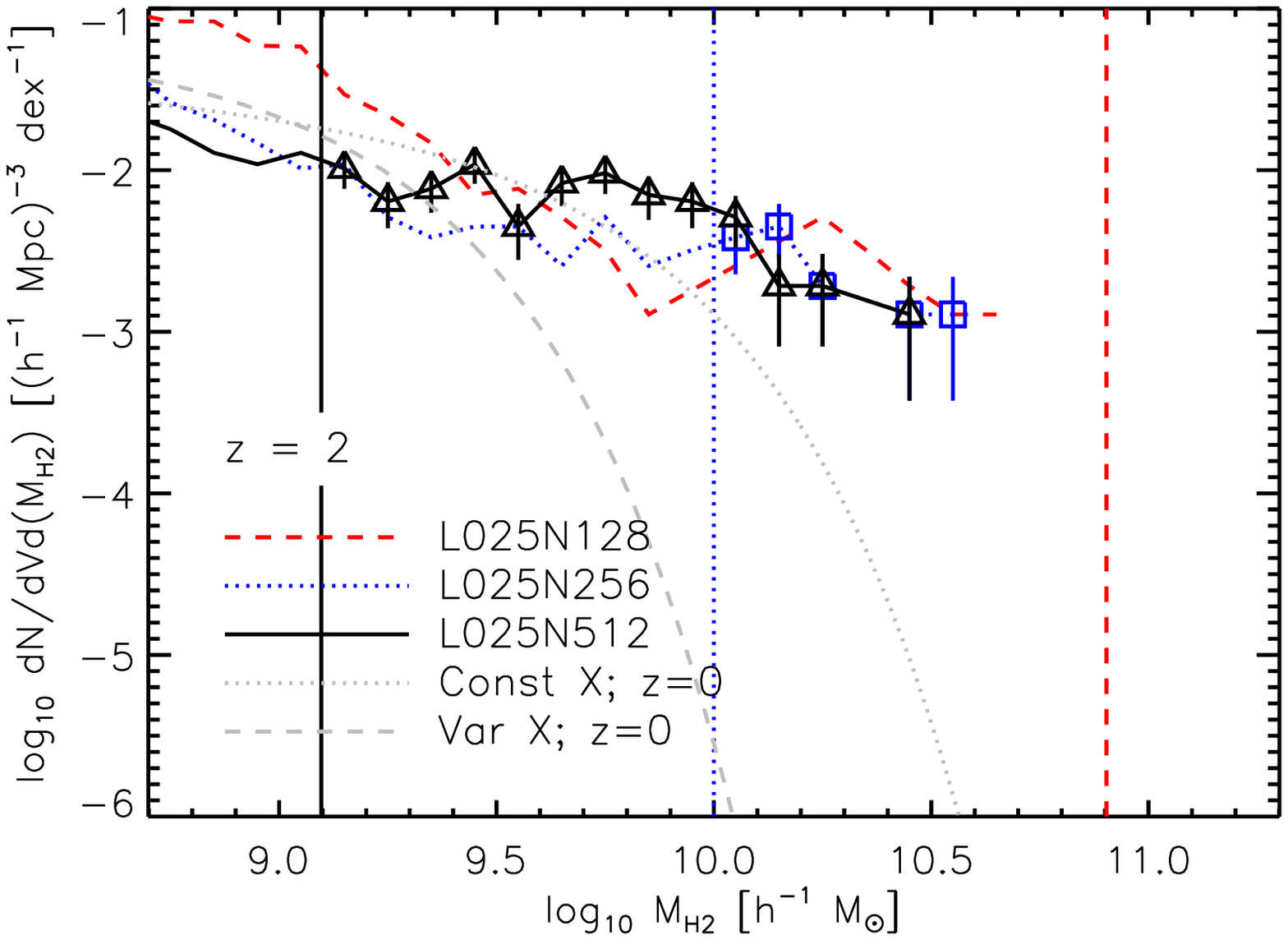, scale=0.45}
  \caption{Molecular $\rm H_{2}$ mass functions for simulations with different resolution.
  	The top panel is for the $50\hMpc$ simulation box at $z=0$ and the
    bottom panel is for the smaller $25 \hMpc$ box at $z=2$. Resolution limits are shown as vertical lines and
    we assume a Poisson error for each data-point. We have
    also included the $z=0$ data-points from~\citet{Popping:09} in the top panel. As in Fig.~\ref{fig:restest_z0_hi},
    our resolution limit applied to their simulation (orange line) suggests that their results may suffer from numerical effects.
    The grey curves are the $z=0$ $\rm H_{2}$ mass function results from the CO mass function of~\citet{Keres:03} 
    when using a constant ($X$-factor) conversion or the varying scheme proposed by~\citet{Obreschkow:09a} 
    (denoted with the dotted and dashed lines, respectively).}
  \label{fig:restest_z2_htwo}
\end{figure}

Finally, we have also tested the sensitivity of the molecular mass function to the effects of large-scale structure
by reducing the simulation volume at constant mass resolution. As with the HI case, there appears to be 
excellent agreement between the overlapping box sizes.

\label{lastpage}
\end{document}

%% file: jdefs.tex
\def\aj{AJ}					% Astronomical Journal
\def\araa{ARA\&A}				% Annual Review of Astron and Astrophys
\def\apj{ApJ}					% Astrophysical Journal
\def\apjl{ApJL}					% Astrophysical Journal, Letters
\def\apjs{ApJS}					% Astrophysical Journal, Supplement
\def\aap{A\&A}					% Astronomy and Astrophysics
\def\mnras{MNRAS}				% Monthly Notices of the RAS
\def\pasp{PASP}					% Publications of the ASP
\def\nat{Nature}				% Nature